\documentclass[footinbib,a4paper,aps,superscriptaddress,reprint,twocolumn,preprintnumbers,amsmath,amssymb,nobalancelastpage,10pt]{revtex4-2}

\usepackage{amsmath}
\usepackage{verbatim}
\usepackage{graphicx}
\usepackage{color}
\usepackage{braket}
\usepackage{yfonts}
\usepackage{soul}
\usepackage[normalem]{ulem}
\usepackage{dsfont}
\usepackage[colorlinks=true,citecolor=blue,linkcolor=blue,urlcolor=blue]{hyperref}

\usepackage{qcircuit}

\begin{document}
	
	\title{Hardware efficient quantum simulation of non-abelian gauge theories \\with qudits on Rydberg platforms}

	\author{Daniel Gonz\'{a}lez-Cuadra}
	\altaffiliation{These authors contributed equally\\
		daniel.gonzalez-cuadra@uibk.ac.at\\
		torsten.zache@uibk.ac.at}
	\affiliation{Institute for Theoretical Physics, University of Innsbruck, 6020 Innsbruck, Austria}
	\affiliation{Institute for Quantum Optics and Quantum Information of the Austrian Academy of Sciences,
		6020 Innsbruck, Austria}
	\author{Torsten V. Zache}
	\altaffiliation{These authors contributed equally\\
		daniel.gonzalez-cuadra@uibk.ac.at\\
		torsten.zache@uibk.ac.at}
	\affiliation{Institute for Theoretical Physics, University of Innsbruck, 6020 Innsbruck, Austria}
	\affiliation{Institute for Quantum Optics and Quantum Information of the Austrian Academy of Sciences,
		6020 Innsbruck, Austria}
	\author{Jose Carrasco}
	\affiliation{Institute for Theoretical Physics, University of Innsbruck, 6020 Innsbruck, Austria}
	\author{Barbara Kraus}
	\affiliation{Institute for Theoretical Physics, University of Innsbruck, 6020 Innsbruck, Austria}
	\author{Peter Zoller}
	\affiliation{Institute for Theoretical Physics, University of Innsbruck, 6020 Innsbruck, Austria}
	\affiliation{Institute for Quantum Optics and Quantum Information of the Austrian Academy of Sciences, 6020 Innsbruck, Austria}
	
	\begin{abstract}
		Non-abelian gauge theories underlie our understanding of fundamental forces in nature, and developing tailored quantum hardware and algorithms to simulate them is an outstanding challenge in the rapidly evolving field of quantum simulation. Here we take an approach where gauge fields, discretized in spacetime, are represented by qudits and are time-evolved in Trotter steps with multi-qudit quantum gates. This maps naturally and hardware-efficiently to an architecture based on Rydberg tweezer arrays, where long-lived internal atomic states represent qudits, and the required quantum gates are performed as holonomic operations supported by a Rydberg blockade mechanism. We illustrate our proposal for a minimal digitization of SU($2$) gauge fields, demonstrating a significant reduction in circuit depth and gate errors in comparison to a traditional qubit-based approach, which puts simulations of non-abelian gauge theories within reach of NISQ devices.
	\end{abstract}
	
	\maketitle

	\paragraph*{Introduction.--}
	
	Quantum field theories form the backbone of the Standard Model of particle physics, where quantized gauge fields mediate the interactions between fundamental particles~\cite{weinberg1995quantum}.
	Lattice gauge theories (LGTs), where fields are discretized on a space-time lattice~\cite{montvay1997quantum},
	provide a convenient framework to study non-perturbative high-energy phenomena, and have been extensively used to extract numerous experimentally relevant predictions~\cite{Aoki_2019}. Despite this success, standard approaches based on Monte Carlo methods are severely limited by the sign problem~\cite{Troyer_2005}, preventing the study of real-time gauge theory dynamics, among other drawbacks. The latter are essential to analyze experimental results in heavy-ion colliders, where open problems in particle physics are currently being addressed~\cite{Brambilla_2014,berges2021qcd}, including the search of new physics beyond the Standard Model.

	\begin{figure}[ht!]
		\centering
		\includegraphics[width=0.95\linewidth]{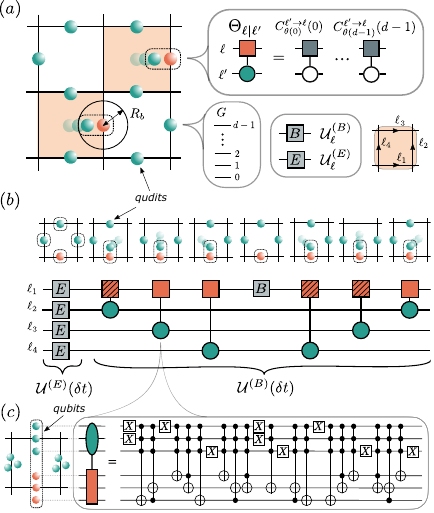}
		\caption{{\bf 
				Gauge field dynamics on qudit vs.~qubit quantum simulator:} (a) Our proposal employs Rydberg atoms trapped in optical tweezers, arranged on the links $\ell$ of a hypercubic lattice. Each atom encodes a qudit using $d$ internal levels,
			where single-qudit gates are realized holonomically.
			To implement the entangling two-qudit gate $\Theta_{\ell|\ell'}$ we first bring pairs of atoms within the Rydberg blockade radius $R_{\rm b}$.
			(b) First order decomposition of a Trotter step, including the four-qudit plaquette interaction, into the native atomic gates $\mathcal{U}^{(E/B)}_\ell$ and $\Theta_{\ell|\ell'}$. (c) For comparison, we show a qubit-based circuit decomposition of $\Theta_{\ell|\ell'}$ for the $Q_8$ group,
			where the number of required atoms is increased by a factor ${\rm log}_2\,8 = 3$, leading to much lower gate fidelities, while our qudit-approach enables a faithful simulation [see Figs.~\ref{fig:fig2}(c) and (d)].}
		\label{fig:fig1}
	\end{figure}
	
	In the recent years, quantum simulators (QS)~\cite{Feynman_1982} have emerged as a promising pathway to circumvent these problems~\cite{Wiese_2013, Zohar_2015, Dalmonte_2016, Banuls_2020, Aidelsburger_2022, Zohar_2022}, leading to several experimental demonstrations where simple LGTs were investigated using digital, analog and variational methods~\cite{Martinez_2016, Schweizer_2019, Kokail_2019, Mil_2020, Yang_2020, Zhou_2021, Nguyen_2021}. For digital QS~\cite{klco2022standard}, in particular, different schemes have been proposed to address high-dimensional non-abelian gauge theories using different platforms, including trapped ions~\cite{Muschik_2017, Paulson_2021, Davoudi_2021a}, ultracold atoms~\cite{Tagliacozzo_2013a,Tagliacozzo_2013b,Zohar_2017a,Zohar_2017b,Bender_2018}, superconducting circuits~\cite{Mezzacapo_2015, Klco_2018, Atas_2021} and cavities~\cite{Armon_2021}. Despite their higher flexibility to simulate complex many-body Hamiltonians compared to the analog approach, crucial in particular for non-abelian theories, a full digital quantum simulation requires access to gate-based quantum computers, which are currently restricted to Noisy Intermediate Scale Quantum (NISQ) devices~\cite{Preskill_2018}, limited in qubit number and circuit depths. Although an impressive effort is currently taking place to reduce the computational complexity using improved quantum software~\cite{Byrnes_2006, Lamm_2019, Alexandru_2019, Ji_2020, Mathis_2020, Kaplan_2020, Brower_2020, Shaw_2020, Kclo_2020, Kclo_2021, Alexandru_2021, Haase_2021, Bauer_2021, Kan_2021, Davoudi_2021b}, simulating relevant LGTs in the NISQ era must be complemented by the development of efficient quantum hardware tailored to the specific algorithmic demands.

	In this Letter, we introduce a qudit architecture based on atoms trapped in optical tweezer arrays and laser excited to Rydberg states~\cite{Kaufman_2021,Saffman_2016,Henriet_2020,Bluvstein_2021,PRXQuantum.2.030322,PhysRevX.11.021036,Madjarov_2020} (Fig.~\ref{fig:fig1}). We co-design the platform to match the requirements to digitally simulate real-time dynamics of non-abelian gauge theories in a hardware-efficient manner. In particular, we show a considerable reduction of experimental resources compared to qubit-based approaches due to a more natural match between the simulating and the simulated degrees of freedom, preserving the local structure of gauge-invariant interactions. 
	
	Although qudit-based quantum simulators can also be implemented with other platforms such as ultracold mixtures~\cite{Kasper_2021}, trapped ions~\cite{Ringbauer_2021} and photonic circuits~\cite{Chi_2022}, multi-dimensional tweezer arrays, both dynamically reconfigurable and locally addressable~\cite{Levine_2019, Madjarov_2020, Ebadi_2021, Scholl_2021, Bluvstein_2021, Young_2022}, satisfy the scalability requirements necessary to address the continuum limit of LGTs. Specifically, here we consider multi-level atoms to encode large gauge-field Hilbert spaces using long-lived qudits~\cite{Wang_2020, Gustafson_2021,Gustafson_2022}. Employing a Rydberg blockade mechanism~\cite{Saffman_2010, Saffman_2016, Henriet_2020}, we develop a native set of holonomic gates~\cite{Zanardi_1999, Sjoqvist_2012, Xu_2012, Feng_2013, Kang_2018}, robust against decoherence, that allows to efficiently simulate the time evolution under a general LGT Hamiltonian. In particular, we show how for the simplest non-trivial digitization of ${\rm SU}(2)$ gauge fields, and including relevant error sources, our qudit approach achieves higher fidelities than a traditional qubit protocol~\footnote{Note that the advantage of using qudits has also been recently demonstrated in the context QAOA \cite{Weggemans2022solvingcorrelation}.}, enabling the quantum simulation of non-abelian gauge theories using NISQ devices~\cite{Bluvstein_2021}. Finally, we note that although larger qudit sizes are required to properly approximate the physics of SU($2$) as relevant for high-energy physics~\cite{Alexandru_2019}, this minimal protocol can readily be employed to study condensed matter systems with non-abelian topological order~\cite{Levin_2005, Xu_2012_2}.
	
	\paragraph*{Qudit quantum computing with Rydberg atoms.--} Atomic systems offer the possibility to encode quantum information in internal states. Here, we go beyond the paradigmatic model of a two-level atomic qubit and consider a collection of $N$ multi-level atoms in state-independent optical traps. For every single atom, we propose to encode a \emph{qudit} with corresponding Hilbert space spanned by $|j\rangle$, $j=0, \dots, d-1$ in $d$ long-lived hyperfine ground states $|F,m_F\rangle$, where large hyperfine manifolds can be accessed e.g. using Erbium~\cite{Trautmann_2021} or Holmium~\cite{Saffman_2008, Hostetter_2015}. In a qubit-based approach, an equivalent Hilbert space of dimension $d^N$ requires control over $N \log_2(d)$ instead of only $N$ atoms as in our case (Fig.~\ref{fig:fig1}(c)). This saving of physical resources is crucial for efficient near-future applications in the NISQ era, since the number of atoms that can be trapped and controlled is limited. For instance, a $4 \times 4$ 2D lattice containing $32$ $d =8$ qudits (which will become relevant later) could be encoded using $32$ atoms, a number that is already available~\cite{Bluvstein_2021}, while almost 100 of them should be used instead in a qubit-based protocol.
	
\begin{figure}
		\centering
		\includegraphics[width=1.0\linewidth]{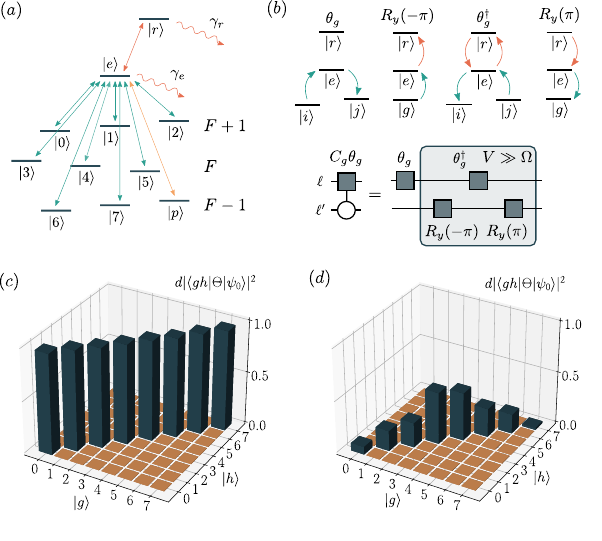}
		\caption{{\bf Qudit enconding and native gates:} (a) Atomic level structure serving as a $Q_8$-register, where the $d = 8$ group basis elements are encoded into different hyperfine manifolds. Single and two-qudit gates are performed holonomically using the auxiliary ground-state level $\ket{p}$ and excited states $\ket{e}$ and $\ket{r}$, with corresponding decay rates $\gamma_{e/r}$. (b) Controlled-permutation gate $C^{\ell'\rightarrow \ell}_{\theta(g)}(g)$ as a sequence of four single qudit-gates in the blockade regime. Quantified by the maximally-entangled state obtained after applying $\Theta_{\ell|\ell^\prime}$ to $\ket{\Psi_0}$, implemented through the (c) qudit and (d) qubit protocols [see the circuit in Fig.~\eqref{fig:fig1}(c)] using the same experimental parameters (see main text), we obtain a state fidelity of $99.6\%$ and $21.4\%$, respectively, demonstrating the clear advantage of a qudit-based decomposition.}
		\label{fig:fig2}
	\end{figure}

	The quantum information stored in each atomic qudit can be efficiently manipulated, e.g., using holonomic operations~\cite{Zanardi_1999,PhysRevA.103.052605}, where arbitrary single-qudit gates $U\in $ SU($d$) can be synthesized via an appropriate sequence of laser pulses, with time-dependent Rabi frequencies $\Omega(t)$. To see this, we first decompose $U$ into a product of at most $d(d-1)/2$ unitaries acting non-trivially only on two atomic levels $(i,j)$~\cite{Li_2013}, and subsequently realize the two-level unitaries via at most three rotations around the $x$ or $y$ axis, denoted by $R^{(i,j)}_{x(y)}(\varphi)$, respectively, utilizing an auxiliary state $|e\rangle$ [see the Supplemental Material (SM)~\cite{SM} for details]. The total time required to implement a general single-qudit gate is upper bounded by $3d(d-1)T/2$, where $T$ is the duration of a laser pulse, estimated below for realistic experimental parameters. We note that the explicit implementation of this scheme should be guided by the constraints imposed by atomic selection rules, see Fig.~\ref{fig:fig2}(a) for an example with $d = 8$.
	
	The main error sources affecting the fidelity of single-qudit gates are the spontaneous decay from the excited state $\ket{e}$ [Fig.~\eqref{fig:fig2}(a)] as well as non-adiabatic state transfer. As we show in \cite{SM}, both can be made negligible by imposing $\Omega \gg 1 / T$, $\gamma_e$, which can be achieved in current experiments. The fast and high fidelity single-qudit gates obtained with our protocol~\cite{SM} contrast with those achievable in a qubit-based approach, where $\mathcal{O}(d^2)$ entangling CNOT gates between {${\rm log}\,d$} qubits are required, leading to larger errors and longer implementation times.
	
	We now come to the main challenge of qudit quantum computing and introduce a new protocol for a general class of controlled-unitary operation, $C_U(j_0) = U \otimes |j_0\rangle \langle j_0|  +\mathds{1} \otimes (\mathds{1}-\ket{j_0}\bra{j_0})$ with $j_0 \in \{0, \dots d-1\}$. Our proposal is based on the Rydberg blockade mechanism~\cite{Saffman_2010, Saffman_2016, Henriet_2020}, which prohibits the simultaneous excitation of two atoms to the Rydberg state $|r\rangle$ when their distance $R$ is below the blockade radius, $ R_\text{b}$, set by the large Rydberg interaction $V \sim 1/R^6 \gg \Omega$.
	This idea gives rise to the following protocol. First, bring the control and target atoms within range $R < R_\text{b}$ and apply the unitary $U$ on target (as outlined above). Now excite the control qudit from $|j_0\rangle$ to the Rydberg state $|r\rangle$ using $S_{(j_0,r)} \equiv R^{(j_0,r)}_y(\pi)$, and subsequently realize $U^\dagger$ on the target by decomposing every two-level rotation as $R^{(i,j)}_{x/y}(\varphi) = S^\dagger_{(i,r)}R^{(j,r)}_{x/y}(\varphi)S_{(i,r)}$ [Fig.~\eqref{fig:fig2}(b)]. To emphasize the involvement of the Rydberg state we denote this gate by $U^\dagger_r$. Finally, applying $S_{j_0,r}^\dagger$ to the control qudit maps the state $|r\rangle$ back to $|j_0\rangle$. Due to the Rydberg blockade, which projects onto the subspace orthogonal to the state $|rr\rangle$ (both atoms in the Rydberg state), this protocol realizes the operator 
	\begin{align}
		(\mathds{1} \otimes S_{(j_0, r)}^\dagger)\left[\mathcal{P}(U^\dagger_r\otimes \mathds{1})\mathcal{P}\right](\mathds{1} \otimes S_{(j_0, r)})(U \otimes \mathds{1} ),
	\end{align}
	with $\mathcal{P} = \mathds{1} - |rr\rangle \langle rr |$. If this operator acts on states where the Rydberg states are not populated, it coincides with $C_U(j_0)$. That is, $U$ is applied to the target qudit \emph{if and only if} the control qudit was in state $|j_0\rangle$. The execution time of this gate is upper bounded by $\left(6d(d-1)+2\right)T$. Apart from the error sources mentioned above, imperfect Rydberg blockade and decay from the Rydberg state can lower the fidelity of controlled-unitary gates. Below we discuss how the latter affect the simulation of non-abelian LGTs for a minimal example with $d = 8$.
	
	In summary, we have established a gate set $\mathcal{G} = \{U,\,C_U(j_0)\}$ consisting of arbitrary single-qudit gates $U\in SU(d)$ and controlled two-qudit gates $C_U(j_0)\in SU(d^2)$. These are naturally available in a qudit register consisting of multi-level Rydberg atoms in a programmable tweezer array. The universal properties of $\mathcal{G}$ for general qudit-based algorithms will be discussed elsewhere~\cite{Carrasco_2022}; here we are specifically interested in the simulation of LGTs as discussed in the following section.
	
	\paragraph*{Gauge field dynamics on a qudit quantum computer.--} In this section, we turn our attention to the digital simulation of real-time gauge theory dynamics with qudits. Most importantly, we will show that the architecture outlined above provides exactly those resources which are required to simulate LGTs on a quantum device. To see this, we take the Hamiltonian lattice approach~\cite{kogut1975hamiltonian}. Given a gauge group $G$ and a hypercubic lattice with $N_\ell$ links $\ell$, we represent the state $|\psi (t)\rangle$ at time $t$ on a so-called $G$-register~\cite{Lamm_2019}, $\ket{\psi(t)} = \sum_{\boldsymbol{g}} \psi_t(\boldsymbol{g})\ket{
		\boldsymbol{g}}$, with $|\boldsymbol{g}\rangle = |g_1\rangle_1 \otimes |g_2\rangle_2 \otimes \dots =  \bigotimes_\ell |g_\ell\rangle_\ell$. Here, every $|g\rangle$ denotes a state labelled by a group element $g\in G$, and the set $\{|g\rangle\}$ forms an orthonormal basis of the local (link) Hilbert space $\mathcal{H}_G$.
	For the relevant cases of $G = $ U($1$) or SU($N$), where $\mathcal{H}_G$ is infinite-dimensional, we replace $G$ with a large, but finite subgroup of itself, which leads to $\mathcal{H}_G \simeq \mathbb{C}^{|G|}$, with dimension given by the order of the group $|G|$, and thus effectively digitizes the many-body wave-function $\psi_t(\boldsymbol{g})$~\cite{Alexandru_2019}. The state $|\psi (t)\rangle$ can then be encoded naturally in a set of $N_\ell$ \emph{qudits} by identifying the computational basis $\{|j\rangle\}$ ($j=0, \dots, d-1$) with the group state basis $\{|g\rangle\}$ ($g\in G$), with $|G|=d$ [Fig.~\ref{fig:fig1}(a)].
	
	The main target of this letter is the time-evolution operator $\mathcal{U}_G (t) = e^{-iH_Gt}$ acting on a given initial state $|\psi(0)\rangle$, i.e. we aim to realize the evolution $|\psi(t)\rangle = \mathcal{U} (t)|\psi(0)\rangle$ in a \emph{hardware-efficient} way. For a general Kogut-Susskind-type LGT, this evolution is generated by a Hamiltonian $H_G = \lambda_E H_E + \lambda_B H_B$ with ``electric'' ($E$) and ``magnetic'' contributions ($B$)~\cite{kogut1975hamiltonian},
	\begin{align}
		\label{eq:KS_hamiltonian}
		H_E = \frac{1}{2}\sum_\ell E_\ell^2  \;, && H_B = \sum_\square \left( \mathcal{U}_\square  + \mathcal{U}^\dagger_\square  \right) \;.
	\end{align}
	While the operator $E_\ell^2$ acts non-trivially only on a single link $\ell$, the plaquette operator
	\begin{align}
		\mathcal{U}_\square = \text{tr} \left[U_{\ell_1} U_{\ell_2} U_{\ell_3}^\dagger U_{\ell_4}^\dagger\right]
	\end{align}
	involves the four links $\ell_{i} $ with $i=1,\ldots,4$ of an elementary plaquette $\square$ (Fig.~\ref{fig:fig1}(b)). To simplify notation we omit here and in the following the links on which  $\mathcal{U}_\square$ is acting on. For SU($N$), the $U_\ell$ are $N\times N$ matrices of operators ($N=1$ for U$(1)$), and $\text{tr}\left[..\right]$ denotes the corresponding trace in $N$ dimensions~\cite{kogut1975hamiltonian}.
	
	We now identify the common challenges in realizing $\mathcal{U}^{(G)}(t)$ for an arbitrary (finite) group $G$.
	In our digital approach, we employ a Trotter decomposition with step size $\delta t$~\cite{Trotter_1959} and error of desired order $\mathcal{O}\left(\delta t^k\right)$~\cite{Hatano_2005}. This reduces the task to realizing the elementary Trotter steps  $\mathcal{U}^{(E/B)}(\delta t) = e^{-i\lambda_{E/B} H_{E/B} \delta t}$, as e.g. $\mathcal{U}^{(G)}(t) = \left(\mathcal{U}^{(E)}(\delta t)\mathcal{U}^{(B)}(\delta t)\right)^{t/\delta t} + \mathcal{O}(\delta t)$ for a first order decomposition. Note that due to the locality of the interactions these steps can be applied in parallel using the local gates $\mathcal{U}^{(E)}_{\ell}(\delta t)= e^{-i\lambda_E E_\ell^2 \delta t}$ and $\mathcal{U}^{(B)}_{\square}(\delta t)= e^{-i\lambda_B \left(\mathcal{U}_\square + \mathcal{U}_\square^\dagger \right) \delta t}$. For any group $G$, the local gates are given by 
	\begin{align}
		\label{eq:group_matrix_elements}
		\mathcal{U}^{(E)}_{\ell}(\delta t)=
		\sum_{h_{\ell}, g_{\ell}\in G}  f^{(E)}(h_{\ell}, g_{\ell},\delta t) \times  |g_{\ell} \rangle_\ell \langle h_{\ell}|
		\;,\\
		\mathcal{U}^{(B)}_{\square}(\delta t)=\sum_{g_{\ell_{1,2,3,4}} \in G}
		f^{(B)}(g_{\ell_1} g_{\ell_2} g^{-1}_{\ell_3}g^{-1}_{\ell_4},\delta t) \times \\ \nonumber
		|g_{\ell_1}, g_{\ell_2},g_{\ell_3}, g_{\ell_4}\rangle \langle g_{\ell_1}, g_{\ell_2}, g_{\ell_3}, g_{\ell_4}| \;,
	\end{align}
	acting trivially on all other links. The group-dependence is encoded in the functions $f^{(E/B)}$ (see SM~\cite{SM} for explicit expressions). Hence, for $|G|=d$,  $\mathcal{U}^{(E)}_{\ell}$ corresponds to a single-qudit gate acting on link $\ell$ while $\mathcal{U}^{(B)}_{\square}$ represents a diagonal four-qudit gate (acting on links $\ell_1,\ldots, \ell_4$).
	
	Let us now decompose the four-qudit diagonal gate,   $\mathcal{U}^{(B)}_{\square}$ into more elementary gates~\cite{Zohar_2017b}.
	We define the two-qudit gate $\Theta_{\ell|\ell'}$, which realizes a controlled group multiplication, by  $\Theta_{\ell|\ell'}|g_\ell\rangle | g_{\ell'}\rangle = |g_\ell g_{\ell'}\rangle | g_{\ell'}\rangle$. As $f^{(B)}$ depends only on the product $g_{\ell_1} g_{\ell_2} g^{-1}_{\ell_3}g^{-1}_{\ell_4} \in G$, $\mathcal{U}^{(B)}_{\square}$ can be implemented by applying $\Theta_{\ell|\ell'}^{(\dagger)}$ between link $\ell_1$ and the other three links followed by the diagonal single-qudit gate $\mathcal{U}^{(B)}_{\ell}(\delta t)| g_\ell \rangle = f^{(B)}(g_\ell,\delta t) | g_\ell \rangle$, and finally undoing the first operations [Fig.~\ref{fig:fig1}(b)]. More explicitly, 
	\begin{equation}
		\label{eq:plaquette_circuit}
		\begin{aligned}
			\mathcal{U}^{(B)}_{\square} = \Theta^\dagger_{\ell_1|\ell_2}  \, \Theta_{\ell_1|\ell_3} \, \Theta_{\ell_1|\ell_4} \, \mathcal{U}^{(B)}_{\ell_1} \, \Theta^\dagger_{\ell_1|\ell_4}  \, \Theta^\dagger_{\ell_1|\ell_3} \, \Theta_{\ell_1|\ell_2} \;.
		\end{aligned}
	\end{equation}
	In summary, it is sufficient to realize the single-qudit gates $\mathcal{U}^{(E/B)}_{\ell}$ and the two-qudit gate $\Theta_{\ell|\ell'}$ for quantum simulating the real-time dynamics of an arbitrary gauge theory. The latter can be further decomposed into a product of controlled-permutation gates [Fig.~\ref{fig:fig1}(a)],
	\begin{align}
		\Theta_{\ell|\ell^\prime} = \sum_{g\in G} \theta_\ell(g) \otimes |g\rangle_{\ell'} \langle g | = \prod_{g\in G} C_{\theta(g)}^{\ell'\rightarrow\ell}(g)\,.
	\end{align}
	Here, $\ell (\ell')$ denotes the control (target) qudit and  $\theta_{\ell}(g)$ is a single-qudit gate implementing the right group multiplication, i.e. $\theta_{\ell}(g)|g_\ell\rangle=|g_\ell g\rangle$, which is just a permutation. This shows that the required gate set reduces to $\{\mathcal{U}_\ell^{(E/B)}, C_{\theta(g)}^{\ell'\rightarrow\ell}(g)\}$, which are precisely the types of gates that are naturally available with the architecture introduced in the previous section. In the spirit of co-design, we have thus identified purpose-made hardware for the digital quantum simulation of LGTs, which is the central result of this letter. Moreover, the possibility of moving the qudits with a programmable tweezer array allows us to perform the required entangling gates in parallel (for example at all even/odd plaquettes in 2D) while avoiding cross-talk from the Rydberg interaction [Fig.~\ref{fig:fig1}(a)], making our protocol scalable in system size.
	
\begin{figure}[t]
		\centering
		\includegraphics[width=1.0\linewidth]{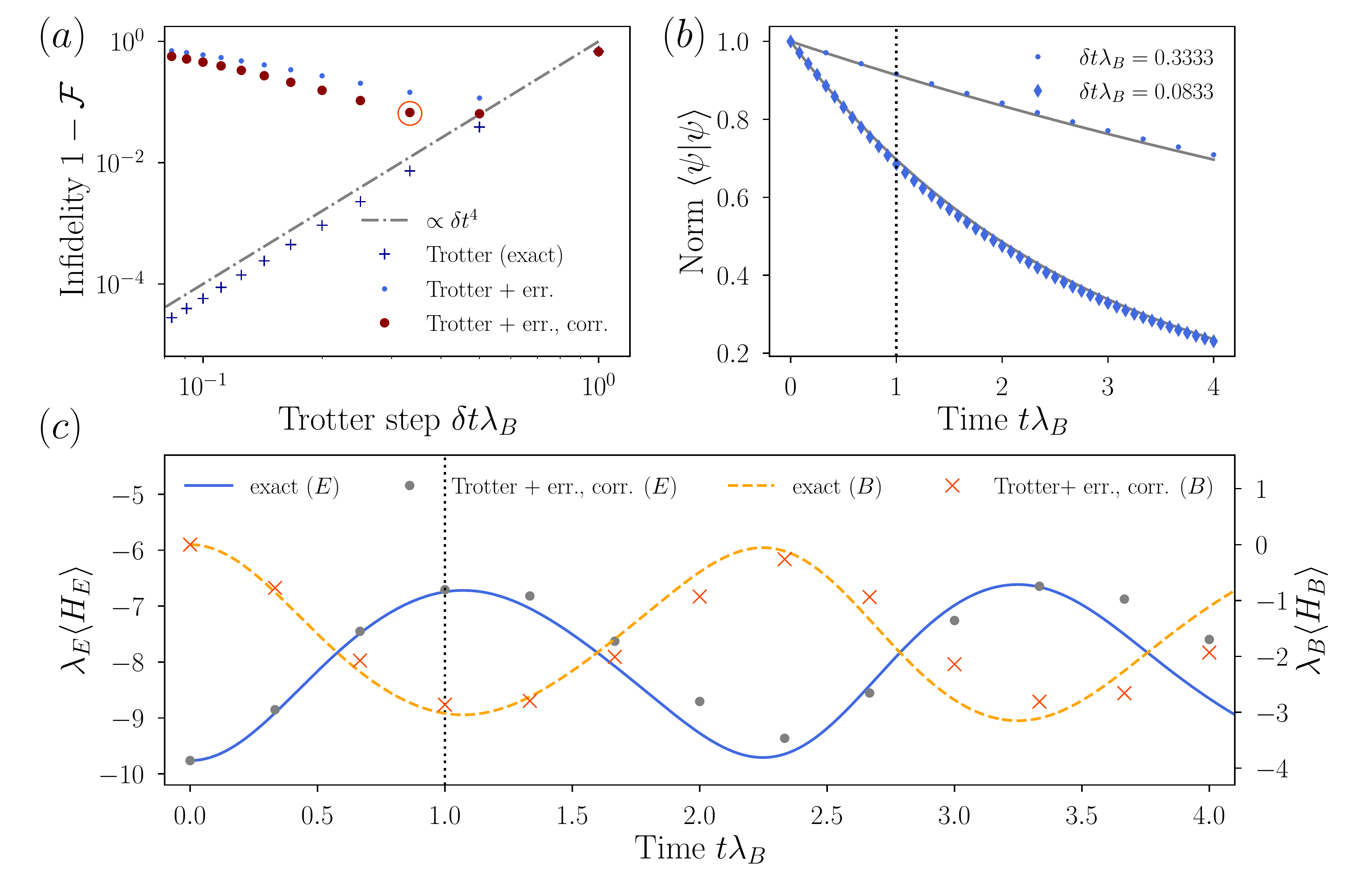}
		\caption{{\bf Fidelity of the simulation:} (a) Infidelity $1-\mathcal{F}$ of the digital simulation at final evolution time $t\lambda_B=1$ as a function of the Trotter step $\delta t\lambda_B$, where an exact second-order Trotter decomposition (crosses) results in the scaling $1-\mathcal{F} \sim (\delta t)^{4}$ (dashed-dotted lines). Including gate errors, the infidelity increases again at small $\delta t \lambda_B$. Correcting for atomic losses due to decoherence, we find that optimal step size $\delta t \lambda_B = 1 / 3$ at the minimum of $1-\mathcal{F} \approx 6.7\%$ (red circle). (b) We measure the losses included in the simulated gates by a decay of the norm $\langle \psi |\psi \rangle$ of the time evolved state. As illustrated for the optimal and a very small Trotter step (symbols), the decay is consistent with a constant loss of $\approx 0.75 \%$ per Trotter step (solid lines).
			(c) Trotterized quench dynamics of a non-abelian $Q_8$ LGT on a single plaquette for $\lambda_E/\lambda_B = 2.88$. The Trotter step corresponds to the ``sweet spot'' indicated in (a). Here, the observables are corrected by a multiplicative time-dependent factor due to the decay shown in (b). }
		\label{fig:fig3}
	\end{figure}

	\paragraph*{Real-time dynamics of $Q_8\subset$ SU($2$).--}
	To be explicit, we now illustrate our approach for the example of the quaternion group $Q_8$, which is the smallest non-abelian subgroup of SU($2$) and requires qudits of size $d=8$~\cite{SM}. Fig.~\ref{fig:fig3}(c) shows Trotter quench dynamics in comparison to the exact result on a single plaquette, demonstrating that the expected interchange between magnetic and electric energies can be observed with a few Trotter steps.
	We now turn to a discussion of the most relevant errors included in this simulation, highlighting the advantage of our proposal in comparison to a traditional qubit-based approach.
	
	Experimental gate errors can be drastically reduced by using qudits instead of qubits, due a substantial reduction of the required entangling operations. We illustrate this fact in Figs.~\ref{fig:fig2}(c) and (d) for the elementary group-multiplication gate $\Theta_{\ell|\ell^\prime}$, where we compare the state fidelity of a maximally-entangled state, $\ket{\psi_1} = 1/\sqrt{d}\sum_g\ket{g}\ket{g}$, prepared from a product state $\ket{\psi_0} = 1/\sqrt{d}\sum_g\ket{0}\ket{g}$ with $\ket{\psi_1} = \Theta \ket{\psi_0}$. Choosing $\Omega T = 3 \cdot 10^2$ , $V/\Omega = 5$ and $\gamma_e / \Omega = \gamma_r / \Omega = 10^{-6}$, we find a state fidelity of $99.6\%$ for the qudit-approach~\footnote{We note that the fidelity of the group-multiplication gate could be further improved through optimal-control methods~\cite{Jandura_2022}, where other experimental imperfections such as phase errors due to Stark shifts could be taken into account.}, in comparison to $21.4\%$ for a qubit-based decomposition (see Fig.~\ref{fig:fig1}(c) and \cite{SM} for details of the employed decompositions~\footnote{Note that, even if the qubit decomposition found is not necessary the optimal one, improved decompositions are not expected to qualitatively change these results.}), demonstrating a clear advantage of the qudits.
	Similarly, the physical time required to implement one Trotter step [Fig.~\ref{fig:fig1}(a)] is drastically reduced by the qudit approach. We estimate a Trotter step time of $\sim 10^3 T$, taking into account a moving velocity below certain threshold to avoid decoherence~\cite{Bluvstein_2021}, and using the structure of the group permutation matrices to reduce the number of pulses required to implement $\Theta_{\ell|\ell^\prime}$ to $2(2d-1)(d-1)$~\cite{SM}. For $\Omega = 2\pi \times 100$ MHz, this leads to a Trotter step time of $\sim 1$ ms in contrast to $\sim 100$~ms~\cite{SM} for a qubit-based approach. In summary, qudits enable the simulation of several Trotter steps within the experimental coherence times of NISQ devices with reasonable fidelity, while an analogous qubit simulation is experimentally unfeasible in the foreseeable future.
	
	In the long run, a faithful quantum simulation of gauge theories in the field theory limit will also require a treatment of systematic errors, such as a finite Trotter step, finite lattice spacing, finite volume and finite subgroup. For brevity, we focus on the finite Trotter step here and briefly comment on other discretization errors in the conclusion. At fixed simulation time $t$, the Trotter error can be systematically reduced by decreasing $\delta t$ at the cost of accumulating experimental gate errors. We quantify this competition for the simulated quantum computation in Fig.~\ref{fig:fig3}(a), where we plot the infidelity $1-\mathcal{F}$ of the evolution as a function of the Trotter step $\delta t$ for $t\lambda_B = 1$. Comparing exact simulations of a second order Trotter decomposition to a simulation including the faulty group-multiplication gate described above, we quantify the overall accuracy by the overlap of the simulated state (Trotter evolved) $|\psi_\text{sim}\rangle$ with the exact result $|\psi_\text{exact}\rangle$, i.e. $\mathcal{F} = |\langle \psi_\text{sim} | \psi_\text{exact}\rangle|^2$. While the power-law behaviour observed for the exact circuit clearly shows the proper convergence of the Trotter expansion, the realistic simulation is limited by a finite decay rate per Trotter step [Fig.~\ref{fig:fig3}(b)]. As a consequence, there is an optimal Trotter step as indicated in Fig.~\ref{fig:fig3}(a), leading to the best overall performance while minimizing the execution time of the simulation.
	
	\paragraph*{Conclusions and outlook.--}
	A prerequisite to quantum simulation of non-abelian LGTs with NISQ devices is hardware efficient encoding and processing with tailored, and scalable quantum hardware. The present work proposes a qudit-based architecture based on atoms stored in tweezer arrays, where single-qudit and entangling gates arising natively in qudit Rydberg-platforms are precisely those required for the simulation of LGTs. We show how our protocol leads to a significantly higher fidelity than a traditional qubit-based approach, which puts non-abelian LGTs within reach of near-term quantum devices. Moreover, the present work can be extended to include dynamical matter~\cite{Gonzalez_2022}, a necessary step towards addressing open questions in and beyond the Standard Model with quantum simulators. Finally, the qudit architecture outlined in this work provides a natural setting for systems with higher spin, for instance for condensed-matter models or quantum chemistry applications~\cite{RevModPhys.92.015003}.

	\paragraph*{Acknowledgments.--} We thank L. Pastori, H. Pichler, T. Olsacher, R. van Bijnen and E. Zohar for valuable discussions. This work was supported by
	the US Air Force Office of Scientific Research (AFOSR)
	via IOE Grant No. FA9550-19-1-7044 LASCEM, the European Union’s Horizon 2020 research and innovation program under Grant Agreement No. 817482 (PASQuanS),
	and by the Simons Collaboration on Ultra-Quantum Matter, which is a grant from the Simons Foundation (651440, P.Z.). JC and BK are grateful for the support of the Austrian Science Fund (FWF): stand alone project P32273-N27 and the SFB BeyondC F~7107-N38.
	
	\bibliography{bibliography}

\begin{thebibliography}{95}%
\makeatletter
\providecommand \@ifxundefined [1]{%
 \@ifx{#1\undefined}
}%
\providecommand \@ifnum [1]{%
 \ifnum #1\expandafter \@firstoftwo
 \else \expandafter \@secondoftwo
 \fi
}%
\providecommand \@ifx [1]{%
 \ifx #1\expandafter \@firstoftwo
 \else \expandafter \@secondoftwo
 \fi
}%
\providecommand \natexlab [1]{#1}%
\providecommand \enquote  [1]{``#1''}%
\providecommand \bibnamefont  [1]{#1}%
\providecommand \bibfnamefont [1]{#1}%
\providecommand \citenamefont [1]{#1}%
\providecommand \href@noop [0]{\@secondoftwo}%
\providecommand \href [0]{\begingroup \@sanitize@url \@href}%
\providecommand \@href[1]{\@@startlink{#1}\@@href}%
\providecommand \@@href[1]{\endgroup#1\@@endlink}%
\providecommand \@sanitize@url [0]{\catcode `\\12\catcode `\$12\catcode
  `\&12\catcode `\#12\catcode `\^12\catcode `\_12\catcode `\%12\relax}%
\providecommand \@@startlink[1]{}%
\providecommand \@@endlink[0]{}%
\providecommand \url  [0]{\begingroup\@sanitize@url \@url }%
\providecommand \@url [1]{\endgroup\@href {#1}{\urlprefix }}%
\providecommand \urlprefix  [0]{URL }%
\providecommand \Eprint [0]{\href }%
\providecommand \doibase [0]{https://doi.org/}%
\providecommand \selectlanguage [0]{\@gobble}%
\providecommand \bibinfo  [0]{\@secondoftwo}%
\providecommand \bibfield  [0]{\@secondoftwo}%
\providecommand \translation [1]{[#1]}%
\providecommand \BibitemOpen [0]{}%
\providecommand \bibitemStop [0]{}%
\providecommand \bibitemNoStop [0]{.\EOS\space}%
\providecommand \EOS [0]{\spacefactor3000\relax}%
\providecommand \BibitemShut  [1]{\csname bibitem#1\endcsname}%
\let\auto@bib@innerbib\@empty
\bibitem [{\citenamefont {Weinberg}(1995)}]{weinberg1995quantum}%
  \BibitemOpen
  \bibfield  {author} {\bibinfo {author} {\bibfnamefont {S.}~\bibnamefont
  {Weinberg}},\ }\href@noop {} {\emph {\bibinfo {title} {The quantum theory of
  fields}}},\ Vol.~\bibinfo {volume} {2}\ (\bibinfo  {publisher} {Cambridge
  university press},\ \bibinfo {year} {1995})\BibitemShut {NoStop}%
\bibitem [{\citenamefont {Montvay}\ and\ \citenamefont
  {M{\"u}nster}(1997)}]{montvay1997quantum}%
  \BibitemOpen
  \bibfield  {author} {\bibinfo {author} {\bibfnamefont {I.}~\bibnamefont
  {Montvay}}\ and\ \bibinfo {author} {\bibfnamefont {G.}~\bibnamefont
  {M{\"u}nster}},\ }\href@noop {} {\emph {\bibinfo {title} {Quantum fields on a
  lattice}}}\ (\bibinfo  {publisher} {Cambridge University Press},\ \bibinfo
  {year} {1997})\BibitemShut {NoStop}%
\bibitem [{\citenamefont {Aoki}\ \emph {et~al.}(2020)\citenamefont {Aoki},
  \citenamefont {Aoki}, \citenamefont {Be{\v c}irevi{\'c}}, \citenamefont
  {Blum}, \citenamefont {Colangelo}, \citenamefont {Collins}, \citenamefont
  {Della~Morte}, \citenamefont {Dimopoulos}, \citenamefont {D{\"u}rr},
  \citenamefont {Fukaya}, \citenamefont {Golterman}, \citenamefont {Gottlieb},
  \citenamefont {Gupta}, \citenamefont {Hashimoto}, \citenamefont {Heller},
  \citenamefont {Herdoiza}, \citenamefont {Horsley}, \citenamefont
  {J{\"u}ttner}, \citenamefont {Kaneko}, \citenamefont {Lin}, \citenamefont
  {Lunghi}, \citenamefont {Mawhinney}, \citenamefont {Nicholson}, \citenamefont
  {Onogi}, \citenamefont {Pena}, \citenamefont {Portelli}, \citenamefont
  {Ramos}, \citenamefont {Sharpe}, \citenamefont {Simone}, \citenamefont
  {Simula}, \citenamefont {Sommer}, \citenamefont {Van~de Water}, \citenamefont
  {Vladikas}, \citenamefont {Wenger},\ and\ \citenamefont
  {Wittig}}]{Aoki_2019}%
  \BibitemOpen
  \bibfield  {author} {\bibinfo {author} {\bibfnamefont {S.}~\bibnamefont
  {Aoki}}, \bibinfo {author} {\bibfnamefont {Y.}~\bibnamefont {Aoki}}, \bibinfo
  {author} {\bibfnamefont {D.}~\bibnamefont {Be{\v c}irevi{\'c}}}, \bibinfo
  {author} {\bibfnamefont {T.}~\bibnamefont {Blum}}, \bibinfo {author}
  {\bibfnamefont {G.}~\bibnamefont {Colangelo}}, \bibinfo {author}
  {\bibfnamefont {S.}~\bibnamefont {Collins}}, \bibinfo {author} {\bibfnamefont
  {M.}~\bibnamefont {Della~Morte}}, \bibinfo {author} {\bibfnamefont
  {P.}~\bibnamefont {Dimopoulos}}, \bibinfo {author} {\bibfnamefont
  {S.}~\bibnamefont {D{\"u}rr}}, \bibinfo {author} {\bibfnamefont
  {H.}~\bibnamefont {Fukaya}}, \bibinfo {author} {\bibfnamefont
  {M.}~\bibnamefont {Golterman}}, \bibinfo {author} {\bibfnamefont
  {S.}~\bibnamefont {Gottlieb}}, \bibinfo {author} {\bibfnamefont
  {R.}~\bibnamefont {Gupta}}, \bibinfo {author} {\bibfnamefont
  {S.}~\bibnamefont {Hashimoto}}, \bibinfo {author} {\bibfnamefont {U.~M.}\
  \bibnamefont {Heller}}, \bibinfo {author} {\bibfnamefont {G.}~\bibnamefont
  {Herdoiza}}, \bibinfo {author} {\bibfnamefont {R.}~\bibnamefont {Horsley}},
  \bibinfo {author} {\bibfnamefont {A.}~\bibnamefont {J{\"u}ttner}}, \bibinfo
  {author} {\bibfnamefont {T.}~\bibnamefont {Kaneko}}, \bibinfo {author}
  {\bibfnamefont {C.~J.~D.}\ \bibnamefont {Lin}}, \bibinfo {author}
  {\bibfnamefont {E.}~\bibnamefont {Lunghi}}, \bibinfo {author} {\bibfnamefont
  {R.}~\bibnamefont {Mawhinney}}, \bibinfo {author} {\bibfnamefont
  {A.}~\bibnamefont {Nicholson}}, \bibinfo {author} {\bibfnamefont
  {T.}~\bibnamefont {Onogi}}, \bibinfo {author} {\bibfnamefont
  {C.}~\bibnamefont {Pena}}, \bibinfo {author} {\bibfnamefont {A.}~\bibnamefont
  {Portelli}}, \bibinfo {author} {\bibfnamefont {A.}~\bibnamefont {Ramos}},
  \bibinfo {author} {\bibfnamefont {S.~R.}\ \bibnamefont {Sharpe}}, \bibinfo
  {author} {\bibfnamefont {J.~N.}\ \bibnamefont {Simone}}, \bibinfo {author}
  {\bibfnamefont {S.}~\bibnamefont {Simula}}, \bibinfo {author} {\bibfnamefont
  {R.}~\bibnamefont {Sommer}}, \bibinfo {author} {\bibfnamefont
  {R.}~\bibnamefont {Van~de Water}}, \bibinfo {author} {\bibfnamefont
  {A.}~\bibnamefont {Vladikas}}, \bibinfo {author} {\bibfnamefont
  {U.}~\bibnamefont {Wenger}},\ and\ \bibinfo {author} {\bibfnamefont
  {H.}~\bibnamefont {Wittig}},\ }\bibfield  {title} {\bibinfo {title} {Flag
  review 2019},\ }\href {https://doi.org/10.1140/epjc/s10052-019-7354-7}
  {\bibfield  {journal} {\bibinfo  {journal} {The European Physical Journal C}\
  }\textbf {\bibinfo {volume} {80}},\ \bibinfo {pages} {113} (\bibinfo {year}
  {2020})}\BibitemShut {NoStop}%
\bibitem [{\citenamefont {Troyer}\ and\ \citenamefont
  {Wiese}(2005)}]{Troyer_2005}%
  \BibitemOpen
  \bibfield  {author} {\bibinfo {author} {\bibfnamefont {M.}~\bibnamefont
  {Troyer}}\ and\ \bibinfo {author} {\bibfnamefont {U.-J.}\ \bibnamefont
  {Wiese}},\ }\bibfield  {title} {\bibinfo {title} {Computational complexity
  and fundamental limitations to fermionic quantum monte carlo simulations},\
  }\href {https://doi.org/10.1103/PhysRevLett.94.170201} {\bibfield  {journal}
  {\bibinfo  {journal} {Phys. Rev. Lett.}\ }\textbf {\bibinfo {volume} {94}},\
  \bibinfo {pages} {170201} (\bibinfo {year} {2005})}\BibitemShut {NoStop}%
\bibitem [{\citenamefont {Brambilla}\ \emph {et~al.}(2014)\citenamefont
  {Brambilla}, \citenamefont {Eidelman}, \citenamefont {Foka}, \citenamefont
  {Gardner}, \citenamefont {Kronfeld}, \citenamefont {Alford}, \citenamefont
  {Alkofer}, \citenamefont {Butenschoen}, \citenamefont {Cohen}, \citenamefont
  {Erdmenger}, \citenamefont {Fabbietti}, \citenamefont {Faber}, \citenamefont
  {Goity}, \citenamefont {Ketzer}, \citenamefont {Lin}, \citenamefont
  {Llanes-Estrada}, \citenamefont {Meyer}, \citenamefont {Pakhlov},
  \citenamefont {Pallante}, \citenamefont {Polikarpov}, \citenamefont
  {Sazdjian}, \citenamefont {Schmitt}, \citenamefont {Snow}, \citenamefont
  {Vairo}, \citenamefont {Vogt}, \citenamefont {Vuorinen}, \citenamefont
  {Wittig}, \citenamefont {Arnold}, \citenamefont {Christakoglou},
  \citenamefont {Di~Nezza}, \citenamefont {Fodor}, \citenamefont {Garcia~i
  Tormo}, \citenamefont {H{\"o}llwieser}, \citenamefont {Janik}, \citenamefont
  {Kalweit}, \citenamefont {Keane}, \citenamefont {Kiritsis}, \citenamefont
  {Mischke}, \citenamefont {Mizuk}, \citenamefont {Odyniec}, \citenamefont
  {Papadodimas}, \citenamefont {Pich}, \citenamefont {Pittau}, \citenamefont
  {Qiu}, \citenamefont {Ricciardi}, \citenamefont {Salgado}, \citenamefont
  {Schwenzer}, \citenamefont {Stefanis}, \citenamefont {von Hippel},\ and\
  \citenamefont {Zakharov}}]{Brambilla_2014}%
  \BibitemOpen
  \bibfield  {author} {\bibinfo {author} {\bibfnamefont {N.}~\bibnamefont
  {Brambilla}}, \bibinfo {author} {\bibfnamefont {S.}~\bibnamefont {Eidelman}},
  \bibinfo {author} {\bibfnamefont {P.}~\bibnamefont {Foka}}, \bibinfo {author}
  {\bibfnamefont {S.}~\bibnamefont {Gardner}}, \bibinfo {author} {\bibfnamefont
  {A.~S.}\ \bibnamefont {Kronfeld}}, \bibinfo {author} {\bibfnamefont {M.~G.}\
  \bibnamefont {Alford}}, \bibinfo {author} {\bibfnamefont {R.}~\bibnamefont
  {Alkofer}}, \bibinfo {author} {\bibfnamefont {M.}~\bibnamefont
  {Butenschoen}}, \bibinfo {author} {\bibfnamefont {T.~D.}\ \bibnamefont
  {Cohen}}, \bibinfo {author} {\bibfnamefont {J.}~\bibnamefont {Erdmenger}},
  \bibinfo {author} {\bibfnamefont {L.}~\bibnamefont {Fabbietti}}, \bibinfo
  {author} {\bibfnamefont {M.}~\bibnamefont {Faber}}, \bibinfo {author}
  {\bibfnamefont {J.~L.}\ \bibnamefont {Goity}}, \bibinfo {author}
  {\bibfnamefont {B.}~\bibnamefont {Ketzer}}, \bibinfo {author} {\bibfnamefont
  {H.~W.}\ \bibnamefont {Lin}}, \bibinfo {author} {\bibfnamefont {F.~J.}\
  \bibnamefont {Llanes-Estrada}}, \bibinfo {author} {\bibfnamefont {H.~B.}\
  \bibnamefont {Meyer}}, \bibinfo {author} {\bibfnamefont {P.}~\bibnamefont
  {Pakhlov}}, \bibinfo {author} {\bibfnamefont {E.}~\bibnamefont {Pallante}},
  \bibinfo {author} {\bibfnamefont {M.~I.}\ \bibnamefont {Polikarpov}},
  \bibinfo {author} {\bibfnamefont {H.}~\bibnamefont {Sazdjian}}, \bibinfo
  {author} {\bibfnamefont {A.}~\bibnamefont {Schmitt}}, \bibinfo {author}
  {\bibfnamefont {W.~M.}\ \bibnamefont {Snow}}, \bibinfo {author}
  {\bibfnamefont {A.}~\bibnamefont {Vairo}}, \bibinfo {author} {\bibfnamefont
  {R.}~\bibnamefont {Vogt}}, \bibinfo {author} {\bibfnamefont {A.}~\bibnamefont
  {Vuorinen}}, \bibinfo {author} {\bibfnamefont {H.}~\bibnamefont {Wittig}},
  \bibinfo {author} {\bibfnamefont {P.}~\bibnamefont {Arnold}}, \bibinfo
  {author} {\bibfnamefont {P.}~\bibnamefont {Christakoglou}}, \bibinfo {author}
  {\bibfnamefont {P.}~\bibnamefont {Di~Nezza}}, \bibinfo {author}
  {\bibfnamefont {Z.}~\bibnamefont {Fodor}}, \bibinfo {author} {\bibfnamefont
  {X.}~\bibnamefont {Garcia~i Tormo}}, \bibinfo {author} {\bibfnamefont
  {R.}~\bibnamefont {H{\"o}llwieser}}, \bibinfo {author} {\bibfnamefont
  {M.~A.}\ \bibnamefont {Janik}}, \bibinfo {author} {\bibfnamefont
  {A.}~\bibnamefont {Kalweit}}, \bibinfo {author} {\bibfnamefont
  {D.}~\bibnamefont {Keane}}, \bibinfo {author} {\bibfnamefont
  {E.}~\bibnamefont {Kiritsis}}, \bibinfo {author} {\bibfnamefont
  {A.}~\bibnamefont {Mischke}}, \bibinfo {author} {\bibfnamefont
  {R.}~\bibnamefont {Mizuk}}, \bibinfo {author} {\bibfnamefont
  {G.}~\bibnamefont {Odyniec}}, \bibinfo {author} {\bibfnamefont
  {K.}~\bibnamefont {Papadodimas}}, \bibinfo {author} {\bibfnamefont
  {A.}~\bibnamefont {Pich}}, \bibinfo {author} {\bibfnamefont {R.}~\bibnamefont
  {Pittau}}, \bibinfo {author} {\bibfnamefont {J.~W.}\ \bibnamefont {Qiu}},
  \bibinfo {author} {\bibfnamefont {G.}~\bibnamefont {Ricciardi}}, \bibinfo
  {author} {\bibfnamefont {C.~A.}\ \bibnamefont {Salgado}}, \bibinfo {author}
  {\bibfnamefont {K.}~\bibnamefont {Schwenzer}}, \bibinfo {author}
  {\bibfnamefont {N.~G.}\ \bibnamefont {Stefanis}}, \bibinfo {author}
  {\bibfnamefont {G.~M.}\ \bibnamefont {von Hippel}},\ and\ \bibinfo {author}
  {\bibfnamefont {V.~I.}\ \bibnamefont {Zakharov}},\ }\bibfield  {title}
  {\bibinfo {title} {Qcd and strongly coupled gauge theories: challenges and
  perspectives},\ }\href {https://doi.org/10.1140/epjc/s10052-014-2981-5}
  {\bibfield  {journal} {\bibinfo  {journal} {The European Physical Journal C}\
  }\textbf {\bibinfo {volume} {74}},\ \bibinfo {pages} {2981} (\bibinfo {year}
  {2014})}\BibitemShut {NoStop}%
\bibitem [{\citenamefont {Berges}\ \emph {et~al.}(2021)\citenamefont {Berges},
  \citenamefont {Heller}, \citenamefont {Mazeliauskas},\ and\ \citenamefont
  {Venugopalan}}]{berges2021qcd}%
  \BibitemOpen
  \bibfield  {author} {\bibinfo {author} {\bibfnamefont {J.}~\bibnamefont
  {Berges}}, \bibinfo {author} {\bibfnamefont {M.~P.}\ \bibnamefont {Heller}},
  \bibinfo {author} {\bibfnamefont {A.}~\bibnamefont {Mazeliauskas}},\ and\
  \bibinfo {author} {\bibfnamefont {R.}~\bibnamefont {Venugopalan}},\
  }\bibfield  {title} {\bibinfo {title} {Qcd thermalization: Ab initio
  approaches and interdisciplinary connections},\ }\href@noop {} {\bibfield
  {journal} {\bibinfo  {journal} {Reviews of Modern Physics}\ }\textbf
  {\bibinfo {volume} {93}},\ \bibinfo {pages} {035003} (\bibinfo {year}
  {2021})}\BibitemShut {NoStop}%
\bibitem [{\citenamefont {Feynman}(1981)}]{Feynman_1982}%
  \BibitemOpen
  \bibfield  {author} {\bibinfo {author} {\bibfnamefont {R.~P.}\ \bibnamefont
  {Feynman}},\ }\bibfield  {title} {\bibinfo {title} {Simulating physics with
  computers},\ }\href {https://doi.org/10.1007/BF02650179} {\bibfield
  {journal} {\bibinfo  {journal} {International Journal of Theoretical
  Physics}\ }\textbf {\bibinfo {volume} {21}},\ \bibinfo {pages} {467}
  (\bibinfo {year} {1981})}\BibitemShut {NoStop}%
\bibitem [{\citenamefont {Wiese}(2013)}]{Wiese_2013}%
  \BibitemOpen
  \bibfield  {author} {\bibinfo {author} {\bibfnamefont {U.-J.}\ \bibnamefont
  {Wiese}},\ }\bibfield  {title} {\bibinfo {title} {Ultracold quantum gases and
  lattice systems: quantum simulation of lattice gauge theories},\ }\href
  {https://doi.org/https://doi.org/10.1002/andp.201300104} {\bibfield
  {journal} {\bibinfo  {journal} {Annalen der Physik}\ }\textbf {\bibinfo
  {volume} {525}},\ \bibinfo {pages} {777} (\bibinfo {year}
  {2013})}\BibitemShut {NoStop}%
\bibitem [{\citenamefont {Zohar}\ \emph {et~al.}(2015)\citenamefont {Zohar},
  \citenamefont {Cirac},\ and\ \citenamefont {Reznik}}]{Zohar_2015}%
  \BibitemOpen
  \bibfield  {author} {\bibinfo {author} {\bibfnamefont {E.}~\bibnamefont
  {Zohar}}, \bibinfo {author} {\bibfnamefont {J.~I.}\ \bibnamefont {Cirac}},\
  and\ \bibinfo {author} {\bibfnamefont {B.}~\bibnamefont {Reznik}},\
  }\bibfield  {title} {\bibinfo {title} {Quantum simulations of lattice gauge
  theories using ultracold atoms in optical lattices},\ }\href
  {https://doi.org/10.1088/0034-4885/79/1/014401} {\bibfield  {journal}
  {\bibinfo  {journal} {Reports on Progress in Physics}\ }\textbf {\bibinfo
  {volume} {79}},\ \bibinfo {pages} {014401} (\bibinfo {year}
  {2015})}\BibitemShut {NoStop}%
\bibitem [{\citenamefont {Dalmonte}\ and\ \citenamefont
  {Montangero}(2016)}]{Dalmonte_2016}%
  \BibitemOpen
  \bibfield  {author} {\bibinfo {author} {\bibfnamefont {M.}~\bibnamefont
  {Dalmonte}}\ and\ \bibinfo {author} {\bibfnamefont {S.}~\bibnamefont
  {Montangero}},\ }\bibfield  {title} {\bibinfo {title} {Lattice gauge theory
  simulations in the quantum information era},\ }\href
  {https://doi.org/10.1080/00107514.2016.1151199} {\bibfield  {journal}
  {\bibinfo  {journal} {Contemporary Physics}\ }\textbf {\bibinfo {volume}
  {57}},\ \bibinfo {pages} {388} (\bibinfo {year} {2016})}\BibitemShut
  {NoStop}%
\bibitem [{\citenamefont {Ba{\~n}uls}\ \emph {et~al.}(2020)\citenamefont
  {Ba{\~n}uls}, \citenamefont {Blatt}, \citenamefont {Catani}, \citenamefont
  {Celi}, \citenamefont {Cirac}, \citenamefont {Dalmonte}, \citenamefont
  {Fallani}, \citenamefont {Jansen}, \citenamefont {Lewenstein}, \citenamefont
  {Montangero}, \citenamefont {Muschik}, \citenamefont {Reznik}, \citenamefont
  {Rico}, \citenamefont {Tagliacozzo}, \citenamefont {Van~Acoleyen},
  \citenamefont {Verstraete}, \citenamefont {Wiese}, \citenamefont {Wingate},
  \citenamefont {Zakrzewski},\ and\ \citenamefont {Zoller}}]{Banuls_2020}%
  \BibitemOpen
  \bibfield  {author} {\bibinfo {author} {\bibfnamefont {M.~C.}\ \bibnamefont
  {Ba{\~n}uls}}, \bibinfo {author} {\bibfnamefont {R.}~\bibnamefont {Blatt}},
  \bibinfo {author} {\bibfnamefont {J.}~\bibnamefont {Catani}}, \bibinfo
  {author} {\bibfnamefont {A.}~\bibnamefont {Celi}}, \bibinfo {author}
  {\bibfnamefont {J.~I.}\ \bibnamefont {Cirac}}, \bibinfo {author}
  {\bibfnamefont {M.}~\bibnamefont {Dalmonte}}, \bibinfo {author}
  {\bibfnamefont {L.}~\bibnamefont {Fallani}}, \bibinfo {author} {\bibfnamefont
  {K.}~\bibnamefont {Jansen}}, \bibinfo {author} {\bibfnamefont
  {M.}~\bibnamefont {Lewenstein}}, \bibinfo {author} {\bibfnamefont
  {S.}~\bibnamefont {Montangero}}, \bibinfo {author} {\bibfnamefont {C.~A.}\
  \bibnamefont {Muschik}}, \bibinfo {author} {\bibfnamefont {B.}~\bibnamefont
  {Reznik}}, \bibinfo {author} {\bibfnamefont {E.}~\bibnamefont {Rico}},
  \bibinfo {author} {\bibfnamefont {L.}~\bibnamefont {Tagliacozzo}}, \bibinfo
  {author} {\bibfnamefont {K.}~\bibnamefont {Van~Acoleyen}}, \bibinfo {author}
  {\bibfnamefont {F.}~\bibnamefont {Verstraete}}, \bibinfo {author}
  {\bibfnamefont {U.-J.}\ \bibnamefont {Wiese}}, \bibinfo {author}
  {\bibfnamefont {M.}~\bibnamefont {Wingate}}, \bibinfo {author} {\bibfnamefont
  {J.}~\bibnamefont {Zakrzewski}},\ and\ \bibinfo {author} {\bibfnamefont
  {P.}~\bibnamefont {Zoller}},\ }\bibfield  {title} {\bibinfo {title}
  {Simulating lattice gauge theories within quantum technologies},\ }\href
  {https://doi.org/10.1140/epjd/e2020-100571-8} {\bibfield  {journal} {\bibinfo
   {journal} {The European Physical Journal D}\ }\textbf {\bibinfo {volume}
  {74}},\ \bibinfo {pages} {165} (\bibinfo {year} {2020})}\BibitemShut
  {NoStop}%
\bibitem [{\citenamefont {Aidelsburger}\ \emph {et~al.}(2022)\citenamefont
  {Aidelsburger}, \citenamefont {Barbiero}, \citenamefont {Bermudez},
  \citenamefont {Chanda}, \citenamefont {Dauphin}, \citenamefont
  {Gonz{\'a}lez-Cuadra}, \citenamefont {Grzybowski}, \citenamefont {Hands},
  \citenamefont {Jendrzejewski}, \citenamefont {J{\"u}nemann}, \citenamefont
  {Juzeli{\=u}nas}, \citenamefont {Kasper}, \citenamefont {Piga}, \citenamefont
  {Ran}, \citenamefont {Rizzi}, \citenamefont {Sierra}, \citenamefont
  {Tagliacozzo}, \citenamefont {Tirrito}, \citenamefont {Zache}, \citenamefont
  {Zakrzewski}, \citenamefont {Zohar},\ and\ \citenamefont
  {Lewenstein}}]{Aidelsburger_2022}%
  \BibitemOpen
  \bibfield  {author} {\bibinfo {author} {\bibfnamefont {M.}~\bibnamefont
  {Aidelsburger}}, \bibinfo {author} {\bibfnamefont {L.}~\bibnamefont
  {Barbiero}}, \bibinfo {author} {\bibfnamefont {A.}~\bibnamefont {Bermudez}},
  \bibinfo {author} {\bibfnamefont {T.}~\bibnamefont {Chanda}}, \bibinfo
  {author} {\bibfnamefont {A.}~\bibnamefont {Dauphin}}, \bibinfo {author}
  {\bibfnamefont {D.}~\bibnamefont {Gonz{\'a}lez-Cuadra}}, \bibinfo {author}
  {\bibfnamefont {P.~R.}\ \bibnamefont {Grzybowski}}, \bibinfo {author}
  {\bibfnamefont {S.}~\bibnamefont {Hands}}, \bibinfo {author} {\bibfnamefont
  {F.}~\bibnamefont {Jendrzejewski}}, \bibinfo {author} {\bibfnamefont
  {J.}~\bibnamefont {J{\"u}nemann}}, \bibinfo {author} {\bibfnamefont
  {G.}~\bibnamefont {Juzeli{\=u}nas}}, \bibinfo {author} {\bibfnamefont
  {V.}~\bibnamefont {Kasper}}, \bibinfo {author} {\bibfnamefont
  {A.}~\bibnamefont {Piga}}, \bibinfo {author} {\bibfnamefont {S.-J.}\
  \bibnamefont {Ran}}, \bibinfo {author} {\bibfnamefont {M.}~\bibnamefont
  {Rizzi}}, \bibinfo {author} {\bibfnamefont {G.}~\bibnamefont {Sierra}},
  \bibinfo {author} {\bibfnamefont {L.}~\bibnamefont {Tagliacozzo}}, \bibinfo
  {author} {\bibfnamefont {E.}~\bibnamefont {Tirrito}}, \bibinfo {author}
  {\bibfnamefont {T.~V.}\ \bibnamefont {Zache}}, \bibinfo {author}
  {\bibfnamefont {J.}~\bibnamefont {Zakrzewski}}, \bibinfo {author}
  {\bibfnamefont {E.}~\bibnamefont {Zohar}},\ and\ \bibinfo {author}
  {\bibfnamefont {M.}~\bibnamefont {Lewenstein}},\ }\bibfield  {title}
  {\bibinfo {title} {Cold atoms meet lattice gauge theory},\ }\href
  {https://doi.org/10.1098/rsta.2021.0064} {\bibfield  {journal} {\bibinfo
  {journal} {Philosophical Transactions of the Royal Society A: Mathematical,
  Physical and Engineering Sciences}\ }\textbf {\bibinfo {volume} {380}},\
  \bibinfo {pages} {20210064} (\bibinfo {year} {2022})}\BibitemShut {NoStop}%
\bibitem [{\citenamefont {Zohar}(2022)}]{Zohar_2022}%
  \BibitemOpen
  \bibfield  {author} {\bibinfo {author} {\bibfnamefont {E.}~\bibnamefont
  {Zohar}},\ }\bibfield  {title} {\bibinfo {title} {Quantum simulation of
  lattice gauge theories in more than one space dimension -- requirements,
  challenges and methods},\ }\href {https://doi.org/10.1098/rsta.2021.0069}
  {\bibfield  {journal} {\bibinfo  {journal} {Philosophical Transactions of the
  Royal Society A: Mathematical, Physical and Engineering Sciences}\ }\textbf
  {\bibinfo {volume} {380}},\ \bibinfo {pages} {20210069} (\bibinfo {year}
  {2022})}\BibitemShut {NoStop}%
\bibitem [{\citenamefont {Martinez}\ \emph {et~al.}(2016)\citenamefont
  {Martinez}, \citenamefont {Muschik}, \citenamefont {Schindler}, \citenamefont
  {Nigg}, \citenamefont {Erhard}, \citenamefont {Heyl}, \citenamefont {Hauke},
  \citenamefont {Dalmonte}, \citenamefont {Monz}, \citenamefont {Zoller},\ and\
  \citenamefont {Blatt}}]{Martinez_2016}%
  \BibitemOpen
  \bibfield  {author} {\bibinfo {author} {\bibfnamefont {E.~A.}\ \bibnamefont
  {Martinez}}, \bibinfo {author} {\bibfnamefont {C.~A.}\ \bibnamefont
  {Muschik}}, \bibinfo {author} {\bibfnamefont {P.}~\bibnamefont {Schindler}},
  \bibinfo {author} {\bibfnamefont {D.}~\bibnamefont {Nigg}}, \bibinfo {author}
  {\bibfnamefont {A.}~\bibnamefont {Erhard}}, \bibinfo {author} {\bibfnamefont
  {M.}~\bibnamefont {Heyl}}, \bibinfo {author} {\bibfnamefont {P.}~\bibnamefont
  {Hauke}}, \bibinfo {author} {\bibfnamefont {M.}~\bibnamefont {Dalmonte}},
  \bibinfo {author} {\bibfnamefont {T.}~\bibnamefont {Monz}}, \bibinfo {author}
  {\bibfnamefont {P.}~\bibnamefont {Zoller}},\ and\ \bibinfo {author}
  {\bibfnamefont {R.}~\bibnamefont {Blatt}},\ }\bibfield  {title} {\bibinfo
  {title} {Real-time dynamics of lattice gauge theories with a few-qubit
  quantum computer},\ }\href {https://doi.org/10.1038/nature18318} {\bibfield
  {journal} {\bibinfo  {journal} {Nature}\ }\textbf {\bibinfo {volume} {534}},\
  \bibinfo {pages} {516} (\bibinfo {year} {2016})}\BibitemShut {NoStop}%
\bibitem [{\citenamefont {Schweizer}\ \emph {et~al.}(2019)\citenamefont
  {Schweizer}, \citenamefont {Grusdt}, \citenamefont {Berngruber},
  \citenamefont {Barbiero}, \citenamefont {Demler}, \citenamefont {Goldman},
  \citenamefont {Bloch},\ and\ \citenamefont {Aidelsburger}}]{Schweizer_2019}%
  \BibitemOpen
  \bibfield  {author} {\bibinfo {author} {\bibfnamefont {C.}~\bibnamefont
  {Schweizer}}, \bibinfo {author} {\bibfnamefont {F.}~\bibnamefont {Grusdt}},
  \bibinfo {author} {\bibfnamefont {M.}~\bibnamefont {Berngruber}}, \bibinfo
  {author} {\bibfnamefont {L.}~\bibnamefont {Barbiero}}, \bibinfo {author}
  {\bibfnamefont {E.}~\bibnamefont {Demler}}, \bibinfo {author} {\bibfnamefont
  {N.}~\bibnamefont {Goldman}}, \bibinfo {author} {\bibfnamefont
  {I.}~\bibnamefont {Bloch}},\ and\ \bibinfo {author} {\bibfnamefont
  {M.}~\bibnamefont {Aidelsburger}},\ }\bibfield  {title} {\bibinfo {title}
  {Floquet approach to $\mathbb{Z}_2$ lattice gauge theories with ultracold
  atoms in optical lattices},\ }\href
  {https://doi.org/10.1038/s41567-019-0649-7} {\bibfield  {journal} {\bibinfo
  {journal} {Nature Physics}\ }\textbf {\bibinfo {volume} {15}},\ \bibinfo
  {pages} {1168} (\bibinfo {year} {2019})}\BibitemShut {NoStop}%
\bibitem [{\citenamefont {Kokail}\ \emph {et~al.}(2019)\citenamefont {Kokail},
  \citenamefont {Maier}, \citenamefont {van Bijnen}, \citenamefont {Brydges},
  \citenamefont {Joshi}, \citenamefont {Jurcevic}, \citenamefont {Muschik},
  \citenamefont {Silvi}, \citenamefont {Blatt}, \citenamefont {Roos},\ and\
  \citenamefont {Zoller}}]{Kokail_2019}%
  \BibitemOpen
  \bibfield  {author} {\bibinfo {author} {\bibfnamefont {C.}~\bibnamefont
  {Kokail}}, \bibinfo {author} {\bibfnamefont {C.}~\bibnamefont {Maier}},
  \bibinfo {author} {\bibfnamefont {R.}~\bibnamefont {van Bijnen}}, \bibinfo
  {author} {\bibfnamefont {T.}~\bibnamefont {Brydges}}, \bibinfo {author}
  {\bibfnamefont {M.~K.}\ \bibnamefont {Joshi}}, \bibinfo {author}
  {\bibfnamefont {P.}~\bibnamefont {Jurcevic}}, \bibinfo {author}
  {\bibfnamefont {C.~A.}\ \bibnamefont {Muschik}}, \bibinfo {author}
  {\bibfnamefont {P.}~\bibnamefont {Silvi}}, \bibinfo {author} {\bibfnamefont
  {R.}~\bibnamefont {Blatt}}, \bibinfo {author} {\bibfnamefont {C.~F.}\
  \bibnamefont {Roos}},\ and\ \bibinfo {author} {\bibfnamefont
  {P.}~\bibnamefont {Zoller}},\ }\bibfield  {title} {\bibinfo {title}
  {Self-verifying variational quantum simulation of lattice models},\ }\href
  {https://doi.org/10.1038/s41586-019-1177-4} {\bibfield  {journal} {\bibinfo
  {journal} {Nature}\ }\textbf {\bibinfo {volume} {569}},\ \bibinfo {pages}
  {355} (\bibinfo {year} {2019})}\BibitemShut {NoStop}%
\bibitem [{\citenamefont {Mil}\ \emph {et~al.}(2020)\citenamefont {Mil},
  \citenamefont {Zache}, \citenamefont {Hegde}, \citenamefont {Xia},
  \citenamefont {Bhatt}, \citenamefont {Oberthaler}, \citenamefont {Hauke},
  \citenamefont {Berges},\ and\ \citenamefont {Jendrzejewski}}]{Mil_2020}%
  \BibitemOpen
  \bibfield  {author} {\bibinfo {author} {\bibfnamefont {A.}~\bibnamefont
  {Mil}}, \bibinfo {author} {\bibfnamefont {T.~V.}\ \bibnamefont {Zache}},
  \bibinfo {author} {\bibfnamefont {A.}~\bibnamefont {Hegde}}, \bibinfo
  {author} {\bibfnamefont {A.}~\bibnamefont {Xia}}, \bibinfo {author}
  {\bibfnamefont {R.~P.}\ \bibnamefont {Bhatt}}, \bibinfo {author}
  {\bibfnamefont {M.~K.}\ \bibnamefont {Oberthaler}}, \bibinfo {author}
  {\bibfnamefont {P.}~\bibnamefont {Hauke}}, \bibinfo {author} {\bibfnamefont
  {J.}~\bibnamefont {Berges}},\ and\ \bibinfo {author} {\bibfnamefont
  {F.}~\bibnamefont {Jendrzejewski}},\ }\bibfield  {title} {\bibinfo {title} {A
  scalable realization of local u(1) gauge invariance in cold atomic
  mixtures},\ }\href {https://doi.org/10.1126/science.aaz5312} {\bibfield
  {journal} {\bibinfo  {journal} {Science}\ }\textbf {\bibinfo {volume}
  {367}},\ \bibinfo {pages} {1128} (\bibinfo {year} {2020})}\BibitemShut
  {NoStop}%
\bibitem [{\citenamefont {Yang}\ \emph {et~al.}(2020)\citenamefont {Yang},
  \citenamefont {Sun}, \citenamefont {Ott}, \citenamefont {Wang}, \citenamefont
  {Zache}, \citenamefont {Halimeh}, \citenamefont {Yuan}, \citenamefont
  {Hauke},\ and\ \citenamefont {Pan}}]{Yang_2020}%
  \BibitemOpen
  \bibfield  {author} {\bibinfo {author} {\bibfnamefont {B.}~\bibnamefont
  {Yang}}, \bibinfo {author} {\bibfnamefont {H.}~\bibnamefont {Sun}}, \bibinfo
  {author} {\bibfnamefont {R.}~\bibnamefont {Ott}}, \bibinfo {author}
  {\bibfnamefont {H.-Y.}\ \bibnamefont {Wang}}, \bibinfo {author}
  {\bibfnamefont {T.~V.}\ \bibnamefont {Zache}}, \bibinfo {author}
  {\bibfnamefont {J.~C.}\ \bibnamefont {Halimeh}}, \bibinfo {author}
  {\bibfnamefont {Z.-S.}\ \bibnamefont {Yuan}}, \bibinfo {author}
  {\bibfnamefont {P.}~\bibnamefont {Hauke}},\ and\ \bibinfo {author}
  {\bibfnamefont {J.-W.}\ \bibnamefont {Pan}},\ }\bibfield  {title} {\bibinfo
  {title} {Observation of gauge invariance in a 71-site bose--hubbard quantum
  simulator},\ }\href {https://doi.org/10.1038/s41586-020-2910-8} {\bibfield
  {journal} {\bibinfo  {journal} {Nature}\ }\textbf {\bibinfo {volume} {587}},\
  \bibinfo {pages} {392} (\bibinfo {year} {2020})}\BibitemShut {NoStop}%
\bibitem [{\citenamefont {{Zhou}}\ \emph {et~al.}(2021)\citenamefont {{Zhou}},
  \citenamefont {{Su}}, \citenamefont {{Halimeh}}, \citenamefont {{Ott}},
  \citenamefont {{Sun}}, \citenamefont {{Hauke}}, \citenamefont {{Yang}},
  \citenamefont {{Yuan}}, \citenamefont {{Berges}},\ and\ \citenamefont
  {{Pan}}}]{Zhou_2021}%
  \BibitemOpen
  \bibfield  {author} {\bibinfo {author} {\bibfnamefont {Z.-Y.}\ \bibnamefont
  {{Zhou}}}, \bibinfo {author} {\bibfnamefont {G.-X.}\ \bibnamefont {{Su}}},
  \bibinfo {author} {\bibfnamefont {J.~C.}\ \bibnamefont {{Halimeh}}}, \bibinfo
  {author} {\bibfnamefont {R.}~\bibnamefont {{Ott}}}, \bibinfo {author}
  {\bibfnamefont {H.}~\bibnamefont {{Sun}}}, \bibinfo {author} {\bibfnamefont
  {P.}~\bibnamefont {{Hauke}}}, \bibinfo {author} {\bibfnamefont
  {B.}~\bibnamefont {{Yang}}}, \bibinfo {author} {\bibfnamefont {Z.-S.}\
  \bibnamefont {{Yuan}}}, \bibinfo {author} {\bibfnamefont {J.}~\bibnamefont
  {{Berges}}},\ and\ \bibinfo {author} {\bibfnamefont {J.-W.}\ \bibnamefont
  {{Pan}}},\ }\bibfield  {title} {\bibinfo {title} {{Thermalization dynamics of
  a gauge theory on a quantum simulator}},\ }\href@noop {} {\bibfield
  {journal} {\bibinfo  {journal} {arXiv e-prints}\ ,\ \bibinfo {eid}
  {arXiv:2107.13563}} (\bibinfo {year} {2021})},\ \Eprint
  {https://arxiv.org/abs/2107.13563} {arXiv:2107.13563 [cond-mat.quant-gas]}
  \BibitemShut {NoStop}%
\bibitem [{\citenamefont {{Nguyen}}\ \emph {et~al.}(2021)\citenamefont
  {{Nguyen}}, \citenamefont {{Tran}}, \citenamefont {{Zhu}}, \citenamefont
  {{Green}}, \citenamefont {{Huerta Alderete}}, \citenamefont {{Davoudi}},\
  and\ \citenamefont {{Linke}}}]{Nguyen_2021}%
  \BibitemOpen
  \bibfield  {author} {\bibinfo {author} {\bibfnamefont {N.~H.}\ \bibnamefont
  {{Nguyen}}}, \bibinfo {author} {\bibfnamefont {M.~C.}\ \bibnamefont
  {{Tran}}}, \bibinfo {author} {\bibfnamefont {Y.}~\bibnamefont {{Zhu}}},
  \bibinfo {author} {\bibfnamefont {A.~M.}\ \bibnamefont {{Green}}}, \bibinfo
  {author} {\bibfnamefont {C.}~\bibnamefont {{Huerta Alderete}}}, \bibinfo
  {author} {\bibfnamefont {Z.}~\bibnamefont {{Davoudi}}},\ and\ \bibinfo
  {author} {\bibfnamefont {N.~M.}\ \bibnamefont {{Linke}}},\ }\bibfield
  {title} {\bibinfo {title} {{Digital Quantum Simulation of the Schwinger Model
  and Symmetry Protection with Trapped Ions}},\ }\href@noop {} {\bibfield
  {journal} {\bibinfo  {journal} {arXiv e-prints}\ ,\ \bibinfo {eid}
  {arXiv:2112.14262}} (\bibinfo {year} {2021})},\ \Eprint
  {https://arxiv.org/abs/2112.14262} {arXiv:2112.14262 [quant-ph]} \BibitemShut
  {NoStop}%
\bibitem [{\citenamefont {Klco}\ \emph {et~al.}(2022)\citenamefont {Klco},
  \citenamefont {Roggero},\ and\ \citenamefont {Savage}}]{klco2022standard}%
  \BibitemOpen
  \bibfield  {author} {\bibinfo {author} {\bibfnamefont {N.}~\bibnamefont
  {Klco}}, \bibinfo {author} {\bibfnamefont {A.}~\bibnamefont {Roggero}},\ and\
  \bibinfo {author} {\bibfnamefont {M.~J.}\ \bibnamefont {Savage}},\ }\bibfield
   {title} {\bibinfo {title} {Standard model physics and the digital quantum
  revolution: thoughts about the interface},\ }\href@noop {} {\bibfield
  {journal} {\bibinfo  {journal} {Reports on Progress in Physics}\ } (\bibinfo
  {year} {2022})}\BibitemShut {NoStop}%
\bibitem [{\citenamefont {Muschik}\ \emph {et~al.}(2017)\citenamefont
  {Muschik}, \citenamefont {Heyl}, \citenamefont {Martinez}, \citenamefont
  {Monz}, \citenamefont {Schindler}, \citenamefont {Vogell}, \citenamefont
  {Dalmonte}, \citenamefont {Hauke}, \citenamefont {Blatt},\ and\ \citenamefont
  {Zoller}}]{Muschik_2017}%
  \BibitemOpen
  \bibfield  {author} {\bibinfo {author} {\bibfnamefont {C.}~\bibnamefont
  {Muschik}}, \bibinfo {author} {\bibfnamefont {M.}~\bibnamefont {Heyl}},
  \bibinfo {author} {\bibfnamefont {E.}~\bibnamefont {Martinez}}, \bibinfo
  {author} {\bibfnamefont {T.}~\bibnamefont {Monz}}, \bibinfo {author}
  {\bibfnamefont {P.}~\bibnamefont {Schindler}}, \bibinfo {author}
  {\bibfnamefont {B.}~\bibnamefont {Vogell}}, \bibinfo {author} {\bibfnamefont
  {M.}~\bibnamefont {Dalmonte}}, \bibinfo {author} {\bibfnamefont
  {P.}~\bibnamefont {Hauke}}, \bibinfo {author} {\bibfnamefont
  {R.}~\bibnamefont {Blatt}},\ and\ \bibinfo {author} {\bibfnamefont
  {P.}~\bibnamefont {Zoller}},\ }\bibfield  {title} {\bibinfo {title} {U(1)
  wilson lattice gauge theories in digital quantum simulators},\ }\href
  {https://doi.org/10.1088/1367-2630/aa89ab} {\bibfield  {journal} {\bibinfo
  {journal} {New Journal of Physics}\ }\textbf {\bibinfo {volume} {19}},\
  \bibinfo {pages} {103020} (\bibinfo {year} {2017})}\BibitemShut {NoStop}%
\bibitem [{\citenamefont {Paulson}\ \emph {et~al.}(2021)\citenamefont
  {Paulson}, \citenamefont {Dellantonio}, \citenamefont {Haase}, \citenamefont
  {Celi}, \citenamefont {Kan}, \citenamefont {Jena}, \citenamefont {Kokail},
  \citenamefont {van Bijnen}, \citenamefont {Jansen}, \citenamefont {Zoller},\
  and\ \citenamefont {Muschik}}]{Paulson_2021}%
  \BibitemOpen
  \bibfield  {author} {\bibinfo {author} {\bibfnamefont {D.}~\bibnamefont
  {Paulson}}, \bibinfo {author} {\bibfnamefont {L.}~\bibnamefont
  {Dellantonio}}, \bibinfo {author} {\bibfnamefont {J.~F.}\ \bibnamefont
  {Haase}}, \bibinfo {author} {\bibfnamefont {A.}~\bibnamefont {Celi}},
  \bibinfo {author} {\bibfnamefont {A.}~\bibnamefont {Kan}}, \bibinfo {author}
  {\bibfnamefont {A.}~\bibnamefont {Jena}}, \bibinfo {author} {\bibfnamefont
  {C.}~\bibnamefont {Kokail}}, \bibinfo {author} {\bibfnamefont
  {R.}~\bibnamefont {van Bijnen}}, \bibinfo {author} {\bibfnamefont
  {K.}~\bibnamefont {Jansen}}, \bibinfo {author} {\bibfnamefont
  {P.}~\bibnamefont {Zoller}},\ and\ \bibinfo {author} {\bibfnamefont {C.~A.}\
  \bibnamefont {Muschik}},\ }\bibfield  {title} {\bibinfo {title} {Simulating
  2d effects in lattice gauge theories on a quantum computer},\ }\href
  {https://doi.org/10.1103/PRXQuantum.2.030334} {\bibfield  {journal} {\bibinfo
   {journal} {PRX Quantum}\ }\textbf {\bibinfo {volume} {2}},\ \bibinfo {pages}
  {030334} (\bibinfo {year} {2021})}\BibitemShut {NoStop}%
\bibitem [{\citenamefont {Davoudi}\ \emph
  {et~al.}(2021{\natexlab{a}})\citenamefont {Davoudi}, \citenamefont {Linke},\
  and\ \citenamefont {Pagano}}]{Davoudi_2021a}%
  \BibitemOpen
  \bibfield  {author} {\bibinfo {author} {\bibfnamefont {Z.}~\bibnamefont
  {Davoudi}}, \bibinfo {author} {\bibfnamefont {N.~M.}\ \bibnamefont {Linke}},\
  and\ \bibinfo {author} {\bibfnamefont {G.}~\bibnamefont {Pagano}},\
  }\bibfield  {title} {\bibinfo {title} {Toward simulating quantum field
  theories with controlled phonon-ion dynamics: A hybrid analog-digital
  approach},\ }\href {https://doi.org/10.1103/PhysRevResearch.3.043072}
  {\bibfield  {journal} {\bibinfo  {journal} {Phys. Rev. Research}\ }\textbf
  {\bibinfo {volume} {3}},\ \bibinfo {pages} {043072} (\bibinfo {year}
  {2021}{\natexlab{a}})}\BibitemShut {NoStop}%
\bibitem [{\citenamefont {Tagliacozzo}\ \emph
  {et~al.}(2013{\natexlab{a}})\citenamefont {Tagliacozzo}, \citenamefont
  {Celi}, \citenamefont {Orland}, \citenamefont {Mitchell},\ and\ \citenamefont
  {Lewenstein}}]{Tagliacozzo_2013a}%
  \BibitemOpen
  \bibfield  {author} {\bibinfo {author} {\bibfnamefont {L.}~\bibnamefont
  {Tagliacozzo}}, \bibinfo {author} {\bibfnamefont {A.}~\bibnamefont {Celi}},
  \bibinfo {author} {\bibfnamefont {P.}~\bibnamefont {Orland}}, \bibinfo
  {author} {\bibfnamefont {M.~W.}\ \bibnamefont {Mitchell}},\ and\ \bibinfo
  {author} {\bibfnamefont {M.}~\bibnamefont {Lewenstein}},\ }\bibfield  {title}
  {\bibinfo {title} {Simulation of non-abelian gauge theories with optical
  lattices},\ }\href {https://doi.org/10.1038/ncomms3615} {\bibfield  {journal}
  {\bibinfo  {journal} {Nature Communications}\ }\textbf {\bibinfo {volume}
  {4}},\ \bibinfo {pages} {2615} (\bibinfo {year}
  {2013}{\natexlab{a}})}\BibitemShut {NoStop}%
\bibitem [{\citenamefont {Tagliacozzo}\ \emph
  {et~al.}(2013{\natexlab{b}})\citenamefont {Tagliacozzo}, \citenamefont
  {Celi}, \citenamefont {Zamora},\ and\ \citenamefont
  {Lewenstein}}]{Tagliacozzo_2013b}%
  \BibitemOpen
  \bibfield  {author} {\bibinfo {author} {\bibfnamefont {L.}~\bibnamefont
  {Tagliacozzo}}, \bibinfo {author} {\bibfnamefont {A.}~\bibnamefont {Celi}},
  \bibinfo {author} {\bibfnamefont {A.}~\bibnamefont {Zamora}},\ and\ \bibinfo
  {author} {\bibfnamefont {M.}~\bibnamefont {Lewenstein}},\ }\bibfield  {title}
  {\bibinfo {title} {Optical abelian lattice gauge theories},\ }\href
  {https://doi.org/https://doi.org/10.1016/j.aop.2012.11.009} {\bibfield
  {journal} {\bibinfo  {journal} {Annals of Physics}\ }\textbf {\bibinfo
  {volume} {330}},\ \bibinfo {pages} {160} (\bibinfo {year}
  {2013}{\natexlab{b}})}\BibitemShut {NoStop}%
\bibitem [{\citenamefont {Zohar}\ \emph
  {et~al.}(2017{\natexlab{a}})\citenamefont {Zohar}, \citenamefont {Farace},
  \citenamefont {Reznik},\ and\ \citenamefont {Cirac}}]{Zohar_2017a}%
  \BibitemOpen
  \bibfield  {author} {\bibinfo {author} {\bibfnamefont {E.}~\bibnamefont
  {Zohar}}, \bibinfo {author} {\bibfnamefont {A.}~\bibnamefont {Farace}},
  \bibinfo {author} {\bibfnamefont {B.}~\bibnamefont {Reznik}},\ and\ \bibinfo
  {author} {\bibfnamefont {J.~I.}\ \bibnamefont {Cirac}},\ }\bibfield  {title}
  {\bibinfo {title} {Digital quantum simulation of ${\mathbb{z}}_{2}$ lattice
  gauge theories with dynamical fermionic matter},\ }\href
  {https://doi.org/10.1103/PhysRevLett.118.070501} {\bibfield  {journal}
  {\bibinfo  {journal} {Phys. Rev. Lett.}\ }\textbf {\bibinfo {volume} {118}},\
  \bibinfo {pages} {070501} (\bibinfo {year} {2017}{\natexlab{a}})}\BibitemShut
  {NoStop}%
\bibitem [{\citenamefont {Zohar}\ \emph
  {et~al.}(2017{\natexlab{b}})\citenamefont {Zohar}, \citenamefont {Farace},
  \citenamefont {Reznik},\ and\ \citenamefont {Cirac}}]{Zohar_2017b}%
  \BibitemOpen
  \bibfield  {author} {\bibinfo {author} {\bibfnamefont {E.}~\bibnamefont
  {Zohar}}, \bibinfo {author} {\bibfnamefont {A.}~\bibnamefont {Farace}},
  \bibinfo {author} {\bibfnamefont {B.}~\bibnamefont {Reznik}},\ and\ \bibinfo
  {author} {\bibfnamefont {J.~I.}\ \bibnamefont {Cirac}},\ }\bibfield  {title}
  {\bibinfo {title} {Digital lattice gauge theories},\ }\href
  {https://doi.org/10.1103/PhysRevA.95.023604} {\bibfield  {journal} {\bibinfo
  {journal} {Phys. Rev. A}\ }\textbf {\bibinfo {volume} {95}},\ \bibinfo
  {pages} {023604} (\bibinfo {year} {2017}{\natexlab{b}})}\BibitemShut
  {NoStop}%
\bibitem [{\citenamefont {Bender}\ \emph {et~al.}(2018)\citenamefont {Bender},
  \citenamefont {Zohar}, \citenamefont {Farace},\ and\ \citenamefont
  {Cirac}}]{Bender_2018}%
  \BibitemOpen
  \bibfield  {author} {\bibinfo {author} {\bibfnamefont {J.}~\bibnamefont
  {Bender}}, \bibinfo {author} {\bibfnamefont {E.}~\bibnamefont {Zohar}},
  \bibinfo {author} {\bibfnamefont {A.}~\bibnamefont {Farace}},\ and\ \bibinfo
  {author} {\bibfnamefont {J.~I.}\ \bibnamefont {Cirac}},\ }\bibfield  {title}
  {\bibinfo {title} {Digital quantum simulation of lattice gauge theories in
  three spatial dimensions},\ }\href@noop {} {\bibfield  {journal} {\bibinfo
  {journal} {New Journal of Physics}\ }\textbf {\bibinfo {volume} {20}},\
  \bibinfo {pages} {093001} (\bibinfo {year} {2018})}\BibitemShut {NoStop}%
\bibitem [{\citenamefont {Mezzacapo}\ \emph {et~al.}(2015)\citenamefont
  {Mezzacapo}, \citenamefont {Rico}, \citenamefont {Sab\'{\i}n}, \citenamefont
  {Egusquiza}, \citenamefont {Lamata},\ and\ \citenamefont
  {Solano}}]{Mezzacapo_2015}%
  \BibitemOpen
  \bibfield  {author} {\bibinfo {author} {\bibfnamefont {A.}~\bibnamefont
  {Mezzacapo}}, \bibinfo {author} {\bibfnamefont {E.}~\bibnamefont {Rico}},
  \bibinfo {author} {\bibfnamefont {C.}~\bibnamefont {Sab\'{\i}n}}, \bibinfo
  {author} {\bibfnamefont {I.~L.}\ \bibnamefont {Egusquiza}}, \bibinfo {author}
  {\bibfnamefont {L.}~\bibnamefont {Lamata}},\ and\ \bibinfo {author}
  {\bibfnamefont {E.}~\bibnamefont {Solano}},\ }\bibfield  {title} {\bibinfo
  {title} {Non-abelian su(2) lattice gauge theories in superconducting
  circuits},\ }\href {https://doi.org/10.1103/PhysRevLett.115.240502}
  {\bibfield  {journal} {\bibinfo  {journal} {Phys. Rev. Lett.}\ }\textbf
  {\bibinfo {volume} {115}},\ \bibinfo {pages} {240502} (\bibinfo {year}
  {2015})}\BibitemShut {NoStop}%
\bibitem [{\citenamefont {Klco}\ \emph {et~al.}(2018)\citenamefont {Klco},
  \citenamefont {Dumitrescu}, \citenamefont {McCaskey}, \citenamefont {Morris},
  \citenamefont {Pooser}, \citenamefont {Sanz}, \citenamefont {Solano},
  \citenamefont {Lougovski},\ and\ \citenamefont {Savage}}]{Klco_2018}%
  \BibitemOpen
  \bibfield  {author} {\bibinfo {author} {\bibfnamefont {N.}~\bibnamefont
  {Klco}}, \bibinfo {author} {\bibfnamefont {E.~F.}\ \bibnamefont
  {Dumitrescu}}, \bibinfo {author} {\bibfnamefont {A.~J.}\ \bibnamefont
  {McCaskey}}, \bibinfo {author} {\bibfnamefont {T.~D.}\ \bibnamefont
  {Morris}}, \bibinfo {author} {\bibfnamefont {R.~C.}\ \bibnamefont {Pooser}},
  \bibinfo {author} {\bibfnamefont {M.}~\bibnamefont {Sanz}}, \bibinfo {author}
  {\bibfnamefont {E.}~\bibnamefont {Solano}}, \bibinfo {author} {\bibfnamefont
  {P.}~\bibnamefont {Lougovski}},\ and\ \bibinfo {author} {\bibfnamefont
  {M.~J.}\ \bibnamefont {Savage}},\ }\bibfield  {title} {\bibinfo {title}
  {Quantum-classical computation of schwinger model dynamics using quantum
  computers},\ }\href {https://doi.org/10.1103/PhysRevA.98.032331} {\bibfield
  {journal} {\bibinfo  {journal} {Phys. Rev. A}\ }\textbf {\bibinfo {volume}
  {98}},\ \bibinfo {pages} {032331} (\bibinfo {year} {2018})}\BibitemShut
  {NoStop}%
\bibitem [{\citenamefont {Atas}\ \emph {et~al.}(2021)\citenamefont {Atas},
  \citenamefont {Zhang}, \citenamefont {Lewis}, \citenamefont {Jahanpour},
  \citenamefont {Haase},\ and\ \citenamefont {Muschik}}]{Atas_2021}%
  \BibitemOpen
  \bibfield  {author} {\bibinfo {author} {\bibfnamefont {Y.~Y.}\ \bibnamefont
  {Atas}}, \bibinfo {author} {\bibfnamefont {J.}~\bibnamefont {Zhang}},
  \bibinfo {author} {\bibfnamefont {R.}~\bibnamefont {Lewis}}, \bibinfo
  {author} {\bibfnamefont {A.}~\bibnamefont {Jahanpour}}, \bibinfo {author}
  {\bibfnamefont {J.~F.}\ \bibnamefont {Haase}},\ and\ \bibinfo {author}
  {\bibfnamefont {C.~A.}\ \bibnamefont {Muschik}},\ }\bibfield  {title}
  {\bibinfo {title} {Su(2) hadrons on a quantum computer via a variational
  approach},\ }\href {https://doi.org/10.1038/s41467-021-26825-4} {\bibfield
  {journal} {\bibinfo  {journal} {Nature Communications}\ }\textbf {\bibinfo
  {volume} {12}},\ \bibinfo {pages} {6499} (\bibinfo {year}
  {2021})}\BibitemShut {NoStop}%
\bibitem [{\citenamefont {Armon}\ \emph {et~al.}(2021)\citenamefont {Armon},
  \citenamefont {Ashkenazi}, \citenamefont {Garc\'{\i}a-Moreno}, \citenamefont
  {Gonz\'alez-Tudela},\ and\ \citenamefont {Zohar}}]{Armon_2021}%
  \BibitemOpen
  \bibfield  {author} {\bibinfo {author} {\bibfnamefont {T.}~\bibnamefont
  {Armon}}, \bibinfo {author} {\bibfnamefont {S.}~\bibnamefont {Ashkenazi}},
  \bibinfo {author} {\bibfnamefont {G.}~\bibnamefont {Garc\'{\i}a-Moreno}},
  \bibinfo {author} {\bibfnamefont {A.}~\bibnamefont {Gonz\'alez-Tudela}},\
  and\ \bibinfo {author} {\bibfnamefont {E.}~\bibnamefont {Zohar}},\ }\bibfield
   {title} {\bibinfo {title} {Photon-mediated stroboscopic quantum simulation
  of a ${\mathbb{z}}_{2}$ lattice gauge theory},\ }\href
  {https://doi.org/10.1103/PhysRevLett.127.250501} {\bibfield  {journal}
  {\bibinfo  {journal} {Phys. Rev. Lett.}\ }\textbf {\bibinfo {volume} {127}},\
  \bibinfo {pages} {250501} (\bibinfo {year} {2021})}\BibitemShut {NoStop}%
\bibitem [{\citenamefont {Preskill}(2018)}]{Preskill_2018}%
  \BibitemOpen
  \bibfield  {author} {\bibinfo {author} {\bibfnamefont {J.}~\bibnamefont
  {Preskill}},\ }\bibfield  {title} {\bibinfo {title} {Quantum {C}omputing in
  the {NISQ} era and beyond},\ }\href
  {https://doi.org/10.22331/q-2018-08-06-79} {\bibfield  {journal} {\bibinfo
  {journal} {{Quantum}}\ }\textbf {\bibinfo {volume} {2}},\ \bibinfo {pages}
  {79} (\bibinfo {year} {2018})}\BibitemShut {NoStop}%
\bibitem [{\citenamefont {Byrnes}\ and\ \citenamefont
  {Yamamoto}(2006)}]{Byrnes_2006}%
  \BibitemOpen
  \bibfield  {author} {\bibinfo {author} {\bibfnamefont {T.}~\bibnamefont
  {Byrnes}}\ and\ \bibinfo {author} {\bibfnamefont {Y.}~\bibnamefont
  {Yamamoto}},\ }\bibfield  {title} {\bibinfo {title} {Simulating lattice gauge
  theories on a quantum computer},\ }\href
  {https://doi.org/10.1103/PhysRevA.73.022328} {\bibfield  {journal} {\bibinfo
  {journal} {Phys. Rev. A}\ }\textbf {\bibinfo {volume} {73}},\ \bibinfo
  {pages} {022328} (\bibinfo {year} {2006})}\BibitemShut {NoStop}%
\bibitem [{\citenamefont {Lamm}\ \emph {et~al.}(2019)\citenamefont {Lamm},
  \citenamefont {Lawrence},\ and\ \citenamefont {Yamauchi}}]{Lamm_2019}%
  \BibitemOpen
  \bibfield  {author} {\bibinfo {author} {\bibfnamefont {H.}~\bibnamefont
  {Lamm}}, \bibinfo {author} {\bibfnamefont {S.}~\bibnamefont {Lawrence}},\
  and\ \bibinfo {author} {\bibfnamefont {Y.}~\bibnamefont {Yamauchi}} (\bibinfo
  {collaboration} {NuQS Collaboration}),\ }\bibfield  {title} {\bibinfo {title}
  {General methods for digital quantum simulation of gauge theories},\ }\href
  {https://doi.org/10.1103/PhysRevD.100.034518} {\bibfield  {journal} {\bibinfo
   {journal} {Phys. Rev. D}\ }\textbf {\bibinfo {volume} {100}},\ \bibinfo
  {pages} {034518} (\bibinfo {year} {2019})}\BibitemShut {NoStop}%
\bibitem [{\citenamefont {Alexandru}\ \emph {et~al.}(2019)\citenamefont
  {Alexandru}, \citenamefont {Bedaque}, \citenamefont {Harmalkar},
  \citenamefont {Lamm}, \citenamefont {Lawrence},\ and\ \citenamefont
  {Warrington}}]{Alexandru_2019}%
  \BibitemOpen
  \bibfield  {author} {\bibinfo {author} {\bibfnamefont {A.}~\bibnamefont
  {Alexandru}}, \bibinfo {author} {\bibfnamefont {P.~F.}\ \bibnamefont
  {Bedaque}}, \bibinfo {author} {\bibfnamefont {S.}~\bibnamefont {Harmalkar}},
  \bibinfo {author} {\bibfnamefont {H.}~\bibnamefont {Lamm}}, \bibinfo {author}
  {\bibfnamefont {S.}~\bibnamefont {Lawrence}},\ and\ \bibinfo {author}
  {\bibfnamefont {N.~C.}\ \bibnamefont {Warrington}} (\bibinfo {collaboration}
  {NuQS Collaboration}),\ }\bibfield  {title} {\bibinfo {title} {Gluon field
  digitization for quantum computers},\ }\href
  {https://doi.org/10.1103/PhysRevD.100.114501} {\bibfield  {journal} {\bibinfo
   {journal} {Phys. Rev. D}\ }\textbf {\bibinfo {volume} {100}},\ \bibinfo
  {pages} {114501} (\bibinfo {year} {2019})}\BibitemShut {NoStop}%
\bibitem [{\citenamefont {Ji}\ \emph {et~al.}(2020)\citenamefont {Ji},
  \citenamefont {Lamm},\ and\ \citenamefont {Zhu}}]{Ji_2020}%
  \BibitemOpen
  \bibfield  {author} {\bibinfo {author} {\bibfnamefont {Y.}~\bibnamefont
  {Ji}}, \bibinfo {author} {\bibfnamefont {H.}~\bibnamefont {Lamm}},\ and\
  \bibinfo {author} {\bibfnamefont {S.}~\bibnamefont {Zhu}} (\bibinfo
  {collaboration} {NuQS Collaboration}),\ }\bibfield  {title} {\bibinfo {title}
  {Gluon field digitization via group space decimation for quantum computers},\
  }\href {https://doi.org/10.1103/PhysRevD.102.114513} {\bibfield  {journal}
  {\bibinfo  {journal} {Phys. Rev. D}\ }\textbf {\bibinfo {volume} {102}},\
  \bibinfo {pages} {114513} (\bibinfo {year} {2020})}\BibitemShut {NoStop}%
\bibitem [{\citenamefont {Mathis}\ \emph {et~al.}(2020)\citenamefont {Mathis},
  \citenamefont {Mazzola},\ and\ \citenamefont {Tavernelli}}]{Mathis_2020}%
  \BibitemOpen
  \bibfield  {author} {\bibinfo {author} {\bibfnamefont {S.~V.}\ \bibnamefont
  {Mathis}}, \bibinfo {author} {\bibfnamefont {G.}~\bibnamefont {Mazzola}},\
  and\ \bibinfo {author} {\bibfnamefont {I.}~\bibnamefont {Tavernelli}},\
  }\bibfield  {title} {\bibinfo {title} {Toward scalable simulations of lattice
  gauge theories on quantum computers},\ }\href
  {https://doi.org/10.1103/PhysRevD.102.094501} {\bibfield  {journal} {\bibinfo
   {journal} {Phys. Rev. D}\ }\textbf {\bibinfo {volume} {102}},\ \bibinfo
  {pages} {094501} (\bibinfo {year} {2020})}\BibitemShut {NoStop}%
\bibitem [{\citenamefont {Kaplan}\ and\ \citenamefont
  {Stryker}(2020)}]{Kaplan_2020}%
  \BibitemOpen
  \bibfield  {author} {\bibinfo {author} {\bibfnamefont {D.~B.}\ \bibnamefont
  {Kaplan}}\ and\ \bibinfo {author} {\bibfnamefont {J.~R.}\ \bibnamefont
  {Stryker}},\ }\bibfield  {title} {\bibinfo {title} {Gauss's law, duality, and
  the hamiltonian formulation of u(1) lattice gauge theory},\ }\href
  {https://doi.org/10.1103/PhysRevD.102.094515} {\bibfield  {journal} {\bibinfo
   {journal} {Phys. Rev. D}\ }\textbf {\bibinfo {volume} {102}},\ \bibinfo
  {pages} {094515} (\bibinfo {year} {2020})}\BibitemShut {NoStop}%
\bibitem [{\citenamefont {{Brower}}\ \emph {et~al.}(2020)\citenamefont
  {{Brower}}, \citenamefont {{Berenstein}},\ and\ \citenamefont
  {{Kawai}}}]{Brower_2020}%
  \BibitemOpen
  \bibfield  {author} {\bibinfo {author} {\bibfnamefont {R.~C.}\ \bibnamefont
  {{Brower}}}, \bibinfo {author} {\bibfnamefont {D.}~\bibnamefont
  {{Berenstein}}},\ and\ \bibinfo {author} {\bibfnamefont {H.}~\bibnamefont
  {{Kawai}}},\ }\bibfield  {title} {\bibinfo {title} {{Lattice Gauge Theory for
  a Quantum Computer}},\ }\href@noop {} {\bibfield  {journal} {\bibinfo
  {journal} {arXiv e-prints}\ ,\ \bibinfo {eid} {arXiv:2002.10028}} (\bibinfo
  {year} {2020})},\ \Eprint {https://arxiv.org/abs/2002.10028}
  {arXiv:2002.10028 [hep-lat]} \BibitemShut {NoStop}%
\bibitem [{\citenamefont {Shaw}\ \emph {et~al.}(2020)\citenamefont {Shaw},
  \citenamefont {Lougovski}, \citenamefont {Stryker},\ and\ \citenamefont
  {Wiebe}}]{Shaw_2020}%
  \BibitemOpen
  \bibfield  {author} {\bibinfo {author} {\bibfnamefont {A.~F.}\ \bibnamefont
  {Shaw}}, \bibinfo {author} {\bibfnamefont {P.}~\bibnamefont {Lougovski}},
  \bibinfo {author} {\bibfnamefont {J.~R.}\ \bibnamefont {Stryker}},\ and\
  \bibinfo {author} {\bibfnamefont {N.}~\bibnamefont {Wiebe}},\ }\bibfield
  {title} {\bibinfo {title} {Quantum {A}lgorithms for {S}imulating the
  {L}attice {S}chwinger {M}odel},\ }\href
  {https://doi.org/10.22331/q-2020-08-10-306} {\bibfield  {journal} {\bibinfo
  {journal} {{Quantum}}\ }\textbf {\bibinfo {volume} {4}},\ \bibinfo {pages}
  {306} (\bibinfo {year} {2020})}\BibitemShut {NoStop}%
\bibitem [{\citenamefont {Klco}\ \emph {et~al.}(2020)\citenamefont {Klco},
  \citenamefont {Savage},\ and\ \citenamefont {Stryker}}]{Kclo_2020}%
  \BibitemOpen
  \bibfield  {author} {\bibinfo {author} {\bibfnamefont {N.}~\bibnamefont
  {Klco}}, \bibinfo {author} {\bibfnamefont {M.~J.}\ \bibnamefont {Savage}},\
  and\ \bibinfo {author} {\bibfnamefont {J.~R.}\ \bibnamefont {Stryker}},\
  }\bibfield  {title} {\bibinfo {title} {Su(2) non-abelian gauge field theory
  in one dimension on digital quantum computers},\ }\href
  {https://doi.org/10.1103/PhysRevD.101.074512} {\bibfield  {journal} {\bibinfo
   {journal} {Phys. Rev. D}\ }\textbf {\bibinfo {volume} {101}},\ \bibinfo
  {pages} {074512} (\bibinfo {year} {2020})}\BibitemShut {NoStop}%
\bibitem [{\citenamefont {Ciavarella}\ \emph {et~al.}(2021)\citenamefont
  {Ciavarella}, \citenamefont {Klco},\ and\ \citenamefont
  {Savage}}]{Kclo_2021}%
  \BibitemOpen
  \bibfield  {author} {\bibinfo {author} {\bibfnamefont {A.}~\bibnamefont
  {Ciavarella}}, \bibinfo {author} {\bibfnamefont {N.}~\bibnamefont {Klco}},\
  and\ \bibinfo {author} {\bibfnamefont {M.~J.}\ \bibnamefont {Savage}},\
  }\bibfield  {title} {\bibinfo {title} {Trailhead for quantum simulation of
  su(3) yang-mills lattice gauge theory in the local multiplet basis},\ }\href
  {https://doi.org/10.1103/PhysRevD.103.094501} {\bibfield  {journal} {\bibinfo
   {journal} {Phys. Rev. D}\ }\textbf {\bibinfo {volume} {103}},\ \bibinfo
  {pages} {094501} (\bibinfo {year} {2021})}\BibitemShut {NoStop}%
\bibitem [{\citenamefont {{Alexandru}}\ \emph {et~al.}(2021)\citenamefont
  {{Alexandru}}, \citenamefont {{Bedaque}}, \citenamefont {{Brett}},\ and\
  \citenamefont {{Lamm}}}]{Alexandru_2021}%
  \BibitemOpen
  \bibfield  {author} {\bibinfo {author} {\bibfnamefont {A.}~\bibnamefont
  {{Alexandru}}}, \bibinfo {author} {\bibfnamefont {P.~F.}\ \bibnamefont
  {{Bedaque}}}, \bibinfo {author} {\bibfnamefont {R.}~\bibnamefont {{Brett}}},\
  and\ \bibinfo {author} {\bibfnamefont {H.}~\bibnamefont {{Lamm}}},\
  }\bibfield  {title} {\bibinfo {title} {{The spectrum of qubitized QCD:
  glueballs in a $S(1080)$ gauge theory}},\ }\href@noop {} {\bibfield
  {journal} {\bibinfo  {journal} {arXiv e-prints}\ ,\ \bibinfo {eid}
  {arXiv:2112.08482}} (\bibinfo {year} {2021})},\ \Eprint
  {https://arxiv.org/abs/2112.08482} {arXiv:2112.08482 [hep-lat]} \BibitemShut
  {NoStop}%
\bibitem [{\citenamefont {Haase}\ \emph {et~al.}(2021)\citenamefont {Haase},
  \citenamefont {Dellantonio}, \citenamefont {Celi}, \citenamefont {Paulson},
  \citenamefont {Kan}, \citenamefont {Jansen},\ and\ \citenamefont
  {Muschik}}]{Haase_2021}%
  \BibitemOpen
  \bibfield  {author} {\bibinfo {author} {\bibfnamefont {J.~F.}\ \bibnamefont
  {Haase}}, \bibinfo {author} {\bibfnamefont {L.}~\bibnamefont {Dellantonio}},
  \bibinfo {author} {\bibfnamefont {A.}~\bibnamefont {Celi}}, \bibinfo {author}
  {\bibfnamefont {D.}~\bibnamefont {Paulson}}, \bibinfo {author} {\bibfnamefont
  {A.}~\bibnamefont {Kan}}, \bibinfo {author} {\bibfnamefont {K.}~\bibnamefont
  {Jansen}},\ and\ \bibinfo {author} {\bibfnamefont {C.~A.}\ \bibnamefont
  {Muschik}},\ }\bibfield  {title} {\bibinfo {title} {A resource efficient
  approach for quantum and classical simulations of gauge theories in particle
  physics},\ }\href {https://doi.org/10.22331/q-2021-02-04-393} {\bibfield
  {journal} {\bibinfo  {journal} {{Quantum}}\ }\textbf {\bibinfo {volume}
  {5}},\ \bibinfo {pages} {393} (\bibinfo {year} {2021})}\BibitemShut {NoStop}%
\bibitem [{\citenamefont {{Bauer}}\ and\ \citenamefont
  {{Grabowska}}(2021)}]{Bauer_2021}%
  \BibitemOpen
  \bibfield  {author} {\bibinfo {author} {\bibfnamefont {C.~W.}\ \bibnamefont
  {{Bauer}}}\ and\ \bibinfo {author} {\bibfnamefont {D.~M.}\ \bibnamefont
  {{Grabowska}}},\ }\bibfield  {title} {\bibinfo {title} {{Efficient
  Representation for Simulating U(1) Gauge Theories on Digital Quantum
  Computers at All Values of the Coupling}},\ }\href@noop {} {\bibfield
  {journal} {\bibinfo  {journal} {arXiv e-prints}\ ,\ \bibinfo {eid}
  {arXiv:2111.08015}} (\bibinfo {year} {2021})},\ \Eprint
  {https://arxiv.org/abs/2111.08015} {arXiv:2111.08015 [hep-ph]} \BibitemShut
  {NoStop}%
\bibitem [{\citenamefont {{Kan}}\ and\ \citenamefont {{Nam}}(2021)}]{Kan_2021}%
  \BibitemOpen
  \bibfield  {author} {\bibinfo {author} {\bibfnamefont {A.}~\bibnamefont
  {{Kan}}}\ and\ \bibinfo {author} {\bibfnamefont {Y.}~\bibnamefont {{Nam}}},\
  }\bibfield  {title} {\bibinfo {title} {{Lattice Quantum Chromodynamics and
  Electrodynamics on a Universal Quantum Computer}},\ }\href@noop {} {\bibfield
   {journal} {\bibinfo  {journal} {arXiv e-prints}\ ,\ \bibinfo {eid}
  {arXiv:2107.12769}} (\bibinfo {year} {2021})},\ \Eprint
  {https://arxiv.org/abs/2107.12769} {arXiv:2107.12769 [quant-ph]} \BibitemShut
  {NoStop}%
\bibitem [{\citenamefont {Davoudi}\ \emph
  {et~al.}(2021{\natexlab{b}})\citenamefont {Davoudi}, \citenamefont
  {Raychowdhury},\ and\ \citenamefont {Shaw}}]{Davoudi_2021b}%
  \BibitemOpen
  \bibfield  {author} {\bibinfo {author} {\bibfnamefont {Z.}~\bibnamefont
  {Davoudi}}, \bibinfo {author} {\bibfnamefont {I.}~\bibnamefont
  {Raychowdhury}},\ and\ \bibinfo {author} {\bibfnamefont {A.}~\bibnamefont
  {Shaw}},\ }\bibfield  {title} {\bibinfo {title} {Search for efficient
  formulations for hamiltonian simulation of non-abelian lattice gauge
  theories},\ }\href {https://doi.org/10.1103/PhysRevD.104.074505} {\bibfield
  {journal} {\bibinfo  {journal} {Phys. Rev. D}\ }\textbf {\bibinfo {volume}
  {104}},\ \bibinfo {pages} {074505} (\bibinfo {year}
  {2021}{\natexlab{b}})}\BibitemShut {NoStop}%
\bibitem [{\citenamefont {Kaufman}\ and\ \citenamefont
  {Ni}(2021)}]{Kaufman_2021}%
  \BibitemOpen
  \bibfield  {author} {\bibinfo {author} {\bibfnamefont {A.~M.}\ \bibnamefont
  {Kaufman}}\ and\ \bibinfo {author} {\bibfnamefont {K.-K.}\ \bibnamefont
  {Ni}},\ }\bibfield  {title} {\bibinfo {title} {Quantum science with optical
  tweezer arrays of ultracold atoms and molecules},\ }\href
  {https://doi.org/10.1038/s41567-021-01357-2} {\bibfield  {journal} {\bibinfo
  {journal} {Nature Physics}\ }\textbf {\bibinfo {volume} {17}},\ \bibinfo
  {pages} {1324} (\bibinfo {year} {2021})}\BibitemShut {NoStop}%
\bibitem [{\citenamefont {Saffman}(2016)}]{Saffman_2016}%
  \BibitemOpen
  \bibfield  {author} {\bibinfo {author} {\bibfnamefont {M.}~\bibnamefont
  {Saffman}},\ }\bibfield  {title} {\bibinfo {title} {Quantum computing with
  atomic qubits and rydberg interactions: progress and challenges},\ }\href
  {https://doi.org/10.1088/0953-4075/49/20/202001} {\bibfield  {journal}
  {\bibinfo  {journal} {Journal of Physics B: Atomic, Molecular and Optical
  Physics}\ }\textbf {\bibinfo {volume} {49}},\ \bibinfo {pages} {202001}
  (\bibinfo {year} {2016})}\BibitemShut {NoStop}%
\bibitem [{\citenamefont {Henriet}\ \emph {et~al.}(2020)\citenamefont
  {Henriet}, \citenamefont {Beguin}, \citenamefont {Signoles}, \citenamefont
  {Lahaye}, \citenamefont {Browaeys}, \citenamefont {Reymond},\ and\
  \citenamefont {Jurczak}}]{Henriet_2020}%
  \BibitemOpen
  \bibfield  {author} {\bibinfo {author} {\bibfnamefont {L.}~\bibnamefont
  {Henriet}}, \bibinfo {author} {\bibfnamefont {L.}~\bibnamefont {Beguin}},
  \bibinfo {author} {\bibfnamefont {A.}~\bibnamefont {Signoles}}, \bibinfo
  {author} {\bibfnamefont {T.}~\bibnamefont {Lahaye}}, \bibinfo {author}
  {\bibfnamefont {A.}~\bibnamefont {Browaeys}}, \bibinfo {author}
  {\bibfnamefont {G.-O.}\ \bibnamefont {Reymond}},\ and\ \bibinfo {author}
  {\bibfnamefont {C.}~\bibnamefont {Jurczak}},\ }\bibfield  {title} {\bibinfo
  {title} {Quantum computing with neutral atoms},\ }\href
  {https://doi.org/10.22331/q-2020-09-21-327} {\bibfield  {journal} {\bibinfo
  {journal} {{Quantum}}\ }\textbf {\bibinfo {volume} {4}},\ \bibinfo {pages}
  {327} (\bibinfo {year} {2020})}\BibitemShut {NoStop}%
\bibitem [{\citenamefont {Bluvstein}\ \emph {et~al.}(2022)\citenamefont
  {Bluvstein}, \citenamefont {Levine}, \citenamefont {Semeghini}, \citenamefont
  {Wang}, \citenamefont {Ebadi}, \citenamefont {Kalinowski}, \citenamefont
  {Keesling}, \citenamefont {Maskara}, \citenamefont {Pichler}, \citenamefont
  {Greiner}, \citenamefont {Vuleti{\'c}},\ and\ \citenamefont
  {Lukin}}]{Bluvstein_2021}%
  \BibitemOpen
  \bibfield  {author} {\bibinfo {author} {\bibfnamefont {D.}~\bibnamefont
  {Bluvstein}}, \bibinfo {author} {\bibfnamefont {H.}~\bibnamefont {Levine}},
  \bibinfo {author} {\bibfnamefont {G.}~\bibnamefont {Semeghini}}, \bibinfo
  {author} {\bibfnamefont {T.~T.}\ \bibnamefont {Wang}}, \bibinfo {author}
  {\bibfnamefont {S.}~\bibnamefont {Ebadi}}, \bibinfo {author} {\bibfnamefont
  {M.}~\bibnamefont {Kalinowski}}, \bibinfo {author} {\bibfnamefont
  {A.}~\bibnamefont {Keesling}}, \bibinfo {author} {\bibfnamefont
  {N.}~\bibnamefont {Maskara}}, \bibinfo {author} {\bibfnamefont
  {H.}~\bibnamefont {Pichler}}, \bibinfo {author} {\bibfnamefont
  {M.}~\bibnamefont {Greiner}}, \bibinfo {author} {\bibfnamefont
  {V.}~\bibnamefont {Vuleti{\'c}}},\ and\ \bibinfo {author} {\bibfnamefont
  {M.~D.}\ \bibnamefont {Lukin}},\ }\bibfield  {title} {\bibinfo {title} {A
  quantum processor based on coherent transport of entangled atom arrays},\
  }\href {https://doi.org/10.1038/s41586-022-04592-6} {\bibfield  {journal}
  {\bibinfo  {journal} {Nature}\ }\textbf {\bibinfo {volume} {604}},\ \bibinfo
  {pages} {451} (\bibinfo {year} {2022})}\BibitemShut {NoStop}%
\bibitem [{\citenamefont {Cohen}\ and\ \citenamefont
  {Thompson}(2021)}]{PRXQuantum.2.030322}%
  \BibitemOpen
  \bibfield  {author} {\bibinfo {author} {\bibfnamefont {S.~R.}\ \bibnamefont
  {Cohen}}\ and\ \bibinfo {author} {\bibfnamefont {J.~D.}\ \bibnamefont
  {Thompson}},\ }\bibfield  {title} {\bibinfo {title} {Quantum computing with
  circular rydberg atoms},\ }\href
  {https://doi.org/10.1103/PRXQuantum.2.030322} {\bibfield  {journal} {\bibinfo
   {journal} {PRX Quantum}\ }\textbf {\bibinfo {volume} {2}},\ \bibinfo {pages}
  {030322} (\bibinfo {year} {2021})}\BibitemShut {NoStop}%
\bibitem [{\citenamefont {Guardado-Sanchez}\ \emph {et~al.}(2021)\citenamefont
  {Guardado-Sanchez}, \citenamefont {Spar}, \citenamefont {Schauss},
  \citenamefont {Belyansky}, \citenamefont {Young}, \citenamefont {Bienias},
  \citenamefont {Gorshkov}, \citenamefont {Iadecola},\ and\ \citenamefont
  {Bakr}}]{PhysRevX.11.021036}%
  \BibitemOpen
  \bibfield  {author} {\bibinfo {author} {\bibfnamefont {E.}~\bibnamefont
  {Guardado-Sanchez}}, \bibinfo {author} {\bibfnamefont {B.~M.}\ \bibnamefont
  {Spar}}, \bibinfo {author} {\bibfnamefont {P.}~\bibnamefont {Schauss}},
  \bibinfo {author} {\bibfnamefont {R.}~\bibnamefont {Belyansky}}, \bibinfo
  {author} {\bibfnamefont {J.~T.}\ \bibnamefont {Young}}, \bibinfo {author}
  {\bibfnamefont {P.}~\bibnamefont {Bienias}}, \bibinfo {author} {\bibfnamefont
  {A.~V.}\ \bibnamefont {Gorshkov}}, \bibinfo {author} {\bibfnamefont
  {T.}~\bibnamefont {Iadecola}},\ and\ \bibinfo {author} {\bibfnamefont
  {W.~S.}\ \bibnamefont {Bakr}},\ }\bibfield  {title} {\bibinfo {title} {Quench
  dynamics of a fermi gas with strong nonlocal interactions},\ }\href
  {https://doi.org/10.1103/PhysRevX.11.021036} {\bibfield  {journal} {\bibinfo
  {journal} {Phys. Rev. X}\ }\textbf {\bibinfo {volume} {11}},\ \bibinfo
  {pages} {021036} (\bibinfo {year} {2021})}\BibitemShut {NoStop}%
\bibitem [{\citenamefont {Madjarov}\ \emph {et~al.}(2020)\citenamefont
  {Madjarov}, \citenamefont {Covey}, \citenamefont {Shaw}, \citenamefont
  {Choi}, \citenamefont {Kale}, \citenamefont {Cooper}, \citenamefont
  {Pichler}, \citenamefont {Schkolnik}, \citenamefont {Williams},\ and\
  \citenamefont {Endres}}]{Madjarov_2020}%
  \BibitemOpen
  \bibfield  {author} {\bibinfo {author} {\bibfnamefont {I.~S.}\ \bibnamefont
  {Madjarov}}, \bibinfo {author} {\bibfnamefont {J.~P.}\ \bibnamefont {Covey}},
  \bibinfo {author} {\bibfnamefont {A.~L.}\ \bibnamefont {Shaw}}, \bibinfo
  {author} {\bibfnamefont {J.}~\bibnamefont {Choi}}, \bibinfo {author}
  {\bibfnamefont {A.}~\bibnamefont {Kale}}, \bibinfo {author} {\bibfnamefont
  {A.}~\bibnamefont {Cooper}}, \bibinfo {author} {\bibfnamefont
  {H.}~\bibnamefont {Pichler}}, \bibinfo {author} {\bibfnamefont
  {V.}~\bibnamefont {Schkolnik}}, \bibinfo {author} {\bibfnamefont {J.~R.}\
  \bibnamefont {Williams}},\ and\ \bibinfo {author} {\bibfnamefont
  {M.}~\bibnamefont {Endres}},\ }\bibfield  {title} {\bibinfo {title}
  {High-fidelity entanglement and detection of alkaline-earth rydberg atoms},\
  }\href {https://doi.org/10.1038/s41567-020-0903-z} {\bibfield  {journal}
  {\bibinfo  {journal} {Nature Physics}\ }\textbf {\bibinfo {volume} {16}},\
  \bibinfo {pages} {857} (\bibinfo {year} {2020})}\BibitemShut {NoStop}%
\bibitem [{\citenamefont {Kasper}\ \emph {et~al.}(2021)\citenamefont {Kasper},
  \citenamefont {Gonz{\'{a}}lez-Cuadra}, \citenamefont {Hegde}, \citenamefont
  {Xia}, \citenamefont {Dauphin}, \citenamefont {Huber}, \citenamefont
  {Tiemann}, \citenamefont {Lewenstein}, \citenamefont {Jendrzejewski},\ and\
  \citenamefont {Hauke}}]{Kasper_2021}%
  \BibitemOpen
  \bibfield  {author} {\bibinfo {author} {\bibfnamefont {V.}~\bibnamefont
  {Kasper}}, \bibinfo {author} {\bibfnamefont {D.}~\bibnamefont
  {Gonz{\'{a}}lez-Cuadra}}, \bibinfo {author} {\bibfnamefont {A.}~\bibnamefont
  {Hegde}}, \bibinfo {author} {\bibfnamefont {A.}~\bibnamefont {Xia}}, \bibinfo
  {author} {\bibfnamefont {A.}~\bibnamefont {Dauphin}}, \bibinfo {author}
  {\bibfnamefont {F.}~\bibnamefont {Huber}}, \bibinfo {author} {\bibfnamefont
  {E.}~\bibnamefont {Tiemann}}, \bibinfo {author} {\bibfnamefont
  {M.}~\bibnamefont {Lewenstein}}, \bibinfo {author} {\bibfnamefont
  {F.}~\bibnamefont {Jendrzejewski}},\ and\ \bibinfo {author} {\bibfnamefont
  {P.}~\bibnamefont {Hauke}},\ }\bibfield  {title} {\bibinfo {title} {Universal
  quantum computation and quantum error correction with ultracold atomic
  mixtures},\ }\href {https://doi.org/10.1088/2058-9565/ac2d39} {\bibfield
  {journal} {\bibinfo  {journal} {Quantum Science and Technology}\ }\textbf
  {\bibinfo {volume} {7}},\ \bibinfo {pages} {015008} (\bibinfo {year}
  {2021})}\BibitemShut {NoStop}%
\bibitem [{\citenamefont {{Ringbauer}}\ \emph {et~al.}(2021)\citenamefont
  {{Ringbauer}}, \citenamefont {{Meth}}, \citenamefont {{Postler}},
  \citenamefont {{Stricker}}, \citenamefont {{Blatt}}, \citenamefont
  {{Schindler}},\ and\ \citenamefont {{Monz}}}]{Ringbauer_2021}%
  \BibitemOpen
  \bibfield  {author} {\bibinfo {author} {\bibfnamefont {M.}~\bibnamefont
  {{Ringbauer}}}, \bibinfo {author} {\bibfnamefont {M.}~\bibnamefont {{Meth}}},
  \bibinfo {author} {\bibfnamefont {L.}~\bibnamefont {{Postler}}}, \bibinfo
  {author} {\bibfnamefont {R.}~\bibnamefont {{Stricker}}}, \bibinfo {author}
  {\bibfnamefont {R.}~\bibnamefont {{Blatt}}}, \bibinfo {author} {\bibfnamefont
  {P.}~\bibnamefont {{Schindler}}},\ and\ \bibinfo {author} {\bibfnamefont
  {T.}~\bibnamefont {{Monz}}},\ }\bibfield  {title} {\bibinfo {title} {{A
  universal qudit quantum processor with trapped ions}},\ }\href@noop {}
  {\bibfield  {journal} {\bibinfo  {journal} {arXiv e-prints}\ ,\ \bibinfo
  {eid} {arXiv:2109.06903}} (\bibinfo {year} {2021})},\ \Eprint
  {https://arxiv.org/abs/2109.06903} {arXiv:2109.06903 [quant-ph]} \BibitemShut
  {NoStop}%
\bibitem [{\citenamefont {Chi}\ \emph {et~al.}(2022)\citenamefont {Chi},
  \citenamefont {Huang}, \citenamefont {Zhang}, \citenamefont {Mao},
  \citenamefont {Zhou}, \citenamefont {Chen}, \citenamefont {Zhai},
  \citenamefont {Bao}, \citenamefont {Dai}, \citenamefont {Yuan}, \citenamefont
  {Zhang}, \citenamefont {Dai}, \citenamefont {Tang}, \citenamefont {Yang},
  \citenamefont {Li}, \citenamefont {Ding}, \citenamefont {Oxenl{\o}we},
  \citenamefont {Thompson}, \citenamefont {O'Brien}, \citenamefont {Li},
  \citenamefont {Gong},\ and\ \citenamefont {Wang}}]{Chi_2022}%
  \BibitemOpen
  \bibfield  {author} {\bibinfo {author} {\bibfnamefont {Y.}~\bibnamefont
  {Chi}}, \bibinfo {author} {\bibfnamefont {J.}~\bibnamefont {Huang}}, \bibinfo
  {author} {\bibfnamefont {Z.}~\bibnamefont {Zhang}}, \bibinfo {author}
  {\bibfnamefont {J.}~\bibnamefont {Mao}}, \bibinfo {author} {\bibfnamefont
  {Z.}~\bibnamefont {Zhou}}, \bibinfo {author} {\bibfnamefont {X.}~\bibnamefont
  {Chen}}, \bibinfo {author} {\bibfnamefont {C.}~\bibnamefont {Zhai}}, \bibinfo
  {author} {\bibfnamefont {J.}~\bibnamefont {Bao}}, \bibinfo {author}
  {\bibfnamefont {T.}~\bibnamefont {Dai}}, \bibinfo {author} {\bibfnamefont
  {H.}~\bibnamefont {Yuan}}, \bibinfo {author} {\bibfnamefont {M.}~\bibnamefont
  {Zhang}}, \bibinfo {author} {\bibfnamefont {D.}~\bibnamefont {Dai}}, \bibinfo
  {author} {\bibfnamefont {B.}~\bibnamefont {Tang}}, \bibinfo {author}
  {\bibfnamefont {Y.}~\bibnamefont {Yang}}, \bibinfo {author} {\bibfnamefont
  {Z.}~\bibnamefont {Li}}, \bibinfo {author} {\bibfnamefont {Y.}~\bibnamefont
  {Ding}}, \bibinfo {author} {\bibfnamefont {L.~K.}\ \bibnamefont
  {Oxenl{\o}we}}, \bibinfo {author} {\bibfnamefont {M.~G.}\ \bibnamefont
  {Thompson}}, \bibinfo {author} {\bibfnamefont {J.~L.}\ \bibnamefont
  {O'Brien}}, \bibinfo {author} {\bibfnamefont {Y.}~\bibnamefont {Li}},
  \bibinfo {author} {\bibfnamefont {Q.}~\bibnamefont {Gong}},\ and\ \bibinfo
  {author} {\bibfnamefont {J.}~\bibnamefont {Wang}},\ }\bibfield  {title}
  {\bibinfo {title} {A programmable qudit-based quantum processor},\ }\href
  {https://doi.org/10.1038/s41467-022-28767-x} {\bibfield  {journal} {\bibinfo
  {journal} {Nature Communications}\ }\textbf {\bibinfo {volume} {13}},\
  \bibinfo {pages} {1166} (\bibinfo {year} {2022})}\BibitemShut {NoStop}%
\bibitem [{\citenamefont {Levine}\ \emph {et~al.}(2019)\citenamefont {Levine},
  \citenamefont {Keesling}, \citenamefont {Semeghini}, \citenamefont {Omran},
  \citenamefont {Wang}, \citenamefont {Ebadi}, \citenamefont {Bernien},
  \citenamefont {Greiner}, \citenamefont {Vuleti\ifmmode~\acute{c}\else
  \'{c}\fi{}}, \citenamefont {Pichler},\ and\ \citenamefont
  {Lukin}}]{Levine_2019}%
  \BibitemOpen
  \bibfield  {author} {\bibinfo {author} {\bibfnamefont {H.}~\bibnamefont
  {Levine}}, \bibinfo {author} {\bibfnamefont {A.}~\bibnamefont {Keesling}},
  \bibinfo {author} {\bibfnamefont {G.}~\bibnamefont {Semeghini}}, \bibinfo
  {author} {\bibfnamefont {A.}~\bibnamefont {Omran}}, \bibinfo {author}
  {\bibfnamefont {T.~T.}\ \bibnamefont {Wang}}, \bibinfo {author}
  {\bibfnamefont {S.}~\bibnamefont {Ebadi}}, \bibinfo {author} {\bibfnamefont
  {H.}~\bibnamefont {Bernien}}, \bibinfo {author} {\bibfnamefont
  {M.}~\bibnamefont {Greiner}}, \bibinfo {author} {\bibfnamefont
  {V.}~\bibnamefont {Vuleti\ifmmode~\acute{c}\else \'{c}\fi{}}}, \bibinfo
  {author} {\bibfnamefont {H.}~\bibnamefont {Pichler}},\ and\ \bibinfo {author}
  {\bibfnamefont {M.~D.}\ \bibnamefont {Lukin}},\ }\bibfield  {title} {\bibinfo
  {title} {Parallel implementation of high-fidelity multiqubit gates with
  neutral atoms},\ }\href {https://doi.org/10.1103/PhysRevLett.123.170503}
  {\bibfield  {journal} {\bibinfo  {journal} {Phys. Rev. Lett.}\ }\textbf
  {\bibinfo {volume} {123}},\ \bibinfo {pages} {170503} (\bibinfo {year}
  {2019})}\BibitemShut {NoStop}%
\bibitem [{\citenamefont {Ebadi}\ \emph {et~al.}(2021)\citenamefont {Ebadi},
  \citenamefont {Wang}, \citenamefont {Levine}, \citenamefont {Keesling},
  \citenamefont {Semeghini}, \citenamefont {Omran}, \citenamefont {Bluvstein},
  \citenamefont {Samajdar}, \citenamefont {Pichler}, \citenamefont {Ho},
  \citenamefont {Choi}, \citenamefont {Sachdev}, \citenamefont {Greiner},
  \citenamefont {Vuleti{\'c}},\ and\ \citenamefont {Lukin}}]{Ebadi_2021}%
  \BibitemOpen
  \bibfield  {author} {\bibinfo {author} {\bibfnamefont {S.}~\bibnamefont
  {Ebadi}}, \bibinfo {author} {\bibfnamefont {T.~T.}\ \bibnamefont {Wang}},
  \bibinfo {author} {\bibfnamefont {H.}~\bibnamefont {Levine}}, \bibinfo
  {author} {\bibfnamefont {A.}~\bibnamefont {Keesling}}, \bibinfo {author}
  {\bibfnamefont {G.}~\bibnamefont {Semeghini}}, \bibinfo {author}
  {\bibfnamefont {A.}~\bibnamefont {Omran}}, \bibinfo {author} {\bibfnamefont
  {D.}~\bibnamefont {Bluvstein}}, \bibinfo {author} {\bibfnamefont
  {R.}~\bibnamefont {Samajdar}}, \bibinfo {author} {\bibfnamefont
  {H.}~\bibnamefont {Pichler}}, \bibinfo {author} {\bibfnamefont {W.~W.}\
  \bibnamefont {Ho}}, \bibinfo {author} {\bibfnamefont {S.}~\bibnamefont
  {Choi}}, \bibinfo {author} {\bibfnamefont {S.}~\bibnamefont {Sachdev}},
  \bibinfo {author} {\bibfnamefont {M.}~\bibnamefont {Greiner}}, \bibinfo
  {author} {\bibfnamefont {V.}~\bibnamefont {Vuleti{\'c}}},\ and\ \bibinfo
  {author} {\bibfnamefont {M.~D.}\ \bibnamefont {Lukin}},\ }\bibfield  {title}
  {\bibinfo {title} {Quantum phases of matter on a 256-atom programmable
  quantum simulator},\ }\href {https://doi.org/10.1038/s41586-021-03582-4}
  {\bibfield  {journal} {\bibinfo  {journal} {Nature}\ }\textbf {\bibinfo
  {volume} {595}},\ \bibinfo {pages} {227} (\bibinfo {year}
  {2021})}\BibitemShut {NoStop}%
\bibitem [{\citenamefont {Scholl}\ \emph {et~al.}(2021)\citenamefont {Scholl},
  \citenamefont {Schuler}, \citenamefont {Williams}, \citenamefont
  {Eberharter}, \citenamefont {Barredo}, \citenamefont {Schymik}, \citenamefont
  {Lienhard}, \citenamefont {Henry}, \citenamefont {Lang}, \citenamefont
  {Lahaye}, \citenamefont {L{\"a}uchli},\ and\ \citenamefont
  {Browaeys}}]{Scholl_2021}%
  \BibitemOpen
  \bibfield  {author} {\bibinfo {author} {\bibfnamefont {P.}~\bibnamefont
  {Scholl}}, \bibinfo {author} {\bibfnamefont {M.}~\bibnamefont {Schuler}},
  \bibinfo {author} {\bibfnamefont {H.~J.}\ \bibnamefont {Williams}}, \bibinfo
  {author} {\bibfnamefont {A.~A.}\ \bibnamefont {Eberharter}}, \bibinfo
  {author} {\bibfnamefont {D.}~\bibnamefont {Barredo}}, \bibinfo {author}
  {\bibfnamefont {K.-N.}\ \bibnamefont {Schymik}}, \bibinfo {author}
  {\bibfnamefont {V.}~\bibnamefont {Lienhard}}, \bibinfo {author}
  {\bibfnamefont {L.-P.}\ \bibnamefont {Henry}}, \bibinfo {author}
  {\bibfnamefont {T.~C.}\ \bibnamefont {Lang}}, \bibinfo {author}
  {\bibfnamefont {T.}~\bibnamefont {Lahaye}}, \bibinfo {author} {\bibfnamefont
  {A.~M.}\ \bibnamefont {L{\"a}uchli}},\ and\ \bibinfo {author} {\bibfnamefont
  {A.}~\bibnamefont {Browaeys}},\ }\bibfield  {title} {\bibinfo {title}
  {Quantum simulation of 2d antiferromagnets with hundreds of rydberg atoms},\
  }\href {https://doi.org/10.1038/s41586-021-03585-1} {\bibfield  {journal}
  {\bibinfo  {journal} {Nature}\ }\textbf {\bibinfo {volume} {595}},\ \bibinfo
  {pages} {233} (\bibinfo {year} {2021})}\BibitemShut {NoStop}%
\bibitem [{\citenamefont {{Young}}\ \emph {et~al.}(2022)\citenamefont
  {{Young}}, \citenamefont {{Eckner}}, \citenamefont {{Schine}}, \citenamefont
  {{Childs}},\ and\ \citenamefont {{Kaufman}}}]{Young_2022}%
  \BibitemOpen
  \bibfield  {author} {\bibinfo {author} {\bibfnamefont {A.~W.}\ \bibnamefont
  {{Young}}}, \bibinfo {author} {\bibfnamefont {W.~J.}\ \bibnamefont
  {{Eckner}}}, \bibinfo {author} {\bibfnamefont {N.}~\bibnamefont {{Schine}}},
  \bibinfo {author} {\bibfnamefont {A.~M.}\ \bibnamefont {{Childs}}},\ and\
  \bibinfo {author} {\bibfnamefont {A.~M.}\ \bibnamefont {{Kaufman}}},\
  }\bibfield  {title} {\bibinfo {title} {{Tweezer-programmable 2D quantum walks
  in a Hubbard-regime lattice}},\ }\href@noop {} {\bibfield  {journal}
  {\bibinfo  {journal} {arXiv e-prints}\ ,\ \bibinfo {eid} {arXiv:2202.01204}}
  (\bibinfo {year} {2022})},\ \Eprint {https://arxiv.org/abs/2202.01204}
  {arXiv:2202.01204 [quant-ph]} \BibitemShut {NoStop}%
\bibitem [{\citenamefont {Wang}\ \emph {et~al.}(2020)\citenamefont {Wang},
  \citenamefont {Hu}, \citenamefont {Sanders},\ and\ \citenamefont
  {Kais}}]{Wang_2020}%
  \BibitemOpen
  \bibfield  {author} {\bibinfo {author} {\bibfnamefont {Y.}~\bibnamefont
  {Wang}}, \bibinfo {author} {\bibfnamefont {Z.}~\bibnamefont {Hu}}, \bibinfo
  {author} {\bibfnamefont {B.~C.}\ \bibnamefont {Sanders}},\ and\ \bibinfo
  {author} {\bibfnamefont {S.}~\bibnamefont {Kais}},\ }\bibfield  {title}
  {\bibinfo {title} {Qudits and high-dimensional quantum computing},\
  }\bibfield  {journal} {\bibinfo  {journal} {Frontiers in Physics}\ }\textbf
  {\bibinfo {volume} {8}},\ \href {https://doi.org/10.3389/fphy.2020.589504}
  {10.3389/fphy.2020.589504} (\bibinfo {year} {2020})\BibitemShut {NoStop}%
\bibitem [{\citenamefont {Gustafson}(2021)}]{Gustafson_2021}%
  \BibitemOpen
  \bibfield  {author} {\bibinfo {author} {\bibfnamefont {E.~J.}\ \bibnamefont
  {Gustafson}},\ }\bibfield  {title} {\bibinfo {title} {Prospects for
  simulating a qudit-based model of $(1+1)\mathrm{D}$ scalar qed},\ }\href
  {https://doi.org/10.1103/PhysRevD.103.114505} {\bibfield  {journal} {\bibinfo
   {journal} {Phys. Rev. D}\ }\textbf {\bibinfo {volume} {103}},\ \bibinfo
  {pages} {114505} (\bibinfo {year} {2021})}\BibitemShut {NoStop}%
\bibitem [{\citenamefont {{Gustafson}}(2022)}]{Gustafson_2022}%
  \BibitemOpen
  \bibfield  {author} {\bibinfo {author} {\bibfnamefont {E.}~\bibnamefont
  {{Gustafson}}},\ }\bibfield  {title} {\bibinfo {title} {{Noise Improvements
  in Quantum Simulations of sQED using Qutrits}},\ }\href@noop {} {\bibfield
  {journal} {\bibinfo  {journal} {arXiv e-prints}\ ,\ \bibinfo {eid}
  {arXiv:2201.04546}} (\bibinfo {year} {2022})},\ \Eprint
  {https://arxiv.org/abs/2201.04546} {arXiv:2201.04546 [quant-ph]} \BibitemShut
  {NoStop}%
\bibitem [{\citenamefont {Saffman}\ \emph {et~al.}(2010)\citenamefont
  {Saffman}, \citenamefont {Walker},\ and\ \citenamefont
  {M\o{}lmer}}]{Saffman_2010}%
  \BibitemOpen
  \bibfield  {author} {\bibinfo {author} {\bibfnamefont {M.}~\bibnamefont
  {Saffman}}, \bibinfo {author} {\bibfnamefont {T.~G.}\ \bibnamefont
  {Walker}},\ and\ \bibinfo {author} {\bibfnamefont {K.}~\bibnamefont
  {M\o{}lmer}},\ }\bibfield  {title} {\bibinfo {title} {Quantum information
  with rydberg atoms},\ }\href {https://doi.org/10.1103/RevModPhys.82.2313}
  {\bibfield  {journal} {\bibinfo  {journal} {Rev. Mod. Phys.}\ }\textbf
  {\bibinfo {volume} {82}},\ \bibinfo {pages} {2313} (\bibinfo {year}
  {2010})}\BibitemShut {NoStop}%
\bibitem [{\citenamefont {Zanardi}\ and\ \citenamefont
  {Rasetti}(1999)}]{Zanardi_1999}%
  \BibitemOpen
  \bibfield  {author} {\bibinfo {author} {\bibfnamefont {P.}~\bibnamefont
  {Zanardi}}\ and\ \bibinfo {author} {\bibfnamefont {M.}~\bibnamefont
  {Rasetti}},\ }\bibfield  {title} {\bibinfo {title} {Holonomic quantum
  computation},\ }\href
  {https://doi.org/https://doi.org/10.1016/S0375-9601(99)00803-8} {\bibfield
  {journal} {\bibinfo  {journal} {Physics Letters A}\ }\textbf {\bibinfo
  {volume} {264}},\ \bibinfo {pages} {94} (\bibinfo {year} {1999})}\BibitemShut
  {NoStop}%
\bibitem [{\citenamefont {Sj{\"o}qvist}\ \emph {et~al.}(2012)\citenamefont
  {Sj{\"o}qvist}, \citenamefont {Tong}, \citenamefont {Andersson},
  \citenamefont {Hessmo}, \citenamefont {Johansson},\ and\ \citenamefont
  {Singh}}]{Sjoqvist_2012}%
  \BibitemOpen
  \bibfield  {author} {\bibinfo {author} {\bibfnamefont {E.}~\bibnamefont
  {Sj{\"o}qvist}}, \bibinfo {author} {\bibfnamefont {D.~M.}\ \bibnamefont
  {Tong}}, \bibinfo {author} {\bibfnamefont {L.~M.}\ \bibnamefont {Andersson}},
  \bibinfo {author} {\bibfnamefont {B.}~\bibnamefont {Hessmo}}, \bibinfo
  {author} {\bibfnamefont {M.}~\bibnamefont {Johansson}},\ and\ \bibinfo
  {author} {\bibfnamefont {K.}~\bibnamefont {Singh}},\ }\bibfield  {title}
  {\bibinfo {title} {Non-adiabatic holonomic quantum computation},\ }\href
  {https://doi.org/10.1088/1367-2630/14/10/103035} {\bibfield  {journal}
  {\bibinfo  {journal} {New Journal of Physics}\ }\textbf {\bibinfo {volume}
  {14}},\ \bibinfo {pages} {103035} (\bibinfo {year} {2012})}\BibitemShut
  {NoStop}%
\bibitem [{\citenamefont {Xu}\ \emph {et~al.}(2012)\citenamefont {Xu},
  \citenamefont {Zhang}, \citenamefont {Tong}, \citenamefont {Sj\"oqvist},\
  and\ \citenamefont {Kwek}}]{Xu_2012}%
  \BibitemOpen
  \bibfield  {author} {\bibinfo {author} {\bibfnamefont {G.~F.}\ \bibnamefont
  {Xu}}, \bibinfo {author} {\bibfnamefont {J.}~\bibnamefont {Zhang}}, \bibinfo
  {author} {\bibfnamefont {D.~M.}\ \bibnamefont {Tong}}, \bibinfo {author}
  {\bibfnamefont {E.}~\bibnamefont {Sj\"oqvist}},\ and\ \bibinfo {author}
  {\bibfnamefont {L.~C.}\ \bibnamefont {Kwek}},\ }\bibfield  {title} {\bibinfo
  {title} {Nonadiabatic holonomic quantum computation in decoherence-free
  subspaces},\ }\href {https://doi.org/10.1103/PhysRevLett.109.170501}
  {\bibfield  {journal} {\bibinfo  {journal} {Phys. Rev. Lett.}\ }\textbf
  {\bibinfo {volume} {109}},\ \bibinfo {pages} {170501} (\bibinfo {year}
  {2012})}\BibitemShut {NoStop}%
\bibitem [{\citenamefont {Feng}\ \emph {et~al.}(2013)\citenamefont {Feng},
  \citenamefont {Xu},\ and\ \citenamefont {Long}}]{Feng_2013}%
  \BibitemOpen
  \bibfield  {author} {\bibinfo {author} {\bibfnamefont {G.}~\bibnamefont
  {Feng}}, \bibinfo {author} {\bibfnamefont {G.}~\bibnamefont {Xu}},\ and\
  \bibinfo {author} {\bibfnamefont {G.}~\bibnamefont {Long}},\ }\bibfield
  {title} {\bibinfo {title} {Experimental realization of nonadiabatic holonomic
  quantum computation},\ }\href
  {https://doi.org/10.1103/PhysRevLett.110.190501} {\bibfield  {journal}
  {\bibinfo  {journal} {Phys. Rev. Lett.}\ }\textbf {\bibinfo {volume} {110}},\
  \bibinfo {pages} {190501} (\bibinfo {year} {2013})}\BibitemShut {NoStop}%
\bibitem [{\citenamefont {Kang}\ \emph {et~al.}(2018)\citenamefont {Kang},
  \citenamefont {Chen}, \citenamefont {Shi}, \citenamefont {Huang},
  \citenamefont {Song},\ and\ \citenamefont {Xia}}]{Kang_2018}%
  \BibitemOpen
  \bibfield  {author} {\bibinfo {author} {\bibfnamefont {Y.-H.}\ \bibnamefont
  {Kang}}, \bibinfo {author} {\bibfnamefont {Y.-H.}\ \bibnamefont {Chen}},
  \bibinfo {author} {\bibfnamefont {Z.-C.}\ \bibnamefont {Shi}}, \bibinfo
  {author} {\bibfnamefont {B.-H.}\ \bibnamefont {Huang}}, \bibinfo {author}
  {\bibfnamefont {J.}~\bibnamefont {Song}},\ and\ \bibinfo {author}
  {\bibfnamefont {Y.}~\bibnamefont {Xia}},\ }\bibfield  {title} {\bibinfo
  {title} {Nonadiabatic holonomic quantum computation using rydberg blockade},\
  }\href {https://doi.org/10.1103/PhysRevA.97.042336} {\bibfield  {journal}
  {\bibinfo  {journal} {Phys. Rev. A}\ }\textbf {\bibinfo {volume} {97}},\
  \bibinfo {pages} {042336} (\bibinfo {year} {2018})}\BibitemShut {NoStop}%
\bibitem [{Note1()}]{Note1}%
  \BibitemOpen
  \bibinfo {note} {Note that the advantage of using qudits has also been
  recently demonstrated in the context QAOA \cite
  {Weggemans2022solvingcorrelation}.}\BibitemShut {Stop}%
\bibitem [{\citenamefont {Levin}\ and\ \citenamefont {Wen}(2005)}]{Levin_2005}%
  \BibitemOpen
  \bibfield  {author} {\bibinfo {author} {\bibfnamefont {M.~A.}\ \bibnamefont
  {Levin}}\ and\ \bibinfo {author} {\bibfnamefont {X.-G.}\ \bibnamefont
  {Wen}},\ }\bibfield  {title} {\bibinfo {title} {String-net condensation: A
  physical mechanism for topological phases},\ }\href
  {https://doi.org/10.1103/PhysRevB.71.045110} {\bibfield  {journal} {\bibinfo
  {journal} {Phys. Rev. B}\ }\textbf {\bibinfo {volume} {71}},\ \bibinfo
  {pages} {045110} (\bibinfo {year} {2005})}\BibitemShut {NoStop}%
\bibitem [{\citenamefont {Xu}\ and\ \citenamefont {Ludwig}(2012)}]{Xu_2012_2}%
  \BibitemOpen
  \bibfield  {author} {\bibinfo {author} {\bibfnamefont {C.}~\bibnamefont
  {Xu}}\ and\ \bibinfo {author} {\bibfnamefont {A.~W.}\ \bibnamefont
  {Ludwig}},\ }\bibfield  {title} {\bibinfo {title} {Topological quantum
  liquids with quaternion non-abelian statistics},\ }\href@noop {} {\bibfield
  {journal} {\bibinfo  {journal} {Physical Review Letters}\ }\textbf {\bibinfo
  {volume} {108}},\ \bibinfo {pages} {047202} (\bibinfo {year}
  {2012})}\BibitemShut {NoStop}%
\bibitem [{\citenamefont {Trautmann}\ \emph {et~al.}(2021)\citenamefont
  {Trautmann}, \citenamefont {Mark}, \citenamefont {Ilzh\"ofer}, \citenamefont
  {Edri}, \citenamefont {Arrach}, \citenamefont {Maloberti}, \citenamefont
  {Greene}, \citenamefont {Robicheaux},\ and\ \citenamefont
  {Ferlaino}}]{Trautmann_2021}%
  \BibitemOpen
  \bibfield  {author} {\bibinfo {author} {\bibfnamefont {A.}~\bibnamefont
  {Trautmann}}, \bibinfo {author} {\bibfnamefont {M.~J.}\ \bibnamefont {Mark}},
  \bibinfo {author} {\bibfnamefont {P.}~\bibnamefont {Ilzh\"ofer}}, \bibinfo
  {author} {\bibfnamefont {H.}~\bibnamefont {Edri}}, \bibinfo {author}
  {\bibfnamefont {A.~E.}\ \bibnamefont {Arrach}}, \bibinfo {author}
  {\bibfnamefont {J.~G.}\ \bibnamefont {Maloberti}}, \bibinfo {author}
  {\bibfnamefont {C.~H.}\ \bibnamefont {Greene}}, \bibinfo {author}
  {\bibfnamefont {F.}~\bibnamefont {Robicheaux}},\ and\ \bibinfo {author}
  {\bibfnamefont {F.}~\bibnamefont {Ferlaino}},\ }\bibfield  {title} {\bibinfo
  {title} {Spectroscopy of rydberg states in erbium using electromagnetically
  induced transparency},\ }\href
  {https://doi.org/10.1103/PhysRevResearch.3.033165} {\bibfield  {journal}
  {\bibinfo  {journal} {Phys. Rev. Research}\ }\textbf {\bibinfo {volume}
  {3}},\ \bibinfo {pages} {033165} (\bibinfo {year} {2021})}\BibitemShut
  {NoStop}%
\bibitem [{\citenamefont {Saffman}\ and\ \citenamefont
  {M\o{}lmer}(2008)}]{Saffman_2008}%
  \BibitemOpen
  \bibfield  {author} {\bibinfo {author} {\bibfnamefont {M.}~\bibnamefont
  {Saffman}}\ and\ \bibinfo {author} {\bibfnamefont {K.}~\bibnamefont
  {M\o{}lmer}},\ }\bibfield  {title} {\bibinfo {title} {Scaling the
  neutral-atom rydberg gate quantum computer by collective encoding in holmium
  atoms},\ }\href {https://doi.org/10.1103/PhysRevA.78.012336} {\bibfield
  {journal} {\bibinfo  {journal} {Phys. Rev. A}\ }\textbf {\bibinfo {volume}
  {78}},\ \bibinfo {pages} {012336} (\bibinfo {year} {2008})}\BibitemShut
  {NoStop}%
\bibitem [{\citenamefont {Hostetter}\ \emph {et~al.}(2015)\citenamefont
  {Hostetter}, \citenamefont {Pritchard}, \citenamefont {Lawler},\ and\
  \citenamefont {Saffman}}]{Hostetter_2015}%
  \BibitemOpen
  \bibfield  {author} {\bibinfo {author} {\bibfnamefont {J.}~\bibnamefont
  {Hostetter}}, \bibinfo {author} {\bibfnamefont {J.~D.}\ \bibnamefont
  {Pritchard}}, \bibinfo {author} {\bibfnamefont {J.~E.}\ \bibnamefont
  {Lawler}},\ and\ \bibinfo {author} {\bibfnamefont {M.}~\bibnamefont
  {Saffman}},\ }\bibfield  {title} {\bibinfo {title} {Measurement of holmium
  rydberg series through magneto-optical trap depletion spectroscopy},\ }\href
  {https://doi.org/10.1103/PhysRevA.91.012507} {\bibfield  {journal} {\bibinfo
  {journal} {Phys. Rev. A}\ }\textbf {\bibinfo {volume} {91}},\ \bibinfo
  {pages} {012507} (\bibinfo {year} {2015})}\BibitemShut {NoStop}%
\bibitem [{\citenamefont {Xu}\ \emph {et~al.}(2021)\citenamefont {Xu},
  \citenamefont {Zhao}, \citenamefont {Sj\"oqvist},\ and\ \citenamefont
  {Tong}}]{PhysRevA.103.052605}%
  \BibitemOpen
  \bibfield  {author} {\bibinfo {author} {\bibfnamefont {G.~F.}\ \bibnamefont
  {Xu}}, \bibinfo {author} {\bibfnamefont {P.~Z.}\ \bibnamefont {Zhao}},
  \bibinfo {author} {\bibfnamefont {E.}~\bibnamefont {Sj\"oqvist}},\ and\
  \bibinfo {author} {\bibfnamefont {D.~M.}\ \bibnamefont {Tong}},\ }\bibfield
  {title} {\bibinfo {title} {Realizing nonadiabatic holonomic quantum
  computation beyond the three-level setting},\ }\href
  {https://doi.org/10.1103/PhysRevA.103.052605} {\bibfield  {journal} {\bibinfo
   {journal} {Phys. Rev. A}\ }\textbf {\bibinfo {volume} {103}},\ \bibinfo
  {pages} {052605} (\bibinfo {year} {2021})}\BibitemShut {NoStop}%
\bibitem [{\citenamefont {Li}\ \emph {et~al.}(2013)\citenamefont {Li},
  \citenamefont {Roberts},\ and\ \citenamefont {Yin}}]{Li_2013}%
  \BibitemOpen
  \bibfield  {author} {\bibinfo {author} {\bibfnamefont {C.-K.}\ \bibnamefont
  {Li}}, \bibinfo {author} {\bibfnamefont {R.}~\bibnamefont {Roberts}},\ and\
  \bibinfo {author} {\bibfnamefont {X.}~\bibnamefont {Yin}},\ }\bibfield
  {title} {\bibinfo {title} {Decomposition of unitary matrices and quantum
  gates},\ }\href {https://doi.org/10.1142/S0219749913500159} {\bibfield
  {journal} {\bibinfo  {journal} {International Journal of Quantum
  Information}\ }\textbf {\bibinfo {volume} {11}},\ \bibinfo {pages} {1350015}
  (\bibinfo {year} {2013})}\BibitemShut {NoStop}%
\bibitem [{SM()}]{SM}%
  \BibitemOpen
  \href@noop {} {}\bibinfo {note} {See the Supplementary Material for further
  details on the implementation of holonomic gates and the digital quantum
  simulation of the $Q_8$ LGT, including a comparison with a qubit-based
  protocol.}\BibitemShut {Stop}%
\bibitem [{Car()}]{Carrasco_2022}%
  \BibitemOpen
  \href@noop {} {}\bibinfo {note} {J. Carrasco \textit{et al.}, in preparation
  (2022)}\BibitemShut {NoStop}%
\bibitem [{\citenamefont {Kogut}\ and\ \citenamefont
  {Susskind}(1975)}]{kogut1975hamiltonian}%
  \BibitemOpen
  \bibfield  {author} {\bibinfo {author} {\bibfnamefont {J.}~\bibnamefont
  {Kogut}}\ and\ \bibinfo {author} {\bibfnamefont {L.}~\bibnamefont
  {Susskind}},\ }\bibfield  {title} {\bibinfo {title} {Hamiltonian formulation
  of wilson's lattice gauge theories},\ }\href@noop {} {\bibfield  {journal}
  {\bibinfo  {journal} {Physical Review D}\ }\textbf {\bibinfo {volume} {11}},\
  \bibinfo {pages} {395} (\bibinfo {year} {1975})}\BibitemShut {NoStop}%
\bibitem [{\citenamefont {Trotter}(1959)}]{Trotter_1959}%
  \BibitemOpen
  \bibfield  {author} {\bibinfo {author} {\bibfnamefont {H.~F.}\ \bibnamefont
  {Trotter}},\ }\bibfield  {title} {\bibinfo {title} {On the product of
  semi-groups of operators},\ }\href {http://www.jstor.org/stable/2033649}
  {\bibfield  {journal} {\bibinfo  {journal} {Proceedings of the American
  Mathematical Society}\ }\textbf {\bibinfo {volume} {10}},\ \bibinfo {pages}
  {545} (\bibinfo {year} {1959})}\BibitemShut {NoStop}%
\bibitem [{\citenamefont {Hatano}\ and\ \citenamefont
  {Suzuki}(2005)}]{Hatano_2005}%
  \BibitemOpen
  \bibfield  {author} {\bibinfo {author} {\bibfnamefont {N.}~\bibnamefont
  {Hatano}}\ and\ \bibinfo {author} {\bibfnamefont {M.}~\bibnamefont
  {Suzuki}},\ }\bibinfo {title} {Finding exponential product formulas of higher
  orders},\ in\ \href {https://doi.org/10.1007/11526216_2} {\emph {\bibinfo
  {booktitle} {Quantum Annealing and Other Optimization Methods}}},\ \bibinfo
  {editor} {edited by\ \bibinfo {editor} {\bibfnamefont {A.}~\bibnamefont
  {Das}}\ and\ \bibinfo {editor} {\bibfnamefont {B.}~\bibnamefont
  {K.~Chakrabarti}}}\ (\bibinfo  {publisher} {Springer Berlin Heidelberg},\
  \bibinfo {address} {Berlin, Heidelberg},\ \bibinfo {year} {2005})\ pp.\
  \bibinfo {pages} {37--68}\BibitemShut {NoStop}%
\bibitem [{Note2()}]{Note2}%
  \BibitemOpen
  \bibinfo {note} {We note that the fidelity of the group-multiplication gate
  could be further improved through optimal-control methods~\cite
  {Jandura_2022}, where other experimental imperfections such as phase errors
  due to Stark shifts could be taken into account.}\BibitemShut {Stop}%
\bibitem [{Note3()}]{Note3}%
  \BibitemOpen
  \bibinfo {note} {Note that, even if the qubit decomposition found is not
  necessary the optimal one, improved decompositions are not expected to
  qualitatively change these results.}\BibitemShut {Stop}%
\bibitem [{Gon()}]{Gonzalez_2022}%
  \BibitemOpen
  \href@noop {} {}\bibinfo {note} {D. Gonz{\'a}lez-Cuadra \textit{et al.}, in
  preparation (2022)}\BibitemShut {NoStop}%
\bibitem [{\citenamefont {McArdle}\ \emph {et~al.}(2020)\citenamefont
  {McArdle}, \citenamefont {Endo}, \citenamefont {Aspuru-Guzik}, \citenamefont
  {Benjamin},\ and\ \citenamefont {Yuan}}]{RevModPhys.92.015003}%
  \BibitemOpen
  \bibfield  {author} {\bibinfo {author} {\bibfnamefont {S.}~\bibnamefont
  {McArdle}}, \bibinfo {author} {\bibfnamefont {S.}~\bibnamefont {Endo}},
  \bibinfo {author} {\bibfnamefont {A.}~\bibnamefont {Aspuru-Guzik}}, \bibinfo
  {author} {\bibfnamefont {S.~C.}\ \bibnamefont {Benjamin}},\ and\ \bibinfo
  {author} {\bibfnamefont {X.}~\bibnamefont {Yuan}},\ }\bibfield  {title}
  {\bibinfo {title} {Quantum computational chemistry},\ }\href
  {https://doi.org/10.1103/RevModPhys.92.015003} {\bibfield  {journal}
  {\bibinfo  {journal} {Rev. Mod. Phys.}\ }\textbf {\bibinfo {volume} {92}},\
  \bibinfo {pages} {015003} (\bibinfo {year} {2020})}\BibitemShut {NoStop}%
\bibitem [{\citenamefont {Weggemans}\ \emph {et~al.}(2022)\citenamefont
  {Weggemans}, \citenamefont {Urech}, \citenamefont {Rausch}, \citenamefont
  {Spreeuw}, \citenamefont {Boucherie}, \citenamefont {Schreck}, \citenamefont
  {Schoutens}, \citenamefont {Min{\'{a}}{\v{r}}},\ and\ \citenamefont
  {Speelman}}]{Weggemans2022solvingcorrelation}%
  \BibitemOpen
  \bibfield  {author} {\bibinfo {author} {\bibfnamefont {J.~R.}\ \bibnamefont
  {Weggemans}}, \bibinfo {author} {\bibfnamefont {A.}~\bibnamefont {Urech}},
  \bibinfo {author} {\bibfnamefont {A.}~\bibnamefont {Rausch}}, \bibinfo
  {author} {\bibfnamefont {R.}~\bibnamefont {Spreeuw}}, \bibinfo {author}
  {\bibfnamefont {R.}~\bibnamefont {Boucherie}}, \bibinfo {author}
  {\bibfnamefont {F.}~\bibnamefont {Schreck}}, \bibinfo {author} {\bibfnamefont
  {K.}~\bibnamefont {Schoutens}}, \bibinfo {author} {\bibfnamefont
  {J.}~\bibnamefont {Min{\'{a}}{\v{r}}}},\ and\ \bibinfo {author}
  {\bibfnamefont {F.}~\bibnamefont {Speelman}},\ }\bibfield  {title} {\bibinfo
  {title} {Solving correlation clustering with {QAOA} and a {R}ydberg qudit
  system: a full-stack approach},\ }\href
  {https://doi.org/10.22331/q-2022-04-13-687} {\bibfield  {journal} {\bibinfo
  {journal} {{Quantum}}\ }\textbf {\bibinfo {volume} {6}},\ \bibinfo {pages}
  {687} (\bibinfo {year} {2022})}\BibitemShut {NoStop}%
\bibitem [{\citenamefont {{Jandura}}\ and\ \citenamefont
  {{Pupillo}}(2022)}]{Jandura_2022}%
  \BibitemOpen
  \bibfield  {author} {\bibinfo {author} {\bibfnamefont {S.}~\bibnamefont
  {{Jandura}}}\ and\ \bibinfo {author} {\bibfnamefont {G.}~\bibnamefont
  {{Pupillo}}},\ }\bibfield  {title} {\bibinfo {title} {{Time-Optimal Two- and
  Three-Qubit Gates for Rydberg Atoms}},\ }\href@noop {} {\bibfield  {journal}
  {\bibinfo  {journal} {arXiv e-prints}\ ,\ \bibinfo {eid} {arXiv:2202.00903}}
  (\bibinfo {year} {2022})},\ \Eprint {https://arxiv.org/abs/2202.00903}
  {arXiv:2202.00903 [quant-ph]} \BibitemShut {NoStop}%
\bibitem [{\citenamefont {Pedersen}\ \emph {et~al.}(2007)\citenamefont
  {Pedersen}, \citenamefont {M{\o}ller},\ and\ \citenamefont
  {M{\o}lmer}}]{Pedersen_2007}%
  \BibitemOpen
  \bibfield  {author} {\bibinfo {author} {\bibfnamefont {L.~H.}\ \bibnamefont
  {Pedersen}}, \bibinfo {author} {\bibfnamefont {N.~M.}\ \bibnamefont
  {M{\o}ller}},\ and\ \bibinfo {author} {\bibfnamefont {K.}~\bibnamefont
  {M{\o}lmer}},\ }\bibfield  {title} {\bibinfo {title} {Fidelity of quantum
  operations},\ }\href
  {https://doi.org/https://doi.org/10.1016/j.physleta.2007.02.069} {\bibfield
  {journal} {\bibinfo  {journal} {Physics Letters A}\ }\textbf {\bibinfo
  {volume} {367}},\ \bibinfo {pages} {47} (\bibinfo {year} {2007})}\BibitemShut
  {NoStop}%
\bibitem [{\citenamefont {Nam}\ \emph {et~al.}(2018)\citenamefont {Nam},
  \citenamefont {Ross}, \citenamefont {Su}, \citenamefont {Childs},\ and\
  \citenamefont {Maslov}}]{Nam_2018}%
  \BibitemOpen
  \bibfield  {author} {\bibinfo {author} {\bibfnamefont {Y.}~\bibnamefont
  {Nam}}, \bibinfo {author} {\bibfnamefont {N.~J.}\ \bibnamefont {Ross}},
  \bibinfo {author} {\bibfnamefont {Y.}~\bibnamefont {Su}}, \bibinfo {author}
  {\bibfnamefont {A.~M.}\ \bibnamefont {Childs}},\ and\ \bibinfo {author}
  {\bibfnamefont {D.}~\bibnamefont {Maslov}},\ }\bibfield  {title} {\bibinfo
  {title} {Automated optimization of large quantum circuits with continuous
  parameters},\ }\href {https://doi.org/10.1038/s41534-018-0072-4} {\bibfield
  {journal} {\bibinfo  {journal} {npj Quantum Information}\ }\textbf {\bibinfo
  {volume} {4}},\ \bibinfo {pages} {23} (\bibinfo {year} {2018})}\BibitemShut
  {NoStop}%
\bibitem [{\citenamefont {Barenco}\ \emph {et~al.}(1995)\citenamefont
  {Barenco}, \citenamefont {Bennett}, \citenamefont {Cleve}, \citenamefont
  {DiVincenzo}, \citenamefont {Margolus}, \citenamefont {Shor}, \citenamefont
  {Sleator}, \citenamefont {Smolin},\ and\ \citenamefont
  {Weinfurter}}]{barenco1995elementary}%
  \BibitemOpen
  \bibfield  {author} {\bibinfo {author} {\bibfnamefont {A.}~\bibnamefont
  {Barenco}}, \bibinfo {author} {\bibfnamefont {C.~H.}\ \bibnamefont
  {Bennett}}, \bibinfo {author} {\bibfnamefont {R.}~\bibnamefont {Cleve}},
  \bibinfo {author} {\bibfnamefont {D.~P.}\ \bibnamefont {DiVincenzo}},
  \bibinfo {author} {\bibfnamefont {N.}~\bibnamefont {Margolus}}, \bibinfo
  {author} {\bibfnamefont {P.}~\bibnamefont {Shor}}, \bibinfo {author}
  {\bibfnamefont {T.}~\bibnamefont {Sleator}}, \bibinfo {author} {\bibfnamefont
  {J.~A.}\ \bibnamefont {Smolin}},\ and\ \bibinfo {author} {\bibfnamefont
  {H.}~\bibnamefont {Weinfurter}},\ }\bibfield  {title} {\bibinfo {title}
  {Elementary gates for quantum computation},\ }\href@noop {} {\bibfield
  {journal} {\bibinfo  {journal} {Physical review A}\ }\textbf {\bibinfo
  {volume} {52}},\ \bibinfo {pages} {3457} (\bibinfo {year}
  {1995})}\BibitemShut {NoStop}%
\bibitem [{\citenamefont {Tong}\ \emph {et~al.}(2022)\citenamefont {Tong},
  \citenamefont {Albert}, \citenamefont {McClean}, \citenamefont {Preskill},\
  and\ \citenamefont {Su}}]{Tong_2022}%
  \BibitemOpen
  \bibfield  {author} {\bibinfo {author} {\bibfnamefont {Y.}~\bibnamefont
  {Tong}}, \bibinfo {author} {\bibfnamefont {V.~V.}\ \bibnamefont {Albert}},
  \bibinfo {author} {\bibfnamefont {J.~R.}\ \bibnamefont {McClean}}, \bibinfo
  {author} {\bibfnamefont {J.}~\bibnamefont {Preskill}},\ and\ \bibinfo
  {author} {\bibfnamefont {Y.}~\bibnamefont {Su}},\ }\bibfield  {title}
  {\bibinfo {title} {Provably accurate simulation of gauge theories and bosonic
  systems},\ }\href {https://doi.org/10.22331/q-2022-09-22-816} {\bibfield
  {journal} {\bibinfo  {journal} {{Quantum}}\ }\textbf {\bibinfo {volume}
  {6}},\ \bibinfo {pages} {816} (\bibinfo {year} {2022})}\BibitemShut {NoStop}%
\end{thebibliography}%

\clearpage

\appendix
\onecolumngrid
\section*{Supplemental Material to ``Qudit-based quantum simulation of non-abelian gauge theories with Rydberg atoms"}

In this Supplementary Material we discuss further details for the holonomic implementation of single-qudit gates and the properties of the quaternion group, including its gauge invariant Hamiltonian and the pulse sequence required to implement the corresponding group-multiplication gate. We also include a comparison with a qubit-based approach, showing an explicit gate count for the quaternion LGT as well as an error estimate.\\

\twocolumngrid

\section{Single-qudit gates}

\subsection{Holonomic implementation}

\begin{figure}[t]
	\centering
	\includegraphics[width=1.0\linewidth]{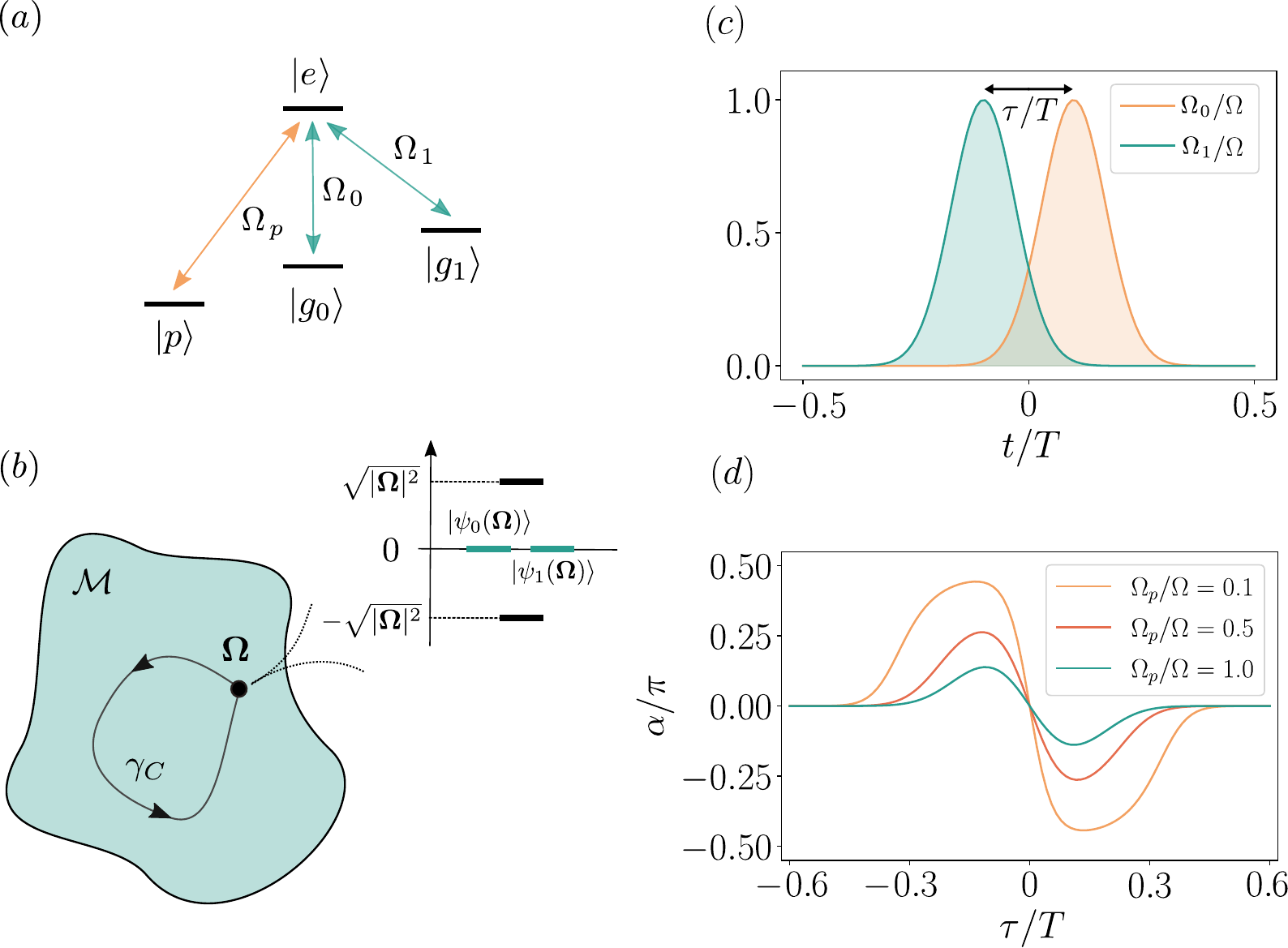}
	\caption{{\bf Single-qudit holonomic gates:} (a) Atomic structure used to perform $u \in {\rm SU}(2)$ operations on $\mathcal{H}$, a two-level system formed by $\ket{g_0}$ and $\ket{g_1}$. These states, as well as an auxiliary state $\ket{p}$, are coupled to an excited state $\ket{e}$ with laser pulses characterized by the Rabi frequencies $\mathbf{\Omega} = (\Omega_{0}, \Omega_{1}, \Omega_{p}) \in \mathcal{M}$. (b) Unitary operations $u$ are performed holonomically in $\mathcal{H}_j$ by adiabatically modifying $\mathbf{\Omega}$ in $\mathcal{M}$ following close loops $\gamma_C$. At every point along the loop, the level structure present two degenerate dark states $\psi_{a=0,1}(\mathbf{\Omega})$ separated from the rest by a gap $\Delta(\mathbf{\Omega})$. (c) Two overlapping Gaussian pulses separated by a time delay $\tau$. (d) Rotation angle $\alpha$ as a function of $\tau / T$ for different values of $\Omega_p/\Omega$. Note that larger values of $\alpha$ can be reached by applying $k$ consecutive pulses with $\alpha_k = \alpha / k$.}
	\label{fig:fig1SM}
\end{figure}

We specify here how to implement general single-qudit holonomic gates, less prone to errors~\cite{Zanardi_1999}, using the experimentally available Hamiltonian introduced below. As discussed in the seminal work~\cite{Zanardi_1999}, the notion of generalized Berry phase (i.e., non-Abelian holonomy) can be used to perform universal quantum computations. The computational space is the $d$-fold degenerate subspace of a family of Hamiltonians ({\em dark states}) parametrized by a manifold $\mathcal M$ representing the couplings between subsystems. This subspace can be taken to be the degenerate null-space of the family of Hamiltonians to be introduced. Adiabatic closed paths in $\mathcal M$ induce special unitary transformations ${\rm SU}(d)$ in the computational subspace. Here we make this construction explicit by introducing a concrete family of Hamiltonians. Given a target unitary in ${\rm SU}(d)$, we find a path in terms of Gaussian pulses that implements it.

A general single-qudit operation, $U \in {\rm SU}(d)$ can be decomposed in terms of at most $d(d-1) / 2$ operators acting non-trivially only on two consecutive levels~\cite{Li_2013}. This is, $U = \prod_{k}\tilde{U}_k$, with $\tilde{U}_k = \mathds{1}_{j_k - 1}\oplus U_k \oplus \mathds{1}_{d - j_k - 1}$, where $U_k \in {\rm SU}(2)$ acts on a two-dimensional subspace $\mathcal{H}_j$ spanned by $\{\ket{g_{j_k}}, \ket{g_{j_k + 1}}\}$. A general unitary acting on two levels can be performed holonomically with the help of two extra auxiliary atomic levels, $\ket{e_{j_k}}$ and $\ket{p_{j_k}}$ [Fig.~\ref{fig:fig1SM}(a)]. Let us consider the following 4-level atomic Hamiltonian,
\begin{equation}
	\begin{aligned}
		H_{j_k}(\mathbf{\Omega}^{(j_k)}) = \frac{1}{2}\ket{e_{j_k}}&\Big(\Omega^{(j_k)}_{0}\bra{g_{j_k}} + \Omega^{(j_k)}_{1} \bra{g_{j_k+1}}\\
		&+ \Omega^{(j_k)}_{p} \bra{p_{j_k}} \Big) + \text{H.c.},
	\end{aligned}
\end{equation}
where $\mathbf{\Omega}^{(j_k)} = (\Omega^{(j_k)}_{0}, \Omega^{(j_k)}_{1}, \Omega^{(j_k)}_{p})$ denotes a vector of Rabi frequencies that belongs to the parameter manifold $\mathcal{M}$ of the Hamiltonian. In the following, we drop the index $j_k$ to simplify the notation. At every point $\mathbf{\Omega} \in \mathcal{M}$, the eigenstates of $H(\mathbf{\Omega})$ consist of two zero-energy dark states and two states with energies $\pm \Delta(\mathbf{\Omega}) = \pm \sqrt{|\mathbf{\Omega}|^2}$, where a constant non-zero value of $\Omega_p$ guarantees that the gap $\Delta(\mathbf{\Omega})$ remains open. In this situation, one can perform ${\rm SU}(2)$ operations, $u:\,\mathcal{H}_j \to \mathcal{H}_j$, via closed loops in the parameter space, $\gamma_C: [t_0, t_1] \to \mathcal{M}$, with $\gamma(t_0) = \gamma(t_1)$ [Fig.~\ref{fig:fig1SM}(b)]. More specifically, if the loop is traversed adiabatically, $u$ depends only on the geometric properties of the path and is given by ~\cite{Zanardi_1999}
\begin{equation}
	u = {\rm \mathbf{P}}\, {\rm exp} \left(-i\int_{\gamma_C}\sum_\mu A^\mu {\rm d}\Omega_\mu\right),   
\end{equation}
where $(A^\mu)_{ab} = \bra{\psi_a(\mathbf{\Omega})}\partial/\partial \Omega_\mu\ket{\psi_b(\mathbf{\Omega})}$ is the connection and $\ket{\psi_{a=0,1}(\mathbf{\Omega})}$ are the corresponding dark states at every point in $\mathcal{M}$.

By using Gaussian pulses for $\Omega_{a = 0, 1}$ of the form
\begin{equation}
	\label{eq:gaussian_pulses}
	\Omega_{a}(t) = \Omega \, \text{e}^{-(t - \tau_{a})^2/(T/10)^2} \text{e}^{-i\varphi_{a}}
\end{equation}
that overlap in time, where $T$ is the pulse time window, and keeping a constant $\Omega_p \neq 0$ to maintain an open energy gap, the following gates are implemented 
\begin{equation}
	\label{eq:u_delta_gamma}
	u(\alpha,\delta)=
	\begin{pmatrix}
		\cos\alpha & e^{-i\delta}\sin\alpha\\
		-e^{i\delta}\sin\alpha & \cos\alpha
	\end{pmatrix}\,.
\end{equation}
Here, $\delta = \varphi_{1} - \varphi_{0}$ is given by the phase difference between the two laser pulses and $\alpha$ depends on its overlap in time, which can be controlled by $\tau = \tau_{1} - \tau_{0}$ [Fig.~\ref{fig:fig1SM}(c)]. The latter can be tuned between $-\pi / 2$ and $\pi / 2$ by modifying the time delay $\tau$ between the pulses [Fig.~\ref{fig:fig1SM}(d)]. In particular, the case $\delta = \pi/2$ ($\pi$) corresponds to arbitrary rotations around the $X$ ($Y$) axis, respectively. Any ${\rm SU}(2)$ operation can be written in terms of at most three of these rotations. In summary, any single-qudit operation can be implemented using at most $3d(d-1)/2$ pairs of pulses.

\subsection{Experimental errors}

\begin{figure}[t]
	\centering
	\includegraphics[width=0.9\linewidth]{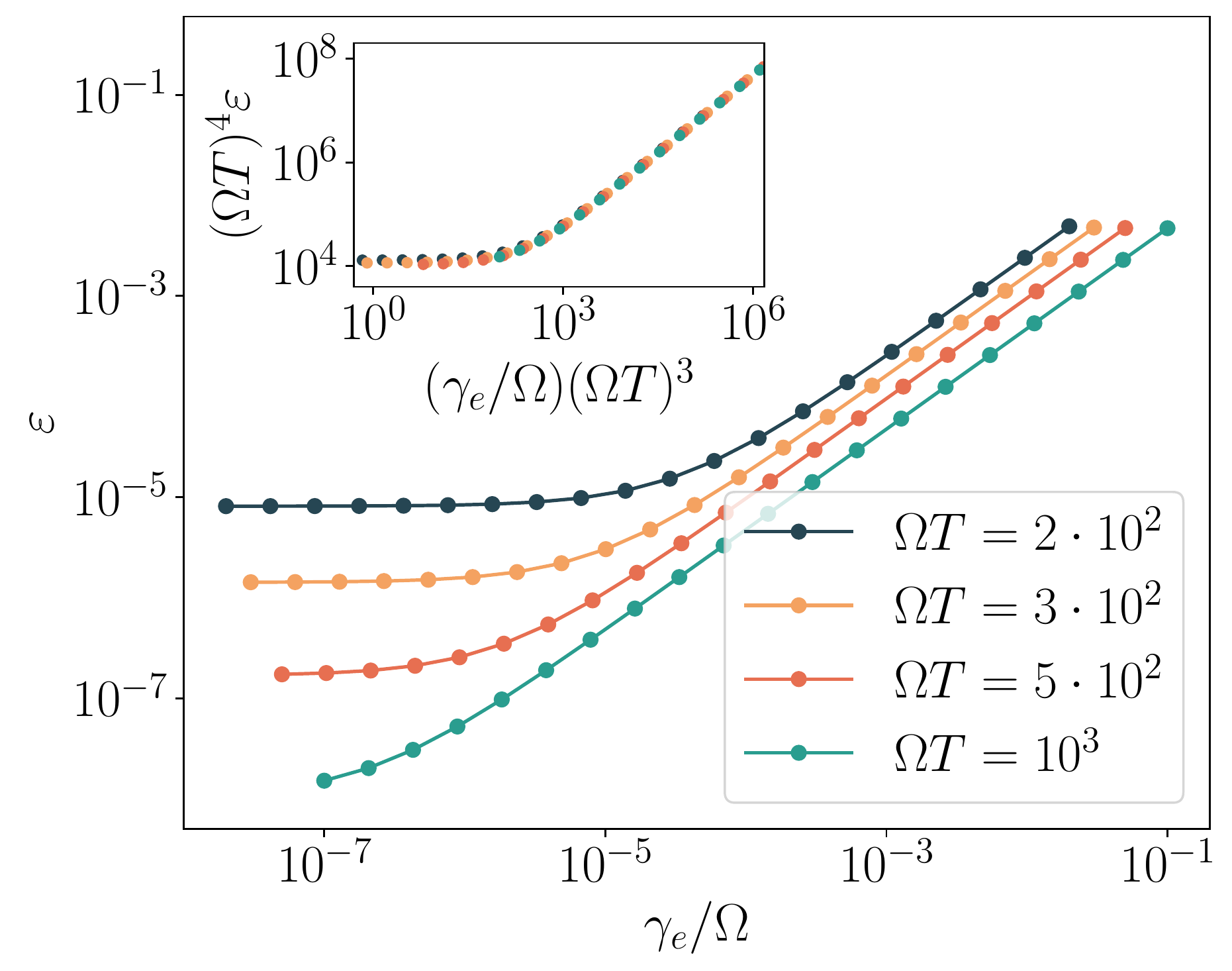}
	\caption{{\bf Single-qudit gate error:} Infidelity $\epsilon$ for the single-qudit magnetic gate $\mathcal{U}^{(B)}_\ell$ required to simulate the $Q_8$ LGT (see main text) for a Trotter time step $\delta t = 0.1$ as a function of $\gamma_e / \Omega$ and $\Omega T$. Inset: The error follows a simple scaling relation, consisting of a power-law and a saturation regime.}
	\label{fig:fig2SM}
\end{figure}

Let us now discuss possible error sources for the single-qudit gates. We emphasize that our protocol involves an excited state $\ket{e}$ with decay rate $\gamma_{e}$. Since $\ket{e}$ is coupled resonantly to the ground-state manifold, the holonomic implementation provides robustness against decoherence caused by spontaneous decay from such state. On the other hand, we impose $\Omega \gg 1 / T$ to avoid errors caused by non-adiabatic state transfer.
This is illustrated in Fig.~\ref{fig:fig2SM}, where we show the simulated error $\epsilon = 1 - \mathcal{F}$ of the single-qudit magnetic gate $\mathcal{U}^{(B)}_\ell$ corresponding to the $Q_8$ LGT, as described below. Here $\mathcal{F}$ denotes the average gate fidelity~\cite{Pedersen_2007}, calculated using the expression
\begin{equation}
	\mathcal{F} = \frac{1}{d(d + 1)}\left[{\rm Tr}(M M^\dagger) + |{\rm Tr}(M)|^2\right],
\end{equation}
with $M = PU^\dagger U_{\rm atom}P$, where $U$ is the target unitary acting non-trvially only on the $d$-dimensional target subspace, $U_{\rm atom}$ is the simulated unitary obtained by solving the time-dependent Schr{\"o}dinger equation for the full pulse sequence on the $(d + 2)$-level atomic system and $P$ is the projection on the qudit subspace.

In the numerical simulation, we take the decay into account by adding non-hermitian terms to the Hamiltonian, $-i\gamma_{e}/2\ket{e}\bra{e}$, such that the evolution contains non-unitary exponentially-decaying terms. As shown in Fig.~\ref{fig:fig2SM}, we find that for a given value of $\gamma_e / \Omega$ the error can be minimized by increasing $\Omega T$ until a saturation point is reached. This sets a mininum clock speed $1 / T$ for the quantum processor that may be increased by allowing for larger holonomic errors. More generally, we expect that both the gate fidelity as well as the clock speed can be further increased using optimal control methods, where other error sources can be taken into account~\cite{Jandura_2022}.

The above results show that single-qudit errors as small as $\sim 10^{-6}$ can be obtained for $\Omega T = 3 \cdot 10^2$ and $\gamma / \Omega = 10^{-6}$. Taking a typical experimental value for the Rabi frequency, $\Omega = 2\pi \times 100$ MHz, we obtain $T\sim 1$~$\mu$s. This leads to an implementation time for general single-qudit gates with $d = 8$ of $\sim 0.1$ ms. Below we take this result into account to estimate the implementation time of a Trotter step for the real-time evolution of the $Q_8$ LGT.

\section{The quaternion group $Q_8$}

\subsection{Hamiltonian LGT}

In this section, we provide more details about the gauge field digitization via the subgroup $Q_8\subset$ SU(2). 
In particular, the general algorithm employed in the main text requires specific facts about the group multiplication as well as the representation theory of $Q_8$, which we summarize below.

The subgroup $Q_8\subset$ SU(2) consists of eight elements represented (up to a sign) by the Pauli matrices $\sigma^{x,y,z}$ and the identity matrix $\sigma^0$,
\begin{align}
	Q_8 = \{\pm \sigma^0, \pm \sigma^z, \pm \sigma^y, \pm \sigma^x \} = \{\pm 1, \pm I, \pm J , \pm K\} \;.
\end{align}
Above we have identified the group elements with the unit quaternions satisfying $I^2 = J^2 = K^2 = IJK = -1$, which explains the common name ``quaternion group'' for $Q_8$.
Explicitly, the group multiplication table reads:
\begin{center}
	\begin{tabular}{ c | c  c  c  c  c  c  c  c }
		& $1$ & $-1$ & $I$ & $-I$ & $J$ & $-J$ & $K$ & $-K$\\ 
		\hline
		$1$ & $1$& $-1$&  $I$&  $-I$&  $J$& $-J$& $K$& $-K$\\  
		$-1$ & $-1$& $1$&  $-I$&  $I$& $-J$ & $J$& $-K$& $K$\\  
		$I$  & $I$& $-I$&  $-1$&  $1$&  $K$& $-K$& $-J$& $J$\\  
		$-I$  & $-I$& $I$&  $1$& $-1$ &  $-K$& $K$& $J$& $-J$\\  
		$J$  & $J$& $-J$&  $-K$&  $K$&  $-1$& $1$& $I$& $-I$\\  
		$-J$  & $-J$& $J$&  $K$&  $-K$&  $1$& $-1$& $-I$& $I$\\  
		$K$  & $K$& $-K$&  $J$&  $-J$&  $-I$& $I$& $-1$& $1$\\  
		$-K$  & $-K$& $K$&  $-J$&  $J$&  $I$& $-I$& $1$& $-1$ 
	\end{tabular}
\end{center}

The group has five conjugacy classes $\{1\}$, $\{-1\}$, $\{\pm I\}$, $\{\pm J\}$, $\{\pm K\}$ and hence five inequivalent  irreducible representations (irreps). These are the trivial irrep $\mathbf{1}$, three one-dimensional ``alternating'' irreps $\mathbf{I}$, $\mathbf{J}$, $\mathbf{K}$, obtained by factoring the normal subgroups generated by $I,J,K$, respectively, and finally $Q_8$ inherits the two-dimensional ``fundamental'' representation $\mathbf{2}$ from SU(2), which is the only one relevant for our purposes. For completeness, the full character table reads:
\begin{center}
	\begin{tabular}{ c | c  c  c  c  c }
		$\chi$ & $\{1\}$ & $\{-1\}$ & $\{\pm I\}$ & $\{\pm J\}$ & $\{\pm K\}$\\ 
		\hline
		$\mathbf{1}$ & 1 & 1 & 1 & 1 & 1\\  
		$\mathbf{I}$ & 1 & 1 & 1 & -1 & -1  \\
		$\mathbf{J}$ & 1 & 1 & -1 & 1 & -1  \\
		$\mathbf{K}$ & 1 & 1 & -1 & -1 & 1  
		\\
		$\mathbf{2}$ & 2 & -2 & 0 & 0 & 0
	\end{tabular}
\end{center}

In a LGT with gauge group $Q_8$, every link carries an eight-dimensional Hilbert space spanned by orthonormal states $\{|j\rangle \}_{j=1}^8\equiv \{|\pm 1\rangle, |\pm I\rangle, |\pm J\rangle, |\pm K\rangle \}$. Group multiplication with an element $h \in Q_8$ corresponds, as for any finite group, to a permutation of the group elements. The unitary operator $\theta(h)$ corresponding to the permutation associated to the element $h \in Q_8$ is given by 
\begin{align}\label{eq:groupmult}
	\theta_{g',g}(h) = \langle g' | \theta(h) | g \rangle = \delta_{g', gh} \;.
\end{align}
The corresponding permutation matrix can be directly read off from the group multiplication table. The operators $\theta(g)$ enter the controlled group multiplication gate $\Theta_{\ell|\ell^\prime}$, which we use to implement the plaquette term. The magnetic gate itself further involves the diagonal matrix
\begin{align}\label{eq:UB}
	\mathcal{U}^{(B)}_{g',g} = \langle g' | \mathcal{U}^{(B)} | g \rangle = \delta_{g', g} f^{(B)}(g) \;, 
\end{align}
where the ``magnetic'' function $f^{(B)}(g) = e^{-2i\lambda_B  \chi^{(\mathbf{2})}(g) \delta t}$ is determined by the character $\chi^{(\mathbf{2})}(g)$ of the irrep $\mathbf{2}$. Here, $\delta t$ denotes the Trotter step size and the coupling $\lambda_B = -\frac{1}{2ag^2}$ is related to the spatial lattice spacing $a$ and the Yang-Mills coupling $g^2$ . Finally, the electric term is derived from analytic continuation of a transfer matrix of the Euclidean path integral in the standard (Wilson) formulation of SU(2) LGT. Explicitly, the matrix $\mathcal{U}^{(E)}_{g',g} = \langle g' | \mathcal{U}^{(E)} | g \rangle = f^{(E)}(g', g)$ and thus the ``electric'' function $f^{(E)}$ is implicitly defined by
\begin{align}\label{eq:UE}
	\mathcal{U}^{(E)}= \exp \left( i \delta t \log T^{(E)} \right) \;, && T^{(E)}_{g', g} = e^{\frac{2}{\lambda_E a_t} \chi^{(\mathbf{2})}(g'g^{-1})} \;,
\end{align}
where the character is evaluated at the product of the group elements $g'$ and the inverse $g^{-1}$ of $g$. Here, the coupling $\lambda_E = \frac{g^2}{a}$ is the one from the Kogut-Susskind formulation, but the transfer matrix necessarily involves a finite temporal lattice spacing $a_t$ because no finite subgroup possesses a continuum limit. For the small system discussed in the main text, we have set $a = a_t = 1$ and work with corresponding dimensionless quantities. We conclude this section by noting that Eqs.~\eqref{eq:groupmult},\eqref{eq:UB},\eqref{eq:UE} directly generalize to arbitrary finite subgroups. In the general case, one only needs to replace $2\chi^{(\mathbf{2})} \rightarrow 2 \text{Re} \left[ \chi^{(\text{fund.})}\right]$ with $\chi^{\text{fund.}}$ the character of the fundamental representation of the gauge group of interest and replace the appropriate group-specific details.

\subsection{Pulse sequence}

We present here an explicit decomposition of the permutation matrices $\theta(h)$ (see Eq.~\eqref{eq:groupmult}) corresponding to $Q_8$ in terms of two-level rotations. Although  permutation matrices in ${\rm SU}(d)$ possess only $d$ non-zero elements, a general decomposition still requires $d(d-1)/2$ two-level unitaries if these are restricted to act between fixed consecutive levels. This is the case for the general encoding of a qudit into $d$ atomic levels belonging to a single hyperfine manifold, due to the constraints imposed by atomic selection rules to connect different levels. However, an all-to-all connectivity between the different $d$ levels is possible for particular cases if we use different hyperfine manifolds (see Fig. 2 of the main text for an example with $d = 8$). For an all-to-all connectivity, permutation matrices can be then decomposed using only $d-1$ two-level unitaries which, for the case of $Q_8$, correspond to
\begin{equation}
	\begin{aligned}
		&\theta(-1) = R^{(0,1)}_{y,2\pi}R^{(1,2)}_{y,\pi}R^{(2,3)}_{y,\pi}R^{(4,5)}_{y,\pi}R^{(5,6)}_{y,2\pi}R^{(6,7)}_{y,\pi} \;,\\
		&\theta(I) = R^{(0,2)}_{y,\pi}R^{(2,1)}_{y,\pi}R^{(1,3)}_{y,\pi}R^{(3,4)}_{y,2\pi}R^{(6,7)}_{y,\pi}R^{(4,6)}_{y,\pi}R^{(7,5)}_{y,\pi}\;,\\
		&\theta(-I) = R^{(0,3)}_{y,\pi}R^{(3,1)}_{y,\pi}R^{(1,2)}_{y,\pi}R^{(2,4)}_{y,2\pi}R^{(4,6)}_{y,\pi}R^{(6,5)}_{y,\pi}R^{(5,7)}_{y,\pi}\;,\\
		&\theta(J) = R^{(0,4)}_{y,\pi}R^{(4,1)}_{y,\pi}R^{(1,5)}_{y,\pi}R^{(5,2)}_{y,2\pi}R^{(2,6)}_{y,\pi}R^{(6,3)}_{y,\pi}R^{(3,7)}_{y,\pi}\;,\\
		&\theta(-J) = R^{(0,5)}_{y,\pi}R^{(5,1)}_{y,\pi}R^{(1,4)}_{y,\pi}R^{(4,2)}_{y,2\pi}R^{(6,7)}_{y,\pi}R^{(2,6)}_{y,\pi}R^{(7,3)}_{y,\pi}\;,\\
		&\theta(K) = R^{(0,6)}_{y,\pi}R^{(6,1)}_{y,\pi}R^{(1,7)}_{y,\pi}R^{(7,2)}_{y,2\pi}R^{(4,5)}_{y,\pi}R^{(2,4)}_{y,\pi}R^{(5,3)}_{y,\pi}\;,\\
		&\theta(-K) = R^{(0,7)}_{y,\pi}R^{(7,1)}_{y,\pi}R^{(1,6)}_{y,\pi}R^{(6,2)}_{y,2\pi}R^{(2,4)}_{y,\pi}R^{(4,3)}_{y,\pi}R^{(3,5)}_{y,\pi} \;.
	\end{aligned}
\end{equation}
This allows us to implement the group-multiplication gate using in total only $2(2d-1)(d-1)$ pairs of pulses.

\subsection{Comparison with a qubit-based approach}

To demonstrate the usefulness of a native qudit realization, we provide an explicit decomposition of the required gates in terms of standard universal single- and two-qubit gates. Here, we focus on the cost of the controlled group multiplication gates that enter the decomposition of the plaquette interaction as one of the main challenges to realize non-abelian lattice gauge theories. We follow a straightforward decomposition of the group multiplication operations into single- and two-qubit gates, not aiming to prove optimality of the decomposition, which could be further improved~\cite{Nam_2018}.

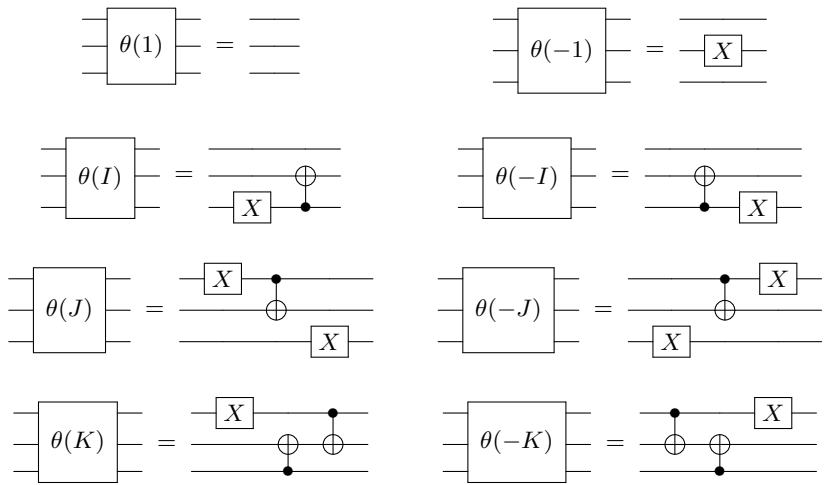
\begin{figure*}[t]
	\centering
	\begin{tabular}{cc}
		\vspace{0.2cm}
		\hspace{0.5cm}
		\Qcircuit @C=1em @R=0.2em {
			& \multigate{2}{\theta(1)} & \qw &  & & \qw & \qw \\
			& \ghost{\theta(1)} & \qw  & =  & &  \qw & \qw\\
			& \ghost{\theta(1)} & \qw  &  &  & \qw & \qw
		}&
		\vspace{0.2cm}
		\hspace{0.5cm}
		\Qcircuit @C=1em @R=0.2em {
			& \multigate{2}{\theta(-1)} & \qw &  & & \qw & \qw \\
			& \ghost{\theta(-1)} & \qw  & =  & & \gate{X} & \qw\\
			& \ghost{\theta(-1)} & \qw  &  & & \qw & \qw
		}
		\\
		\vspace{0.2cm}
		\hspace{0.5cm}
		\Qcircuit @C=1em @R=0.2em {
			& \multigate{2}{\theta(I)} & \qw &  & & \qw & \qw & \qw \\
			& \ghost{\theta(I)} & \qw  & =  & & \qw & \targ & \qw\\
			& \ghost{\theta(I)} & \qw  &  & & \gate{X} & \ctrl{-1} & \qw
		} & 
		\vspace{0.2cm}
		\hspace{0.5cm}
		\Qcircuit @C=1em @R=0.2em {
			& \multigate{2}{\theta(-I)} & \qw &  & & \qw & \qw & \qw & \qw \\
			& \ghost{\theta(-I)} & \qw  & =  & & \qw & \targ & \qw & \qw\\
			& \ghost{\theta(-I)} & \qw  &  & & \qw & \ctrl{-1} & \gate{X} & \qw
		} \\
		\vspace{0.2cm}
		\hspace{0.5cm}
		\Qcircuit @C=1em @R=0.2em {
			& \multigate{2}{\theta(J)} & \qw &  & & \gate{X} & \ctrl{1} & \qw & \qw\\
			& \ghost{\theta(J)} & \qw  & =  & & \qw & \targ & \qw & \qw\\
			& \ghost{\theta(J)} & \qw  &  & & \qw & \qw & \gate{X} &  \qw 
		} &
		\vspace{0.2cm}
		\hspace{0.5cm}
		\Qcircuit @C=1em @R=0.2em {
			& \multigate{2}{\theta(-J)} & \qw &  & & \qw & \ctrl{1} &\gate{X} & \qw \\
			& \ghost{\theta(-J)} & \qw  & =  & & \qw & \targ & \qw & \qw  \\
			& \ghost{\theta(-J)} & \qw  &  & & \gate{X} &  \qw & \qw & \qw  
		} \\
		\vspace{0.2cm}
		\hspace{0.5cm}
		\Qcircuit @C=1em @R=0.2em {
			& \multigate{2}{\theta(K)} & \qw &  & & \gate{X} & \qw &  \ctrl{1} & \qw\\
			& \ghost{\theta(K)} & \qw  & =  & & \qw & \targ w  & \targ & \qw\\
			& \ghost{\theta(K)} & \qw  &  & & \qw & \ctrl{-1}   & \qw & \qw
		}& 
		\vspace{0.2cm}
		\hspace{0.5cm}
		\Qcircuit @C=1em @R=0.2em {
			& \multigate{2}{\theta(-K)} & \qw &  & & \ctrl{1} & \qw  &  \gate{X} & \qw\\
			& \ghost{\theta(-K)} & \qw  & =   & & \targ   & \targ  & \qw & \qw\\
			& \ghost{\theta(-K)} & \qw  &  &     & \qw  &  \ctrl{-1} & \qw  & \qw 
		}
	\end{tabular}
	\caption{\label{fig:group multiplication circuits} Circuit decompositions of the group multiplication gates $\theta(g)$ in terms of single-qubit Pauli-X and two-qubit CNOT gates. The three lines in each diagram correspond to the three qubits $s_1$, $s_2$ and $s_3$ from top to bottom.}
\end{figure*}

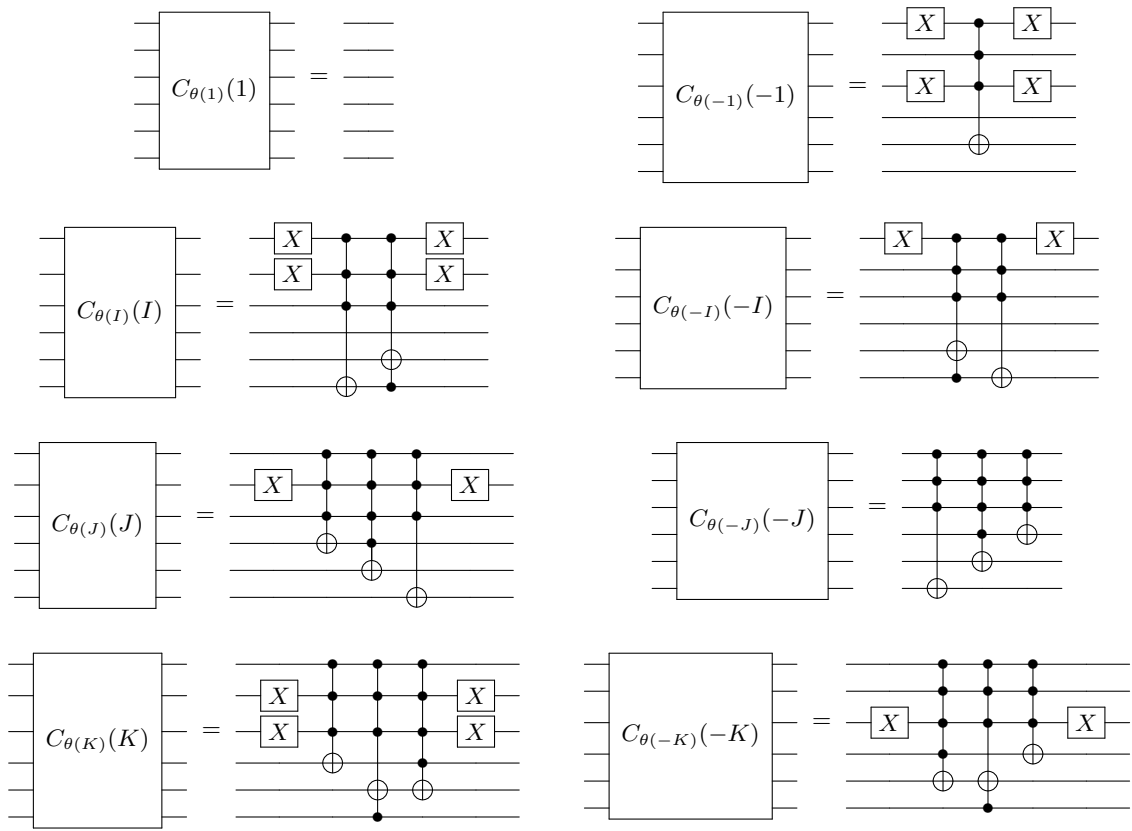
\begin{figure*}[t]
	\centering
	\begin{tabular}{cc}
		\vspace{0.2cm}
		\hspace{0.5cm}
		\Qcircuit @C=1em @R=0.2em {
			& \multigate{5}{C_{\theta(1)}(1)} & \qw &  & & \qw & \qw \\
			& \ghost{C_{\theta(1)}(1)} & \qw  &   & &  \qw & \qw\\
			& \ghost{C_{\theta(1)}(1)} & \qw  & =  & &  \qw & \qw\\
			& \ghost{C_{\theta(1)}(1)} & \qw  &   & &  \qw & \qw\\
			& \ghost{C_{\theta(1)}(1)} & \qw  &   & &  \qw & \qw\\
			& \ghost{C_{\theta(1)}(1)} & \qw  &  &  & \qw & \qw
		}&
		\vspace{0.2cm}
		\hspace{0.5cm}
		\Qcircuit @C=1em @R=0.2em {
			& \multigate{5}{C_{\theta(-1)}(-1)} & \qw &  & & \gate{X} & \ctrl{1} & \gate{X} & \qw \\
			& \ghost{C_{\theta(-1)}(-1)} & \qw  &   & &  \qw & \ctrl{1} & \qw & \qw\\
			& \ghost{C_{\theta(-1)}(-1)} & \qw  & =  & & \gate{X} & \ctrl{2} & \gate{X} & \qw\\
			& \ghost{C_{\theta(-1)}(-1)} & \qw  &   & &  \qw & \qw & \qw & \qw\\
			& \ghost{C_{\theta(-1)}(-1)} & \qw  &   & &  \qw & \targ & \qw & \qw\\
			& \ghost{C_{\theta(-1)}(-1)} & \qw  &  &  & \qw & \qw & \qw & \qw
		}\\
		\vspace{0.2cm}
		\hspace{0.5cm}
		\Qcircuit @C=1em @R=0.2em {
			& \multigate{5}{C_{\theta(I)}(I)} & \qw &  & & \gate{X}  & \ctrl{1}  & \ctrl{1} & \gate{X} & \qw \\
			& \ghost{C_{\theta(I)}(I)} & \qw  &   & &  \gate{X}  & \ctrl{1}  & \ctrl{1} & \gate{X} & \qw\\
			& \ghost{C_{\theta(I)}(I)} & \qw  & =  & & \qw  & \ctrl{3} & \ctrl{2} & \qw & \qw\\
			& \ghost{C_{\theta(I)}(I)} & \qw  &   & &  \qw & \qw & \qw & \qw & \qw\\
			& \ghost{C_{\theta(I)}(I)} & \qw  &   & &  \qw & \qw & \targ & \qw & \qw\\
			& \ghost{C_{\theta(I)}(I)} & \qw  &  &  & \qw & \targ & \ctrl{-1} & \qw  & \qw
		} &
		\vspace{0.2cm}
		\hspace{0.5cm}
		\Qcircuit @C=1em @R=0.2em {
			& \multigate{5}{C_{\theta(-I)}(-I)} & \qw &  & & \gate{X}  &  \ctrl{1}  & \ctrl{1} & \gate{X} & \qw \\
			& \ghost{C_{\theta(-I)}(-I)} & \qw  &   & &  \qw    & \ctrl{1} & \ctrl{1}  & \qw & \qw\\
			& \ghost{C_{\theta(-I)}(-I)} & \qw  & =  & & \qw  &  \ctrl{2} & \ctrl{3} & \qw & \qw\\
			& \ghost{C_{\theta(-I)}(-I)} & \qw  &   & &  \qw  & \qw & \qw & \qw & \qw \\
			& \ghost{C_{\theta(-I)}(-I)} & \qw  &   & &  \qw  & \targ & \qw & \qw & \qw \\
			& \ghost{C_{\theta(-I)}(-I)} & \qw  &  &  & \qw  & \ctrl{-1} & \targ & \qw  & \qw 
		} \\
		\vspace{0.2cm}
		\hspace{0.5cm}
		\Qcircuit @C=1em @R=0.2em {
			& \multigate{5}{C_{\theta(J)}(J)} & \qw &  & & \qw  & \ctrl{1}  & \ctrl{1}  & \ctrl{1} & \qw & \qw \\
			& \ghost{C_{\theta(J)}(J)} & \qw  &   & &  \gate{X}  & \ctrl{1}  & \ctrl{1} &  \ctrl{1} & \gate{X} & \qw\\
			& \ghost{C_{\theta(J)}(J)} & \qw  & =  & & \qw & \ctrl{1} & \ctrl{2} & \ctrl{3} & \qw & \qw\\
			& \ghost{C_{\theta(J)}(J)} & \qw  &   & &  \qw & \targ & \ctrl{1} & \qw & \qw & \qw\\
			& \ghost{C_{\theta(J)}(J)} & \qw  &   & &  \qw & \qw & \targ & \qw & \qw & \qw\\
			& \ghost{C_{\theta(J)}(J)} & \qw  &  &  & \qw  & \qw  & \qw & \targ & \qw & \qw 
		} &
		\vspace{0.2cm}
		\hspace{0.5cm}
		\Qcircuit @C=1em @R=0.2em {
			& \multigate{5}{C_{\theta(-J)}(-J)} & \qw &  &  & \ctrl{1}  & \ctrl{1} & \ctrl{1} & \qw \\
			& \ghost{C_{\theta(-J)}(-J)} & \qw  &   &   & \ctrl{1} & \ctrl{1} & \ctrl{1} &  \qw\\
			& \ghost{C_{\theta(-J)}(-J)} & \qw  & =  &   &  \ctrl{3} & \ctrl{1} &  \ctrl{1}  & \qw\\
			& \ghost{C_{\theta(-J)}(-J)} & \qw  &   & &  \qw & \ctrl{1} & \targ & \qw  \\
			& \ghost{C_{\theta(-J)}(-J)} & \qw  &   & &   \qw & \targ & \qw & \qw\\
			& \ghost{C_{\theta(-J)}(-J)} & \qw  &  &  & \targ & \qw & \qw  & \qw
		} \\
		\vspace{0.2cm}
		\hspace{0.5cm}
		\Qcircuit @C=1em @R=0.2em {
			& \multigate{5}{C_{\theta(K)}(K)} & \qw &  & & \qw  & \ctrl{1}  & \ctrl{1} & \ctrl{1}  & \qw & \qw \\
			& \ghost{C_{\theta(K)}(K)} & \qw  &   & &  \gate{X}  & \ctrl{1}  & \ctrl{1} & \ctrl{1} & \gate{X} & \qw\\
			& \ghost{C_{\theta(K)}(K)} & \qw  &  = & &  \gate{X}  & \ctrl{1}  & \ctrl{2} & \ctrl{1} & \gate{X} & \qw\\
			& \ghost{C_{\theta(K)}(K)} & \qw  &   & &  \qw & \targ & \qw & \ctrl{1} &\qw & \qw \\
			& \ghost{C_{\theta(K)}(K)} & \qw  &   & &  \qw & \qw & \targ  & \targ & \qw & \qw\\
			& \ghost{C_{\theta(K)}(K)} & \qw  &  &  & \qw & \qw & \ctrl{-1} & \qw  & \qw & \qw 
		} &
		\vspace{0.2cm}
		\hspace{0.5cm}
		\Qcircuit @C=1em @R=0.2em {
			& \multigate{5}{C_{\theta(-K)}(-
				K)} & \qw &  & & \qw  & \ctrl{1}  & \ctrl{1} & \ctrl{1}  & \qw & \qw \\
			& \ghost{C_{\theta(-K)}(-
				K)} & \qw  &   & &  \qw  & \ctrl{1}  & \ctrl{1} & \ctrl{1} & \qw & \qw\\
			& \ghost{C_{\theta(-K)}(-
				K)} & \qw  &  = & &  \gate{X}  & \ctrl{1}  & \ctrl{2} & \ctrl{1} & \gate{X} & \qw\\
			& \ghost{C_{\theta(-K)}(-
				K)} & \qw  &   & &  \qw & \ctrl{1} & \qw & \targ &\qw & \qw \\
			& \ghost{C_{\theta(-K)}(-
				K)} & \qw  &   & &  \qw & \targ & \targ  & \qw & \qw & \qw\\
			& \ghost{C_{\theta(-K)}(-
				K)} & \qw  &  &  & \qw & \qw & \ctrl{-1} & \qw  & \qw & \qw 
		} 
	\end{tabular}
	\caption{\label{fig:controlled group multiplication circuits} Circuit decompositions of the controlled gates $C_{\theta(g)}(g)$ in terms of single-qubit Pauli-X and three- and four-qubit Toffoli gates. The six lines in each diagram correspond to the three control qubits, followed by the three target qubits, ordered from top to bottom.}
\end{figure*}

As a first step, we need to choose a computational basis to represent the $8=2^3$ elements of $Q_8$ with three qubits. To minimize the number of entangling operations, it is convenient to ``preserve'' as many subgroups as possible on subsets of the qubits. Apart from the trivial subgroup $\{1\}$, $Q_8$ has a $\mathbb{Z}_2 \simeq \{\pm 1\}$ subgroup and three $\mathbb{Z}_4 \simeq \{\pm 1 , \pm \alpha\}$ subgroups with $\alpha \in \{I,J,K\}$. This motivates to identify the states $|g\rangle$ with three-qubit states $|s_1 s_2 s_3\rangle$ with $s_j\in \{0,1\}$ such that the $\mathbb{Z}_2$ corresponds to a single qubit and two of the three $\mathbb{Z}_4$ subgroups corresponds to the two pairs of qubits that consist of the previously chosen $\mathbb{Z}_2$ qubit and one of the remaining three qubits. One explicit encoding is given by the following list:
\begin{subequations}
	\begin{align}
		g &\leftrightarrow s_1 s_2 s_3 \nonumber\\ 
		+1 &\leftrightarrow 000 \\
		-1 &\leftrightarrow 010 \\
		+I &\leftrightarrow 001 \\
		-I &\leftrightarrow 011 \\
		+J &\leftrightarrow 101 \\
		-J &\leftrightarrow 111 \\
		+K &\leftrightarrow 100 \\
		-K &\leftrightarrow 110 
	\end{align}
\end{subequations}
With this mapping every group-multiplication gate $\theta(g)$ can be realized in terms of CNOT and Pauli-X gates (with at most two each). To find the explicit circuits analytically, note that the $\mathbb{Z}_4$ subgroups are generated by cyclic permutation matrices that implement ``addition mod 4'', which decomposes into a single CNOT gate followed by a single Pauli-X gate according to the identity
\begin{align}
	\begin{pmatrix}
		0 & 1 & 0 & 0\\
		0 & 0 & 1 & 0\\
		0 & 0 & 0 & 1\\
		1 & 0 & 0 & 0
	\end{pmatrix} =  
	\begin{pmatrix}
		1 & 0 & 0 & 0\\
		0 & 0 & 0 & 1\\
		0 & 0 & 1 & 0\\
		0 & 1 & 0 & 0
	\end{pmatrix} 
	\begin{pmatrix}
		0 & 1 & 0 & 0\\
		1 & 0 & 0 & 0\\
		0 & 0 & 0 & 1\\
		0 & 0 & 1 & 0
	\end{pmatrix} \;.
\end{align}
Using this fact, it is straightforward to find the quantum circuits realizing $\theta(g)$ for any $g$ shown in Fig.~\ref{fig:group multiplication circuits}. With these circuits, we can decompose the required group-multiplication gate $\Theta$ (Eq. 7 in the main text). The involved six-qubit gates $C_{\theta(g)}(g)$ are shown in Fig.~\ref{fig:controlled group multiplication circuits}, decomposed into single-qubit Pauli-X gates, as well as the $n$-qubit controlled gates such as the Toffoli gate and its $n$-qubit generalizations. The full circuit is depicted in Fig. 1(a) of the main text.

\begin{figure*}\centering
	\scalebox{0.7}{
		\begin{tabular}{c}
			\vspace{0.2cm}
			\hspace{0.5cm}
			\Qcircuit @C=1em @R=0.8em {
				& \ctrl{1} & \qw &  & & \ctrl{2} & \ctrl{1} & \qw & \ctrl{1} & \qw & \qw \\
				& \ctrl{1} & \qw  & =  & &  \qw & \targ & \ctrl{1} & \targ & \ctrl{1} & \qw\\
				& \gate{U} & \qw  &  & &  \gate{V} & \qw & \gate{V^\dagger} & \qw & \gate{V} & \qw
			}\\
			\vspace{0.2cm}
			\hspace{0.5cm}
			\Qcircuit @C=1em @R=0.8em {
				& \ctrl{1} & \qw &  & & \ctrl{3} & \ctrl{1} & \qw & \ctrl{1} & \qw & \qw & \qw & \ctrl{2} & \qw & \qw  & \qw  & \ctrl{2}  & \qw & \qw \\
				& \ctrl{1} & \qw  & =  & &  \qw & \targ & \ctrl{2} & \targ & \ctrl{2} & \ctrl{1} & \qw & \qw & \qw  & \ctrl{1} & \qw & \qw  & \qw  & \qw \\
				& \ctrl{1} & \qw &  & & \qw & \qw & \qw & \qw & \qw & \targ & \ctrl{1} & \targ & \ctrl{1} & \targ  & \ctrl{1} & \targ & \ctrl{1}  & \qw \\
				& \gate{U} & \qw  &  & &  \gate{V} & \qw & \gate{V^\dagger} & \qw & \gate{V} & \qw & \gate{V^\dagger} & \qw & \gate{V} & \qw & \gate{V^\dagger} & \qw  & \gate{V} & \qw
			}\\
			\vspace{0.2cm}
			\hspace{0.5cm}
			\Qcircuit @C=1em @R=0.8em {
				& \ctrl{1} & \qw &  & & \ctrl{4} & \ctrl{1} & \qw & \ctrl{1} & \qw & \qw & \qw & \ctrl{2} & \qw & \qw  & \qw  & \ctrl{2}  & \qw & \qw & \qw & \ctrl{3} & \qw & \qw & \qw & \ctrl{3} & \qw & \qw & \qw & \ctrl{3} & \qw  & \qw & \qw & \ctrl{3} & \qw & \qw\\
				& \ctrl{1} & \qw  &   & &  \qw & \targ & \ctrl{3} & \targ & \ctrl{3} & \ctrl{1} & \qw & \qw & \qw  & \ctrl{1} & \qw & \qw  & \qw  & \qw  & \qw & \qw & \qw & \ctrl{2} & \qw & \qw & \qw & \qw & \qw & \qw & \qw & \ctrl{2} & \qw & \qw & \qw & \qw\\
				& \ctrl{1} & \qw & = & & \qw & \qw & \qw & \qw & \qw & \targ & \ctrl{2} & \targ & \ctrl{2} & \targ  & \ctrl{2} & \targ & \ctrl{2}  &  \ctrl{1} & \qw & \qw & \qw & \qw & \qw & \qw & \qw & \ctrl{1} & \qw & \qw & \qw & \qw & \qw & \qw & \qw & \qw\\
				& \ctrl{1} & \qw  &  & &  \qw & \qw & \qw & \qw & \qw & \qw & \qw & \qw & \qw & \qw & \qw & \qw  & \qw &  \targ & \ctrl{1} & \targ & \ctrl{1} & \targ & \ctrl{1} & \targ & \ctrl{1} & \targ & \ctrl{1} & \targ & \ctrl{1} & \targ & \ctrl{1} & \targ & \ctrl{1} & \qw
				\\
				& \gate{U} & \qw  &  & &  \gate{V} & \qw & \gate{V^\dagger} & \qw & \gate{V} & \qw & \gate{V^\dagger} & \qw & \gate{V} & \qw & \gate{V^\dagger} & \qw  & \gate{V} & \qw & \gate{V^\dagger} & \qw & \gate{V} & \qw & \gate{V^\dagger} & \qw & \gate{V} & \qw & \gate{V^\dagger} & \qw & \gate{V} & \qw & \gate{V^\dagger} & \qw  & \gate{V} & \qw
			}
		\end{tabular}
	}
	\caption{\label{fig:Toffoli and magnetic} Circuit decompositions of $n$-controlled unitary gates with $n=2,3$ and $4$ in terms of ordinary controlled unitary two-qubit gates with $V^{2^{n-1}} = U$, leading to a quantum cost of $5,13$ and $29$, respectively. For the case $U=X$, i.e. generalized Toffoli gates, it is sufficient to realize two-qubit CNOTs and two-qubit controlled phase gates with $V = 
		\begin{pmatrix} 1 & 0 \\ 0 & e^{i \pi \times 2^{1-n}}
		\end{pmatrix}$ via a Hadamard transformation $H$ of the target qubit since $HXH = Z$.}
\end{figure*}
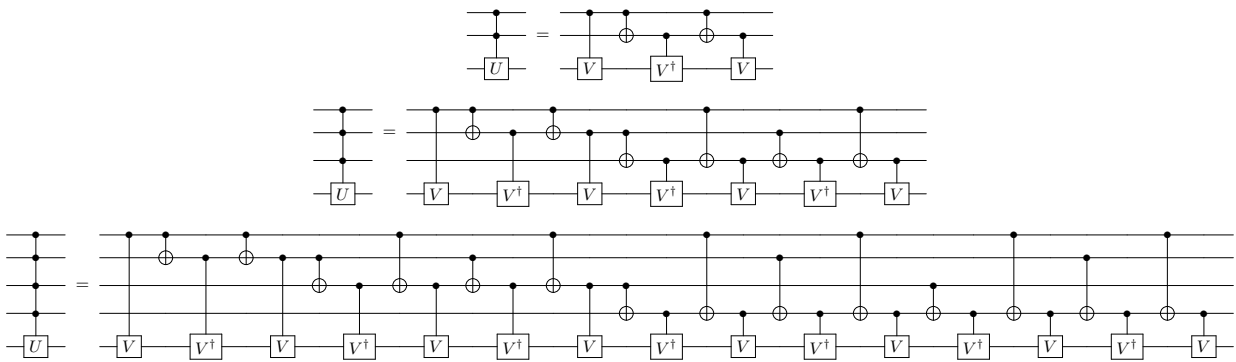

In summary, we find that $9$ four-qubit and $8$ five-qubit Toffoli gates are required for the group multiplication gate.
Using standard decompositions into single qubit and general controlled-unitary two-qubit gates~\cite{barenco1995elementary}, illustrated in Fig.~\ref{fig:Toffoli and magnetic}, this translates into $9\times 13 + 8\times29 = 349$ entangling gates for a single group-multiplication gate. With this decomposition and Eq.~(6) of the main text, together with the decomposition of the required single-qudit ``magnetic'' gate, which can be realized via a $2$-controlled phase gate acting on three qubits, we obtain $5+6\times349 = 2099$ for a single Trotter step of the plaquette interaction. 
We note that even though there might exist more efficient decompositions using less two-qubit gates than the one presented here, we expect that the required number will still be of the same order of magnitude.

Let us now compare the state fidelity of the maximally-entangled state $\ket{\psi_1} = 1/\sqrt{d}\sum_g\ket{g}\ket{g}$ prepared by applying the circuit associated to $\Theta_{\ell|\ell^\prime}$ to the product state $\ket{\psi_0} = 1/\sqrt{d}\sum_g\ket{0}\ket{g}$ using both approaches. As shown in the main text, our qudit protocol leads to a fidelity of $99.6\%$ using realistic experimental parameters. Using the same parameters and a similar protocol based on holonomic operations supported by the Rydberg blockade, we also simulate a single 
two-qubit control gate, obtaining a fidelity of about $99.9\%$. In both cases, we only considered errors in the implementation of the corresponding entangling gates, while single-qubit as well as single-qudit gates are considered without errors.
As discussed in the main text, the errors considered here encompass loss from the excited states $\ket{e}$ and $\ket{r}$, imperfect holonomic operations due to a finite pulse speed, as well as a finite Rydberg interaction strength.
It is clear that, although the error is smaller for the entangling qubit gate, it does not compensate the overhead in circuit depth. In particular, we used circuits related to the ones described above to implement the group-multiplication gate and obtained a state fidelity of $21.4\%$, showing the clear advantage of using qudits to simulate LGTs. For simplicity, here we decomposed all two-qubit control gates into one (for CNOTs) or two (for general two-qubit control gates) elementary single-qubit gates and controlled-$iY$ gates. Note that this increases the number of entangling operations to $9\times (6+2\times 7) + 8\times (14+2\times 15) = 532$, see Fig.~\ref{fig:Toffoli and magnetic}.

Finally, in a Rydberg platform the atoms have to be moved within the Rydberg blockade radius before and after implementing each entangling gate. As shown in \cite{Bluvstein_2021}, in order to avoid decoherence this process should take place at a finite velocity. As indicated in the main text for the implementation of the group-multiplication gate with qudits, this moving time is of the same order of magnitude as the implementation time for the required pulse sequence, leading to a total implementation time of $\sim 1$ ms per Trotter step. In a qubit-based approach, however, the atoms have to be moved back and forth for every entangling two-qubit gate. This leads to an overhead hundreds of times larger, such that only $\sim 1)$ steps can be implemented within the typical experimental coherence time ($\sim 100$ ms), showing again the advantage of a qudit-based approach.

\subsection{Larger groups with ``macro-qudits''}
To faithfully reproduce the physics of continuous gauge groups, such as $SU(N)$, it will be necessary to consider subgroups $G\subset SU(N)$ with large size $D=|G|$ (see, e.g.~\cite{Alexandru_2021} and references therein for recent developments). In this subsection, we give a brief outlook how this can be achieved by blocking several qudits into one composite ``macro-qudit''.

To be specific, consider the case of $SU(2)$. Besides the quaternion group $Q_8$ with $|Q_8|=8$, $SU(2)$ has three relevant subgroups $G$, namely the binary tetrahedral $2T$, octahedral $2O$ and icosahedral $2I$ groups with sizes $24,48$ and $120$, respectively. Since these groups contain themselves $Q_8$ as a subgroup, it seems beneficial to preserve this structure by using one qudit of size $d_1 = 8$, together with a second qudit of size $d_2=3,5$ or $15$, respectively, to represent the local states of the gauge field. That is, if we allow for qudits of size $d\le15$, the three larger subgroups of $SU(2)$ can be represented by a macro-qudit of size $D=d_1 d_2$ that consists of two smaller qudits, i.e. $|g\rangle = |g_1\rangle |g_2 \rangle$ for $g=g_1g_2$, with $g_1 \in G$ and $g_2 \in Q_8$.
For $G=2T$ and $2O$, where $Q_8$ is also normal in $G$, it immediately follows that the operator $\theta(h)$ for $h = (h_1,e)$ with $h_1\in Q_8$ and $e$ the identity element in $G/Q_8$, acts only on the first qudit, $\theta(h)|g_1\rangle  |g_2 \rangle = |g_1 h_1\rangle |g_2\rangle$. In this way, the advantage of using qudits instead of qubits for $|g_1\rangle$ directly carries over to such a macro-qudit, at least for all operations that involve group-multiplication within the $Q_8$ subgroup. 

This simple example illustrates that the qudit advantage can be preserved when blocking into appropriate macro-qudits. While a more thorough comparison of all the required gates for qubit- and qudit-based approaches goes beyond the scope of this work (in particular in the absence of normal subgroups), we plan to study this question in the future.
In this context, it is also important to assess the quality of the approximation for increasing subgroup size, also in comparison to alternative discretizations based on representation basis truncation~\cite{Tong_2022}, as well as the general $SU(N)$  case.

\end{document}